\documentclass[twocolumn,nofootinbib,superscriptaddress,preprintnumbers,floatfix]{revtex4-1}

\pdfoutput=1

\usepackage{amsmath}
\usepackage{amsfonts}
\usepackage{graphicx}
\usepackage[small]{subfigure}
\usepackage[hyperfootnotes=false]{hyperref}
\usepackage{braket}
\usepackage{cancel}
\usepackage{xspace}
\usepackage{fmtcount}

\usepackage{color}
\usepackage{multirow}

\usepackage{marginnote}

\newcommand{\OMIT}[1]{}

\newcommand{\head}[1]{
\vspace{0.3cm}
\noindent {\bf \underline {#1}}
\vspace{0.1cm}
}

\newcommand{\eq}[1]{Eq.~\eqref{eq:#1}}
\newcommand{\eqs}[2]{Eqs.~\eqref{eq:#1} and \eqref{eq:#2}}

\newcommand{\msbar}{\overline{\textrm{MS}}}

\newcommand{\ord}[1]{\mathcal{O}(#1)}

\newcommand{\df}{\mathrm{d}}

\newcommand{\nn}{\nonumber}

\newcommand{\mi}{{\mu}}
\newcommand{\as}{\alpha_s}

\newcommand{\e}{\epsilon}

\newcommand{\V}{V}

\allowdisplaybreaks[4]

\DeclareRobustCommand{\Sec}[1]{Sec.~\ref{#1}}

\DeclareRobustCommand{\App}[1]{App.~\ref{#1}}
\DeclareRobustCommand{\Tab}[1]{Table~\ref{#1}}

\DeclareRobustCommand{\Fig}[1]{Fig.~\ref{#1}}
\DeclareRobustCommand{\Figs}[2]{Figs.~\ref{#1} and \ref{#2}}
\DeclareRobustCommand{\Eq}[1]{Eq.~(\ref{#1})}
\DeclareRobustCommand{\Eqs}[2]{Eqs.~(\ref{#1}) and (\ref{#2})}
\DeclareRobustCommand{\Ref}[1]{Ref.~\cite{#1}}

\newcommand{\be}{\begin{equation}}
\newcommand{\ee}{\end{equation}}

\def\cL{\mathcal{L}}

\begin{document}

\preprint{\vbox{\hbox{UWTHPH 2014-37}\hbox{MIT--CTP 4596}\hbox{LPN14-123}
}}

\title{C-parameter Distribution at N${}^3$LL$^\prime$ including Power Corrections\vspace{0.2cm}}

\author{Andr\'e H.~Hoang}
\affiliation{University of Vienna, Faculty of Physics, Boltzmanngasse 5, A-1090 Wien, Austria}
\affiliation{Erwin Schr\"odinger International Institute for Mathematical Physics,
University of Vienna, Boltzmanngasse 9, A-1090 Vienna, Austria}
\author{Daniel W.~Kolodrubetz}
\affiliation{Center for Theoretical Physics, Massachusetts Institute of
  Technology, Cambridge, MA 02139, USA\vspace{0.2cm}}
\author{Vicent Mateu}
\affiliation{University of Vienna, Faculty of Physics, Boltzmanngasse 5, A-1090 Wien, Austria}
\author{Iain W.~Stewart\,\vspace{0.2cm}}
\affiliation{Center for Theoretical Physics, Massachusetts Institute of
  Technology, Cambridge, MA 02139, USA\vspace{0.2cm}}

\begin{abstract}
We compute the $e^+ e^-$ \mbox{C-parameter} distribution using the Soft-Collinear Effective Theory with a resummation to N${}^3$LL$^\prime$ accuracy of the most singular partonic terms. This includes the  known fixed-order QCD results up to $\ord {\alpha_s^3}$, a numerical determination of the two loop non-logarithmic term of the soft function, and all logarithmic terms in the jet and soft functions up to three loops. Our result holds for $C$ in the peak, tail, and far tail regions. Additionally, we treat hadronization effects using a field theoretic nonperturbative soft function, with moments $\Omega_n$. In order to eliminate an $\ord{\Lambda_{\rm QCD}}$ renormalon ambiguity in the soft function, we switch from the $\overline {\rm MS}$ to a short distance ``Rgap" scheme to define the leading power correction parameter $\Omega_1$. We show how to simultaneously account for running effects in $\Omega_1$ due to renormalon subtractions and hadron-mass effects, enabling power correction universality between C-parameter and thrust to be tested in our setup. We discuss in detail the impact of resummation and renormalon subtractions on the convergence. In the relevant fit region for $\alpha_s(m_Z)$ and $\Omega_1$, the perturbative uncertainty in our cross section is $\simeq 2.5\%$ at $Q=m_Z$.
\end{abstract}

\maketitle

\section{Introduction}
\label{sec:intro}
The study of event shape distributions in $e^+\, e^-$ colliders has served as an
excellent avenue to understand the structure of jets in QCD. Currently they also
provide an important testing ground for new achievements in theoretical formalism, that can then also be extended to applications at hadron colliders. Moreover, event shapes provide accurate determinations of the strong coupling constant; see for example Refs.~\cite{Bethke:2012jm} and \cite{Pich:2013sqa}. Experimentally they have been measured with high accuracy, and there exist perturbative computations at $\mathcal{O}(\alpha_s^3)$~\cite{GehrmannDeRidder:2007bj, GehrmannDeRidder:2007hr, GehrmannDeRidder:2009dp,Ridder:2014wza, Weinzierl:2008iv,Weinzierl:2009ms}. The use of the Soft-Collinear Effective Theory (SCET)~\cite{Bauer:2000ew, Bauer:2000yr, Bauer:2001ct, Bauer:2001yt} has simplified the analysis of factorization theorems for event shapes~\cite{Fleming:2007qr,Fleming:2007xt,Schwartz:2007ib, Bauer:2008dt}, enabling a resummation of large perturbative logs at next-to-next-to-next-to-leading log (N${}^3$LL$^\prime$) order and high-precision analyses for thrust and the heavy jet mass distributions~\cite{Becher:2008cf, Chien:2010kc, Abbate:2010xh, Abbate:2012jh}. The corresponding factorization theorems have recently been extended to oriented event shapes in~\cite{Mateu:2013gya} and to account for the effects of virtual and real secondary massive quark radiation in Refs.~\cite{Gritschacher:2013pha,Pietrulewicz:2014qza}.

Our main motivations for studying the C-parameter distribution are to:
\begin{itemize}
\item[a)] Extend the theoretical precision of the logarithmic resummation for C-parameter from next-to-leading log (NLL)$\,+\,{\cal O}(\alpha_s^3)$ to N$^3$LL\,+\,${\cal O}(\alpha_s^3)$.
\item [b)] Implement the leading power correction $\Omega_1$ using only field theory and with sufficient theoretical precision to provide a serious test of universality between C-parameter and thrust.
\item [c)] Determine $\alpha_s(m_Z)$ using N$^3$LL$^\prime$\,+\,${\cal O}(\alpha_s^3)+\Omega_1$ theoretical precision for $C$, to make this independent extraction competitive with the thrust analysis carried out at this level in Refs.~\cite{Abbate:2010xh,Abbate:2012jh}.
\end{itemize}
In this article we present the theoretical calculation and analysis that yields a N$^3$LL$^\prime$\,+\,${\cal O}(\alpha_s^3)+\Omega_1$ cross section for C-parameter, and we analyze its convergence and perturbative uncertainties.  A numerical analysis that obtains $\alpha_s(m_Z)$ from a fit to a global \mbox{C-parameter} dataset and investigates the power correction universality will be presented in a companion paper~\cite{Hoang:2015hka}. Preliminary versions of these results were presented in Refs.~\cite{DanSCET2013,Hoang:2015gta}.

A nice property of C-parameter is that its definition does not involve any minimization procedure, unlike thrust. This makes its determination from data or Monte Carlo simulations computationally inexpensive.
Unfortunately, this does not translate into a simplification of perturbative theoretical computations, which are similar to those for thrust.

The resummation of singular logarithms in \mbox{C-parameter} was first studied by Catani and Webber in Ref.~\cite{Catani:1998sf} using the coherent branching formalism~\cite{Catani:1992ua}, where NLL accuracy was achieved. Making use of SCET in this article, we achieve a resummation at N$^3$LL order. The relation between thrust and \mbox{C-parameter} in SCET discussed here has been used in the Monte Carlo event generator GENEVA \cite{Alioli:2012fc}, where a next-to-next-to-leading log primed (N$^2$LL$'$) C-parameter result was presented. Nonperturbative effects for the C-parameter distribution have been studied by a number of authors: Gardi and Magnea \cite{Gardi:2003iv}, in the context of the dressed gluon approximation; Korchemsky and Tafat \cite{Korchemsky:2000kp}, in the context of a shape function; and Dokshitzer and Webber \cite{Dokshitzer:1995zt}, in the context of the dispersive model.

Catani and Webber\cite{Catani:1998sf} showed that up to NLL the cross sections for thrust and the reduced C-parameter
\begin{align}
 \widetilde C = \frac{C}{6} \,,
\end{align}
are identical. Gardi and Magnea \cite{Gardi:2003iv} showed that this relation breaks down beyond NLL due to soft radiation
at large angles. Using SCET we confirm and extend these observations by demonstrating that the hard and jet functions, along
with all anomalous dimensions, are identical for thrust and $\widetilde{C}$ to all orders in perturbation theory. At any order in perturbation theory, the perturbative non-universality of the singular terms appears only through fixed-order terms in the soft function, which differ starting at $\mathcal{O}(\alpha_s)$.

There is also a universality between the leading power corrections for thrust and C-parameter which has been widely
discussed~\cite{Dokshitzer:1995zt,Akhoury:1995sp,Korchemsky:1994is,Lee:2006nr}. This universality has been proven nonperturbatively in Ref.~\cite{Lee:2006nr} using the field theory definition of the leading power correction with massless particle kinematics. In our notation this relation is
\begin{align} \label{eq:O1c}
\Omega_1^C = \frac{3\pi}{2}\, \Omega_1^\tau  \,.
\end{align}
Here $\Omega_1^e$ is the first moment of the nonperturbative soft functions for the event shape $e$ and, in the tail of the
distribution acts to shift the event shape variable 
\begin{align} \label{eq:shift}
 \hat\sigma(e)\to \hat\sigma(e - \Omega_1^e/Q)\,,
\end{align}
at leading power.
The exact equality in \eq{O1c}
can be spoiled by hadron-mass effects~\cite{Salam:2001bd}, which have been formulated using a field theoretic definition
of the $\Omega_1^e$ parameters in Ref.~\cite{Mateu:2012nk}.  Even though nonzero hadron masses can yield quite large effects
for some event shapes, the universality breaking corrections between thrust and \mbox{C-parameter} are at the $2.5\%$ level and hence for our purposes are small relative to other uncertainties related to determining $\Omega_1$. Since relations like Eq.~(\ref{eq:O1c}) do not hold for higher moments $\Omega_{n>1}^e$ of the nonperturative soft functions, these are generically different for thrust and C-parameter.

\begin{figure}[t!]
\includegraphics[width=0.95\columnwidth]{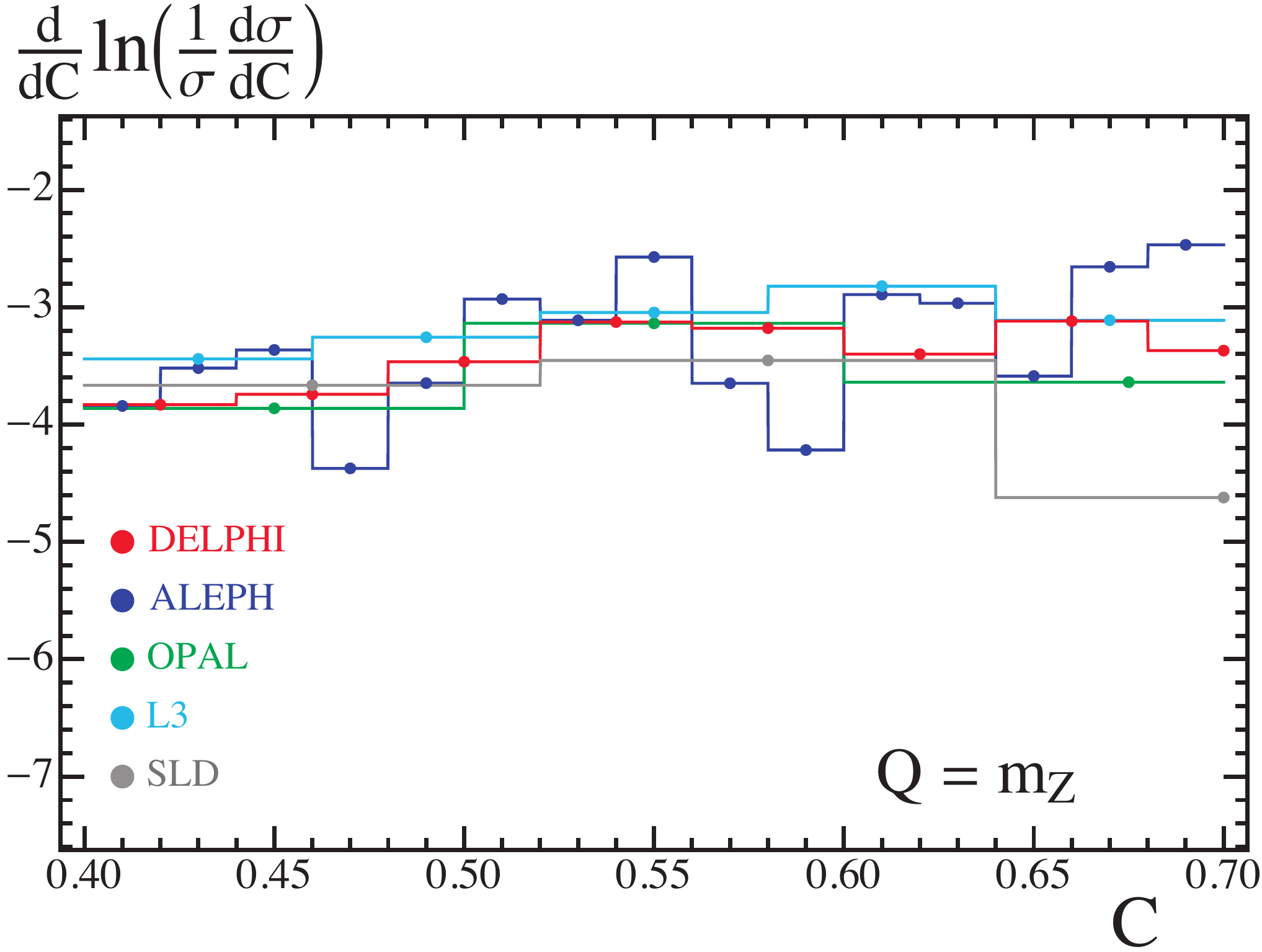}
\caption{\label{fig:Numerical-derivative} Plot of $\df/\df C\, \ln[ (1/\sigma)\df\sigma/\df C]\simeq h'(C)/h(C)$,  computed from experimental data at \mbox{$Q=m_Z$}. The derivative is calculated using the central difference with neighboring points. }
\end{figure}

Following Ref.~\cite{Abbate:2010xh}, a rough estimate of the
impact of power corrections can be obtained from the
experimental data with very little theoretical input.
We write $(1/\sigma)\, \df\sigma/\df C \simeq  h(C-\Omega_1^C/Q) = h(C) - h^\prime(C)\, \Omega_1^C/Q +\ldots$ for the tail region,
and assume the perturbative function $h(C)$ is proportional to $\alpha_s$. Then one can easily derive that if a value $\alpha_s$
is extracted from data by setting $\Omega_1^C=0$ then the change in the extracted value $\delta\alpha_s$ when $\Omega_1^C$
is present will be
\begin{align}
  \frac{\delta\alpha_s}{\alpha_s} \simeq \frac{\Omega_1^C}{Q}\: \frac{h^\prime(C)}{h(C)} \,,
\end{align}
where the slope factor $h'(C)/h(C)$ should be constant at the level of these approximations.  By looking at the experimental results at the Z-pole shown in
Fig.~\ref{fig:Numerical-derivative}, we see that this is true at the level expected from these approximations, finding \mbox{$h'(C)/h(C) \simeq -\,3.3\pm 0.8$}.  This same analysis
for thrust \mbox{$T=1-\tau$} involves a different function $h$ and yields \mbox{$[h'(\tau)/h(\tau)]_\tau \simeq -\,14\pm 4$}~\cite{Abbate:2010xh}.  It is interesting to note that even this very simple analysis
gives a value \mbox{$[\,h'(\tau)/h(\tau)\,]_\tau/[\,h'(C)/ h(C)\,]_C \simeq 4.2$} that is very close to the universality prediction of $3\pi/2 = 4.7$.
In the context of Eq.~(\ref{eq:O1c}), this already implies that in a C-parameter analysis we can anticipate the impact of the power correction in the extraction of the strong coupling to be quite similar to that in
the thrust analysis~\cite{Abbate:2010xh}, where $\delta\alpha_s/\alpha_s \simeq -\,9\%$.

In Ref.~\cite{Catani:1997xc} it was shown that within perturbation theory the C-parameter distribution reaches an infinite
value at a point in the physical spectrum $0 < C < 1$,  despite being an infrared and collinear safe observable.  This happens
for the configuration that distinguishes planar and non-planar partonic events and first occurs at $\mathcal{O}(\alpha_s^2)$ where one has enough partons to create a non-planar event at the value $C_{\rm shoulder} = 3/4$. However, this singularity is
integrable and related to the fact that at $\mathcal{O}(\alpha_s)$ the cross section does not vanish at the three-parton
endpoint $C_{\rm shoulder}$. In Ref.~\cite{Catani:1997xc} this deficiency was cured by performing soft gluon resummation
at $C_{\rm shoulder}$ to achieve a smooth distribution at leading-log (LL) order.
Since $C_{\rm shoulder}$ is far away from the dijet limit (in fact, it is a pure three-jet configuration), we will not include this resummation. In our analysis the shoulder effect is included in the non-singular contributions in fixed-order, and when the partonic distribution is convoluted with a nonperturbative shape function, the shoulder effect is smoothed out, providing a continuous cross section across the entire $C$ range.

This paper is organized as follows: In Sec.~\ref{sec:definition} we review the definition of C-parameter and highlight some
of its properties and relations to other event shapes, and in Sec.~\ref{sec:kinematics} we study simplifications that occur for the calculation of \mbox{C-parameter} when we consider soft and collinear radiation in the dijet limit. These simplifications are necessary to derive the C-parameter factorization theorem and to address the question of which perturbative objects are common with thrust. In Sec.~\ref{sec:factorization-singular} we establish the factorization theorem for $C$ and show how to perform large log resummation using the renormalization group evolution (RGE). Some details on the form of the kinematically power suppressed terms known as nonsingular contributions to the C-parameter distribution are discussed in Sec.~\ref{sec:nonsingular}, and in Sec.~\ref{sec:softtwoloop} we show results for our numerical determination of the non-logarithmic coefficients of the two-loop soft function. Section~\ref{sec:power} contains a presentation of the form of the nonperturbative power corrections and their definition in a renormalon-free scheme. Our method to implement scale variation using profile functions is discussed in Sec.~\ref{sec:profiles}.  In \Sec{sec:results} we present our main results for the C-parameter cross section, including the impact of the resummation slope parameter, the convergence of the resummed perturbation theory, and the perturbative theoretical uncertainties. In Sec.~\ref{sec:conclusions} we give the conclusions and an outlook.

The work contains seven appendices: Appendix~\ref{ap:formulae} provides all needed formulae for the singular cross section beyond those in the main body of the paper. In App.~\ref{ap:MC-comparison} we present a comparison of our SCET prediction, expanded at fixed order, with the numerical results in full QCD at $\mathcal{O}(\alpha_s^2)$ and $\mathcal{O}(\alpha_s^3)$ from EVENT2 and EERAD3, respectively. In App.~\ref{ap:softoneloop} we give a general formula for the one-loop soft function, valid for any event shape which is not recoil sensitive. In App.~\ref{ap:Gijcoefficients} we give analytic expressions for the $B_i$ and $G_{ij}$ coefficients of the fixed-order singular logs up to $\mathcal{O}(\alpha_s^3)$ according to the exponential formula of Sec.~\ref{subsec:Gij}. The R-evolution of the renormalon-free gap parameter is described in App.~\ref{ap:hadronmassR}. Appendix~\ref{ap:ArcTan} is devoted to a discussion of how the Rgap scheme is handled in the shoulder region above $C_{\rm shoulder}=3/4$. Finally, in App.~\ref{ap:subtractionchoice} we show results for the perturbative gap subtraction series based on the C-parameter soft function.
\section{Definition and properties of C-parameter}
\label{sec:definition}
\mbox{C-parameter} is defined in terms of the linearized momentum tensor
\cite{Parisi:1978eg,Donoghue:1979vi},
\begin{align}\label{eq:lintensor}
\Theta^{\alpha\beta}\,=\,\frac{1}{\sum_i |{\vec p}_i|}
\sum_i\frac{p_i^\alpha p_i^\beta}{|{\vec p}_i|}\,,
\end{align}
where $\alpha=1,2,3$ are spacial indices and $i$ sums over all final state particles. Since $\Theta$ is a symmetric positive semi-definite matrix\,\footnote{This property follows trivially from Eq.~(\ref{eq:lintensor}), since for any three-vector $\vec q$ one has that
$q_\alpha q_\beta\Theta^{\alpha\beta}\propto \sum_i({\vec q}\cdot {\vec p}_i)^2/|{\vec p}_i|\ge 0$.},
its eigenvalues are real and non-negative. Let us denote them by
$\lambda_i$, \mbox{$i=1,2,3$}. As $\Theta$ has unit trace, $\sum_i \lambda_i = 1$, which implies that the eigenvalues are bounded $0\le \lambda_i\le 1$.
Without loss of generality we can assume $1\ge \lambda_1 \ge \lambda_2 \ge \lambda_3 \ge 0$. The characteristic polynomial for the eigenvalues of $\Theta$ is:
\begin{align}
x^3 -x^2 + (\lambda_1 \lambda_2 + \lambda_1 \lambda_3 + \lambda_2 \lambda_3)\,x 
- \lambda_1 \lambda_2 \lambda_3 = 0\,.
\end{align}
\mbox{C-parameter} is defined to be proportional to the coefficient of the term linear in $x$:
\begin{align}\label{eq:C-lambda}
C & = 3\, (\lambda_1 \lambda_2 + \lambda_1 \lambda_3 + \lambda_2 \lambda_3)\\
  & = 3\, [\,(\lambda_1+\lambda_2)(1-\lambda_1) - \lambda_2^2\,]\,,\nonumber
\end{align}
where we have used the unit trace property to write $\lambda_3$ in terms of $\lambda_1$
and $\lambda_2$ in order to get the second line. Similarly one defines \mbox{D-parameter}
as $D = 27\,\lambda_1 \lambda_2 \lambda_3$, proportional to the $x$-independent term
in the characteristic equation. Trivially one also finds that
\begin{align}
{\rm Tr}\,[\,\Theta^2\,] = \sum_i \lambda_i^2 = 1 \,-\, \frac{2}{3}\,C\,,
\end{align}
where again we have used $\lambda_3 = 1\,-\,\lambda_1\,-\,\lambda_2$. We can easily compute
${\rm Tr}\,[\,\Theta^2\,]$ using Eq.~(\ref{eq:lintensor})
\begin{align}
{\rm Tr}\,[\,\Theta^2\,] & = \frac{1}{(\sum_i |{\vec p}_i|)^2}
\sum_{ij}\frac{({\vec p}_i\cdot {\vec p}_j)^2}{|{\vec p}_i||{\vec p}_j|}\\\nonumber
& = \frac{1}{(\sum_i |{\vec p}_i|)^2}
\sum_{ij}|{\vec p}_i| |{\vec p}_j| \cos^2 \theta_{ij} \\\nonumber
& = 1 \,-\, \frac{1}{(\sum_i |{\vec p}_i|)^2}
\sum_{ij}|{\vec p}_i| |{\vec p}_j| \sin^2 \theta_{ij}\,.
\end{align}
From the last relation one gets the familiar expression:
 \begin{equation} \label{eq:Ckinematicdef}
 C=\frac{3}{2} \, \frac{\sum_{i,j} | \vec{p}_i | | \vec{p}_j | \sin^2 \theta_{ij}}
 {\left( \sum_i | \vec{p}_i | \right)^2}\,.
 \end{equation}
From Eq.~(\ref{eq:Ckinematicdef}) and the properties of $\lambda_i$, it follows that $C \ge 0$, and from the second line of
Eq.~(\ref{eq:C-lambda}), one finds that $C\le 1$, and the maximum value is achieved for the
symmetric configuration \mbox{$\lambda_1 = \lambda_2 = \lambda_3 = 1/3$}. Hence, \mbox{$0\le C\le 1$}.
Planar events have $\lambda_3 = 0$.  To see this simply consider that the planar event defines
the $x-y$ plane, and then any vector in the $z$ direction is an eigenstate of $\Theta$ with zero eigenvalue.
Hence, planar events have $D=0$ and $C=3\,\lambda_1\,(1-\lambda_1)$, which gives a maximum value for
$\lambda_1 = \lambda_2 = 1/2$, and
one has $0\le C_{\rm planar} \le 3/4$. Thus $C>3/4$ needs at least four particles in the final state.
\mbox{C-parameter} is related to the first non-trivial Fox--Wolfram parameter~\cite{Fox:1978vu}. The Fox--Wolfram event
shapes are defined as follows:
\begin{align}\label{eq:Fox-Wolfram}
H_\ell = \sum_{ij}\frac{| \vec{p}_i | | \vec{p}_j |}{Q^2}\,P_\ell (\cos \theta_{ij})\,.
\end{align}
One has $H_0=1$, $H_1=0$, and
\begin{align}
H_2=1-\frac{3}{2}\,\frac{1}{Q^2}\sum_{ij}| \vec{p}_i | | \vec{p}_j |\sin^2 \theta_{ij}\,,
\end{align}
which is similar to Eq.~(\ref{eq:Ckinematicdef}). It turns out that for massless partonic particles
they are related in a simple way: \mbox{$H_2 \,=\, 1 \,-\, C_{\rm part}$}. As a closing remark, we note that for
massless particles \mbox{C-parameter} can be easily expressed in terms of scalar products with four vectors:
\begin{align}
C_{\rm part} = 3-\frac{3}{2}\,\frac{1}{Q^2}\sum_{ij}\frac{(p_i\cdot p_j)^2}{E_iE_j}\,.
\end{align}
\section{C-Parameter Kinematics in the Dijet Limit}
\label{sec:kinematics}
We now show that in a dijet configuration with only soft\,\footnote{In this paper, for simplicity we denote as soft particles what are sometimes called ultrasoft particles in SCET$_{\rm I}$.}, $n$-collinear, and $\bar n$-collinear particles, the value of \mbox{C-parameter} can, up to corrections of higher power in the SCET counting parameter $\lambda$, be written as the sum of contributions from these three kinds of particles:
\begin{align}\label{eq:C-split}
C_{\rm dijet} = C_n + C_{\bar n} + C_{\rm s} + \mathcal{O}(\lambda^4)\,.
\end{align}
To that end we define
\begin{align}\label{eq:C-decomposition}
C_{\rm dijet} & = C_{n,n} + C_{\bar n,\bar n} + 2\,C_{n, \bar n} +
2\,C_{n, \rm s} + 2\,C_{\bar n, \rm s} + C_{\rm s,s}\,,\nonumber\\
C_{a,b} & \equiv \frac{3}{2}\, \frac{1}{\left( \sum_i | \vec{p}_i | \right)^2}\!\!
 \sum_{i\in a,j\in b} \! | \vec{p}_i || \vec{p}_j | \sin^2 \theta_{ij}\,.
\end{align}
The various factors of $2$ take into account that for $a\neq b$ one has to add the symmetric
term $a \leftrightarrow b$.

The $\rm SCET_I$ power counting rules imply the following scaling for momenta:
$p_{\rm s}^\mu \sim Q\, \lambda^2$,
\mbox{$p_n\sim Q\,(\lambda^2,1,\lambda)$}, \mbox{$p_{\bar n}\sim Q\,(1,\lambda^2,\lambda)$}, where we use the
light-cone components $p^\mu = (p^+,p^-,p_\perp)$. Each one of the terms in \eq{C-decomposition}, as well
as $C_{\rm dijet}$ itself, can be expanded in powers of $\lambda$:
\begin{align}
C_{a,b} = \sum_{i=0} C_{a,b}^{(i)}\,,\quad C_{a,b}^{(i)} \sim \mathcal{O}(\lambda^{i+2})\,.
\end{align}
The power counting implies that $C_{\rm dijet}$ starts at $\mathcal{O}(\lambda^2)$ and
$C_{\rm s, \rm s}$ is a power correction since $C_{s,s}^{(0)}=0$, while $C_{s,s}^{(2)}\ne 0$. All $n-$ and ${\bar n}$-collinear particles together will be denoted as the collinear particles with $c = n \cup \bar n$. The collinear particles have masses much smaller than $Q\,\lambda$ and can be taken as massless at leading power. For soft particles we have perturbative components that can be treated as massless when $Q\lambda^2 \gg \Lambda_{\rm QCD}$, and nonperturbative components that always should be treated as massive. Also at leading order one can use $\sum_i | \vec{p}_i | = \sum_{i \in \rm c} | \vec{p}_i | + \mathcal{O}(\lambda^2) = Q + \mathcal{O}(\lambda^2)$ 
and $| \vec{p}_{i \in \rm c} | = E_{i \in \rm c} + \mathcal{O}(\lambda^2)$.

\vspace*{0.2cm}
Defining $C_s \equiv 2\,C_{n, \rm s} + 2\,C_{\bar n, \rm s}$ we find,
\begin{align} \label{eq:Cs}
C_{s} & \,=\, \frac{3}{Q^2}\!\!\! \sum_{i\in {\rm c},j\in {\rm s}}\!\!\!
| \vec{p}_i | | \vec{p}_j | \sin^2 \theta_{ij}\\\nonumber
& = \frac{3}{Q^2}\!\sum_{i \in \rm c} |\vec{p}_i|
\sum_{j \in \rm s} | \vec{p}_j | \big[\sin^2 \theta_{j} + O(\lambda^2)\,\big]\\\nonumber
& = \frac{3}{Q}\!\sum_{j \in \rm s} \frac{(p_j^\perp)^2}{| \vec{p}_j |} + O(\lambda^4)
= \frac{3}{Q}\sum_{j \in \rm s} \frac{p_j^\perp}{\cosh \eta_j} + O(\lambda^4)\,,
\end{align}
where the last displayed term will be denoted as $C_s^{(0)}$. Here $\theta_j$ is the angle between the three-momenta of a particle and the thrust axis and hence is directly related to the pseudorapidity $\eta_j$. Also $p_j^\perp \equiv | \vec{p}^{\,\perp}_{j} |$ is the magnitude of the three-momentum projection normal to the thrust axis. To get to the second line, we have used that $\sin \theta_{ij} = \sin \theta_{j} \,+\, O(\lambda^2)$, and to get the last line, we have used $\sin \theta_j = p^\perp_j/| \vec{p}_j\, |=1/\cosh\eta_j$. In order to compute the partonic soft function, it is useful to consider $C_s^{(0)}$ for the case of massless particles:
\begin{align}\label{eq:C-soft-massless}
C_s^{(0)}\Big|_{\rm m = 0} = \frac{6}{Q}\sum_{j \in \rm s} \frac{p_j^+ p_j^-}{p_j^+ + p_j^-}\,.
\end{align}

Let us next consider $C_{n,n}$ and $C_{\bar n, \bar n}$. Using energy conservation and momentum conservation in the thrust direction one can show that, up to $\mathcal{O}(\lambda^2)$, \mbox{$E_n = E_{\bar n} = Q/2$}. All $n$-collinear particles are in the plus-hemisphere, and all the $\bar n$-collinear particles are in the minus-hemisphere. Here the plus- and minus-hemispheres are defined by the  thrust axis. For later convenience we define \mbox{$\mathbb{P}_a^\mu = \sum_{i\in a} p_i^\mu$} and $E_a = \mathbb{P}_a^0$ with $a \in \{n, {\bar n}\}$ denoting the set of collinear particles in each hemisphere. We also define $s_a = \mathbb{P}_a^2$.

For $C_{n,n}$ one finds
\begin{align}\label{eq:Cnn}
& C_{n,n} = \frac{3}{2Q^2} \!\!\sum_{i,j\in n}| \vec{p}_i | | \vec{p}_j |
(1-\cos\theta_{ij})(1+\cos\theta_{ij})\\\nonumber
&= [\,1 + O(\lambda^2)\,]\,\frac{3}{Q^2}\!\!\sum_{i,j\in n}p_i\cdot p_j =
\frac{3}{Q^2} \bigg(\sum_{i\in n}p_i\bigg)^{\!\!2} + O(\lambda^4)\\\nonumber
&= \frac{3}{Q^2}\,s_n + O(\lambda^4) = \frac{3}{Q}\sum_{i\in n}p_i^+ + O(\lambda^4) \\\nonumber
&=
\frac{3}{Q}\,\mathbb{P}_n^+ + O(\lambda^4)\,,
\end{align}
and we can identify $C_{n,n}^{(0)}=3\, \mathbb{P}_n^+ /Q$. To get to the second line, we have used that for collinear particles in the same direction $\cos\theta_{ij} = 1 \,+\, \mathcal{O}(\lambda^2)$. In the third line, we use the property that the total perpendicular momenta of each hemisphere is exactly zero and that $\vec 0 = \sum_{i\in +}\vec p_i^{\,\perp} = \sum_{i\in n}\vec p_i^{\,\perp} + \mathcal{O}(\lambda^2)$ and $s_n = Q \, \mathbb{P}_n^+$. In a completely analogous way, we get
\begin{align}\label{eq:Cnbarnbar}
C_{\bar n,\bar n}^{(0)} \, 
=\, \frac{3}{Q}\sum_{i\in \bar n}p_i^- = \frac{3}{Q}\,\mathbb{P}_{\bar n}^-\,.
\end{align}
The last configuration to consider is $C_{n,\bar n}$:
\begin{align}\label{eq:Cnnbar}
2\,C_{n,\bar n}  
&= \frac{3}{Q^2} \!\!\!\sum_{i,j\in n, \bar n}\!\!\! | \vec{p}_i | | \vec{p}_j |
(1-\cos\theta_{ij})(1+\cos\theta_{ij})\\\nonumber
&=\frac{6}{Q^2}\!\!\sum_{i,j\in n}(2E_i E_j - p_i\cdot p_j) 
  [\,1 + O(\lambda^2)\,]\,
  \\\nonumber
&= \frac{3}{Q^2}\, (\,\mathbb{P}_n^+\,\mathbb{P}_{\bar n}^+ +
\mathbb{P}_n^-\,\mathbb{P}_{\bar n}^-
 -2\,\mathbb{P}_n^\perp\cdot \mathbb{P}_{\bar n}^\perp\,)+ O(\lambda^4)\\\nonumber
&= \frac{3}{Q}\,(\,\mathbb{P}_n^+ + \mathbb{P}_{\bar n}^-\,) + O(\lambda^4)
 = C_{n,n}^{(0)} + C_{\bar n,\bar n}^{(0)}+ O(\lambda^4). \nn
\end{align}
In the second equality, we have used that for collinear particles in opposite directions $\cos\theta_{ij} = -\, 1 + \mathcal{O}(\lambda^2)$; in the third equality, we have written $2E = p^+ + p^-$; and in the fourth equality, we have discarded the scalar product of perpendicular momenta since it is $\mathcal{O}(\lambda^4)$ and also used that at leading order $\mathbb{P}_n^- = \mathbb{P}_{\bar n}^+ = Q + \mathcal{O}(\lambda^2)$ and $\mathbb{P}_n^- \sim \lambda^2$, $\mathbb{P}_{\bar n}^+\sim \lambda^2$. For the final equality, we use the results obtained in Eqs.~(\ref{eq:Cnn}) and (\ref{eq:Cnbarnbar}). Because the final result in \Eq{eq:Cnnbar} just doubles those from \Eqs{eq:Cnn}{eq:Cnbarnbar}, we can define
\begin{align} \label{eq:Cndef}
 C_n^{(0)} &\equiv 2\, C_{n,n}^{(0)} \,,
 & C_{\bar n}^{(0)} &\equiv 2\, C_{\bar n,\bar n}^{(0)} \,.
\end{align}
Using Eqs.~(\ref{eq:Cnn}), (\ref{eq:Cnbarnbar}), and (\ref{eq:Cnnbar}), we then have
\begin{align}\label{eq:C-col}
\!C_n^{(0)} \,=\, \frac{6}{Q}\, & \mathbb{P}_n^+\,, \quad
C_{\bar n}^{(0)} \,=\, \frac{6}{Q}\, \mathbb{P}_{\bar n}^-\,.
\end{align}
Equation~(\ref{eq:Cs}) together with \Eq{eq:C-col} finalize the proof of Eq.~(\ref{eq:C-split}). As a final comment, we note that one can express $p^{\pm} = p^\perp \exp(\mp\, \eta)$, and since for $n$-collinear particles $2\cosh \eta = \exp(\eta) [\,1\,+\,{\cal O}(\lambda^2)\,]$ whereas for $\bar n$-collinear particles $2\cosh \eta = \exp(-\,\eta)[\,1\,+\,{\cal O}(\lambda^2)\,]$, one can also write
\begin{equation}
C_n^{(0)} = \frac{3}{Q} \sum_{i \in n} \frac{p_i^\perp}{\cosh \eta_i}\,,\quad
C^{(0)}_{\bar{n}} = \frac{3}{Q}\sum_{i \in \bar{n}} \frac{p_i^\perp}{\cosh \eta_i}\,,
\end{equation}
such that the same master formula applies for soft and collinear particles in the dijet limit, and we can write 
\begin{align}
C^{(0)}_{\rm dijet}= \frac{3}{Q} \sum_{i} \frac{p_i^\perp}{\cosh \eta_i} \,.
\end{align}

\section{Factorization and Resummation}
\label{sec:factorization-singular}
The result in Eq.~(\ref{eq:C-split}) leads to a factorization in terms of hard, jet, and soft functions. The dominant nonperturbative corrections at the order at which we are working come from the soft function and can be factorized with the following formula in the $\overline{\rm MS}$ scheme for the power corrections~\cite{Korchemsky:2000kp,Hoang:2007vb,Ligeti:2008ac}:
\begin{align}\label{eq:singular-nonperturbative}
\frac{1}{\sigma_0}\frac{\df \sigma}{\df C} &= \!\int \!\df p \,\frac{1}{\sigma_0}
\frac{\df \hat\sigma}{\df C}\Big(C-\frac{p}{Q}\Big)F_C(p)\,,\\
\frac{\df \hat\sigma}{\df C} & \,=\, \frac{\df {\hat\sigma}_{\rm s}}{\df C} \,+\,
\frac{\df {\hat\sigma}_{\rm ns}}{\df C}\,.\nn
\end{align}
Here $F_C$ is a shape function describing hadronic effects, and whose first moment $\Omega_1^C$ is the
leading nonperturbative power correction in the tail of the distribution. $\Omega_1^C$ and $\Omega_1^\tau$
are related to each other, as will be discussed further along with other aspects of power corrections in Sec.~\ref{sec:power}. The terms $\df \hat\sigma/\df C$, $\df \hat\sigma_{\rm s}/\df C$, and $\df \hat\sigma_{\rm ns}/\df C$ are the total partonic cross section and the singular and nonsingular contributions, respectively. The latter will be discussed in Sec.~\ref{sec:nonsingular}.

After having shown Eq.~(\ref{eq:C-split}), we can use the general results of Ref.~\cite{Bauer:2008dt} for the factorization theorem for the singular terms of the partonic cross section that splits into a sum of soft and collinear components. One finds
\begin{align}\label{eq:factorization-partonic-singular}
\frac{1}{\sigma_0}\frac{\df \hat\sigma_{\rm s}}{\df C}
  &=\frac{1}{6} \frac{1}{\sigma_0}\frac{\df \hat\sigma_{\rm s}}{\df \widetilde C}
   \\
  &=\frac{Q}{6} H(Q,\mu)\!
\int\! \df s\, J_\tau(s,\mu) \hat S_{\widetilde C}\Big( Q \widetilde C- \frac{s}{Q},\mu\Big)\,,
  \nn
\end{align}
where in order to make the connection to thrust more explicit we have switched to the variable $\widetilde C = C/6$. Here $J_\tau$ is the thrust jet function which is obtained by the convolution of the two hemisphere jet functions and where our definition for $J_\tau$ coincides with that of Ref.~\cite{Abbate:2010xh}. It describes the collinear radiation in the direction of the two jets. Expressions up to $\mathcal{O}(\alpha_s^2)$ and the logarithmic terms determined by its anomalous dimension at three loops are summarized in \App{ap:formulae}.

The hard factor $H$ contains short-distance QCD effects and is obtained from the Wilson
coefficient of the SCET to QCD matching for the vector and axial vector currents. The hard function is the same
for all event shapes, and its expression up to $\mathcal{O}(\alpha_s^3)$
is summarized in \App{ap:formulae}, together with the full anomalous dimension for $H$  at three loops.

The soft function $S_{\widetilde C}$ describes wide-angle soft radiation between the
two jets. It is defined as
\begin{align}
 S_{C}(\ell,\mu) & = \frac{1}{N_c} \big\langle \, 0\,\big|\, {\rm tr}\: \overline{Y}_{\bar n}^T Y_n
    \delta(\ell- Q{\widehat C}) Y_n^\dagger \overline{Y}_{\bar n}^*\, \big|\, 0\,\big\rangle\,,\\
  S_{\widetilde C}(\ell,\mu) 
    & = \frac{1}{N_c}  \big\langle \, 0\,\big|\, {\rm tr}\: \overline{Y}_{\bar n}^T Y_n
    \delta\Big(\ell- \frac{Q{\widehat C}}{6}\Big) Y_n^\dagger \overline{Y}_{\bar n}^*\, \big|\, 0\,\big\rangle
    \nn\\
    & = 6\,S_{C}(6\,\ell,\mu) 
    , \nn
\end{align}
where $Y_n^\dagger$ is a Wilson line in the fundamental representation from $0$ to $\infty$ and $\overline Y_{\bar n}^\dagger$ is a Wilson lines in the anti-fundamental representation from $0$ to $\infty$. Here $\widehat C$ is an operator whose eigenvalues on physical states correspond to the value of \mbox{C-parameter} for that state:
$\widehat C \,|X\rangle = C(X)\, |X\rangle$. Since the hard and jet functions are the same as for
thrust, the anomalous dimension of the C-parameter soft function has to coincide with the anomalous
dimension of the thrust soft function to all orders in $\alpha_s$ by consistency of the RGE. This allows us to determine all logarithmic terms of $S_C$ up to ${\cal O}(\alpha_s^3)$.  Hence one only needs to determine the
non-logarithmic terms of $S_C$. We compute it analytically at one loop and use EVENT2 to numerically determine the two-loop constant, $s_2^{\widetilde C}$. The three-loop constant $s_3^{\widetilde C}$ is currently not known and we estimate it with a Pad\'e, assigning a very conservative error. We vary this constant in our theoretical uncertainty analysis, but it only has a noticeable impact in the peak region.

In Eq.~(\ref{eq:factorization-partonic-singular}) the hard, jet, and
soft functions are evaluated at a common scale $\mu$. There is no
choice that simultaneously minimizes the logarithms of these three
matrix elements. One can use the renormalization group equations to evolve to $\mu$ from
the scales $\mu_H\sim Q$, $\mu_J\sim Q \sqrt{\widetilde C}$, and
$\mu_S\sim Q \widetilde C$ at which logs are minimized in each piece. In this way large
logs of ratios of scales are summed up in the renormalization group
factors:
\begin{align}\label{eq:singular-resummation}
&\frac{1}{\sigma_0}\frac{\df \hat\sigma_{\rm s}}{\df  C} =
\frac{Q}{6} H(Q,\mu_H)\,U_H(Q,\mu_H,\mu)\! \int\! \df s\, \df s^\prime\, \df k\\\nonumber
&\times\!\,
J_\tau(s,\mu_J)\,U_J^\tau(s-s^\prime,\mu,\mu_J)\,U_S^\tau(k,\mu,\mu_S) \\\nonumber
&\times\!
e^{-\,3\pi\frac{\delta(R,\mu_{\!s})}{Q}\frac{\partial}{\partial C}}
\hat S_{\widetilde C}\bigg(\!\frac{Q C-3\pi\bar\Delta(R,\mu_S)}{6}- \frac{s}{Q}-k,\mu_S\!\bigg).
\end{align}
The terms $\delta$ and $\Delta$ are related to the definition of the leading power correction in a
renormalon-free scheme, as explained in Sec.~\ref{sec:power} below.

\section{Nonsingular Terms}
\label{sec:nonsingular}

We include the kinematically power suppressed terms in the C-parameter distribution using the nonsingular
partonic distribution, $\mathrm{d} \hat{\sigma}_{\rm ns} / \mathrm{d}C$. We calculate the nonsingular distribution using
\begin{align} \label{eq:nssubt}
\frac{\df \hat{\sigma}_{\rm ns}}{\df C}(Q,\mu_{\rm ns}) \,=\, \frac{\df \hat{\sigma}_{\text{full}}^{\text{FO}}}{\df C}(Q,\mu_{\rm ns}) \,-\,
 \frac{\df \hat{\sigma}_{s}^{\text{FO}}}{\df C}(Q,\mu_{\rm ns})\,.
\end{align}
Here $\df \hat{\sigma}_{s}^{\text{FO}}/\df C$ is obtained by using Eq.~(\ref{eq:singular-resummation}) with $\mu = \mu_H = \mu_J = \mu_S = \mu_{\rm ns}$.
This nonsingular distribution is independent of the scale $\mu_{\rm ns}$ order by order in perturbation theory as an expansion in $\alpha_s(\mu_{\rm ns})$.
We can identify the nontrivial ingredients in the nonsingular distribution by choosing $\mu_{\rm ns} = Q$ to give
\begin{align} \label{eq:nonsingular-expansion}
\frac{1}{\sigma_0} \frac{\df \hat{\sigma}_{\rm ns}}{\df C} & \,=\, \frac{\alpha_s(Q)}{2\pi}\, f_1(C)\,+\,\left(\frac{\alpha_s(Q)}{2\pi}\right)^{\!\!2}\!\! f_2(C)\\
&+\,\left(\frac{\alpha_s(Q)}{2\pi}\right)^{\!\!3}\!\! f_3(C) \,+\,\ldots\nonumber
\end{align}
We can calculate each $f_i(C)$ using an order-by-order subtraction of the fixed-order singular distribution from the full fixed-order distribution as displayed in \Eq{eq:nssubt}.
\begin{figure}
\begin{center}
\includegraphics[width=0.95\columnwidth]{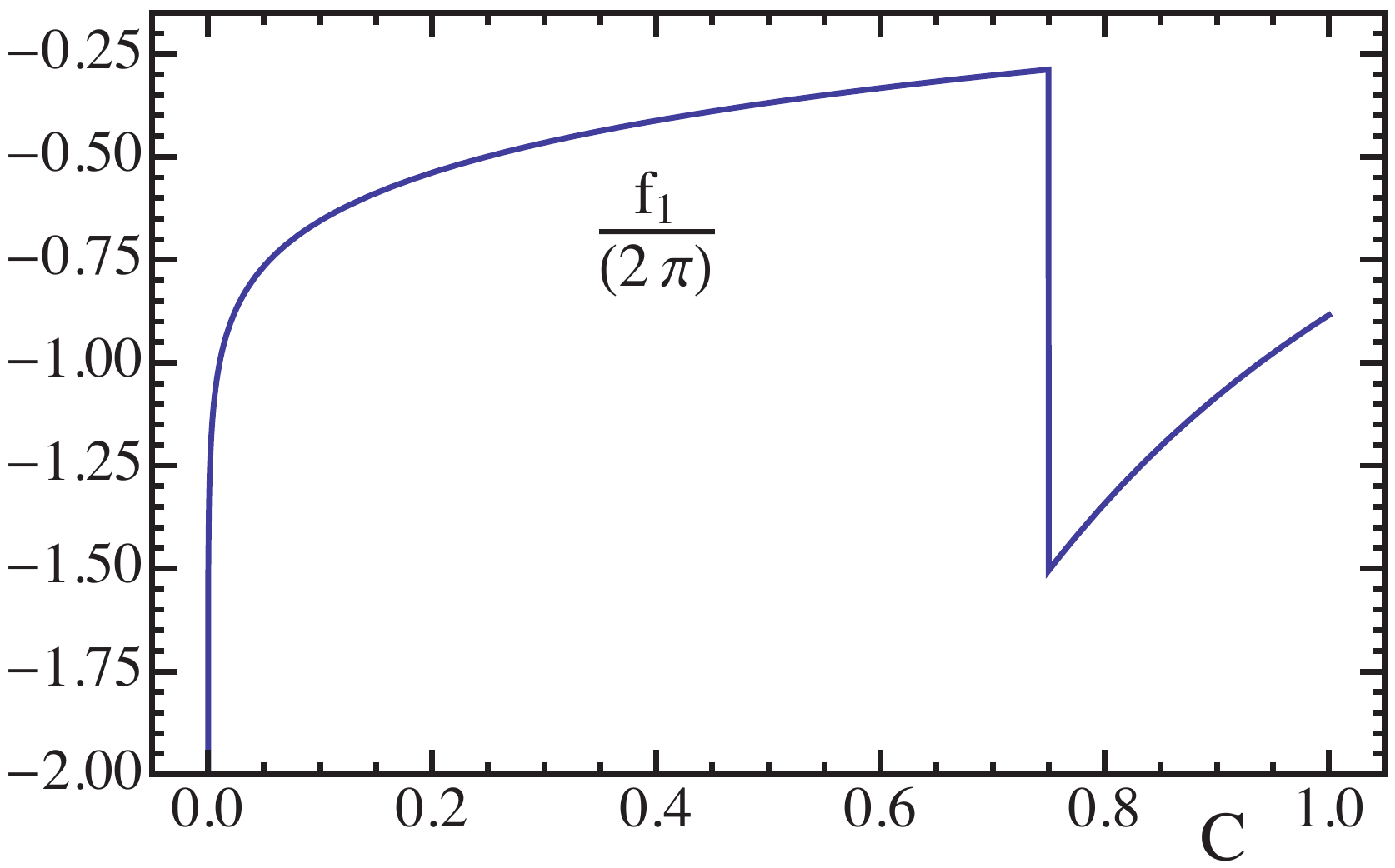}
\caption{$\ord {\alpha_s}$ nonsingular C-parameter distribution, corresponding to Eq.~(\ref{eq:1NS}).}
  \label{fig:1-loopNS}
\end{center}
\end{figure}

\begin{figure}
\begin{center}
\includegraphics[width=0.95\columnwidth]{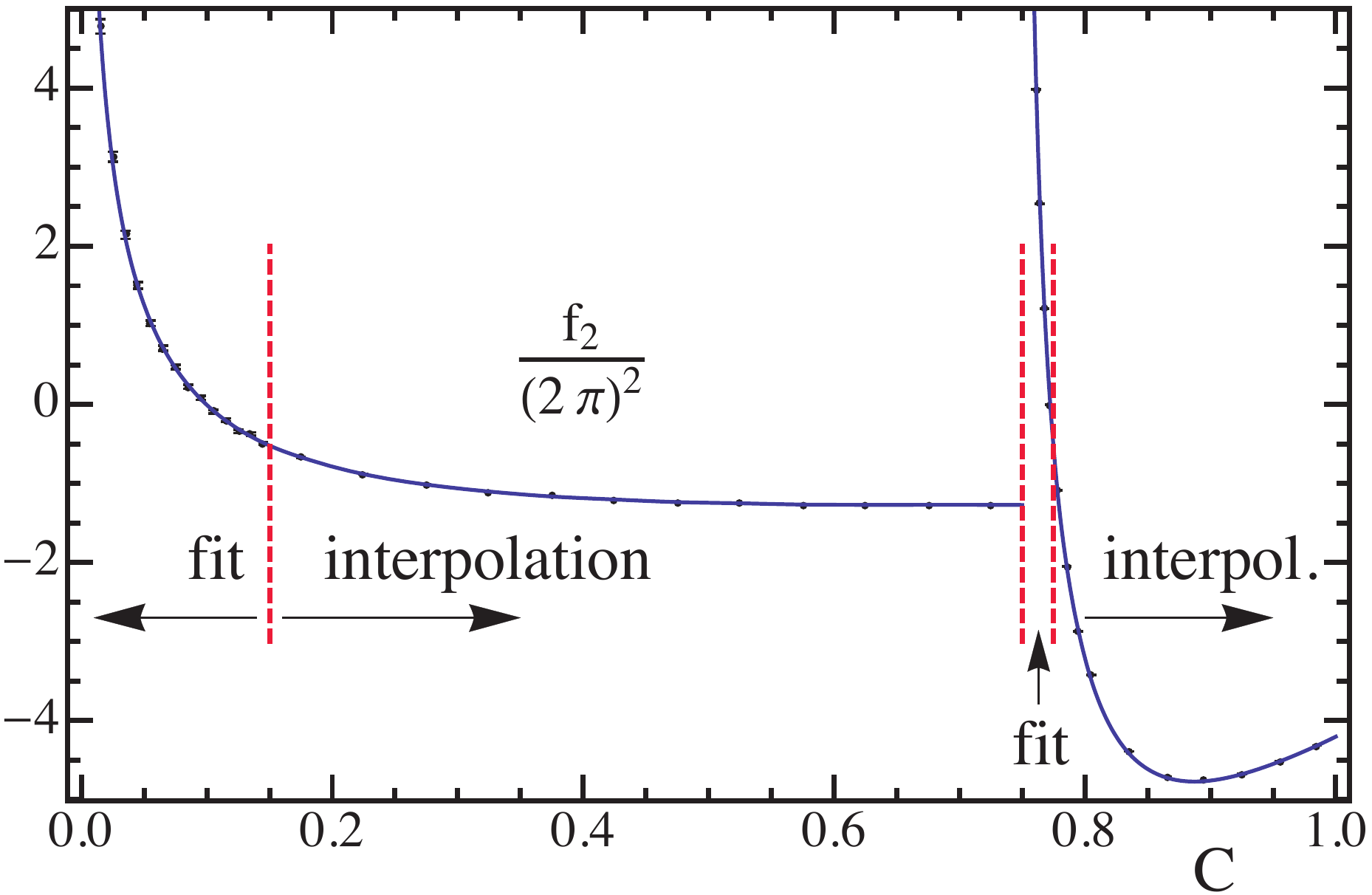}
\caption{$\ord {\alpha_s^2}$ nonsingular C-parameter distribution. The solid line shows our reconstruction, whereas dots with error bars correspond to the EVENT2 output with the singular terms subtracted. Our reconstruction consists of fit functions to the left of the red dashed line at $C = 0.15$ and between the two red dashed lines at $C = 0.75$ and $C = 0.8$ and interpolation functions elsewhere.}
  \label{fig:2-loopNS}
\end{center}
\end{figure}

\begin{figure}
\begin{center}
\includegraphics[width=0.95\columnwidth]{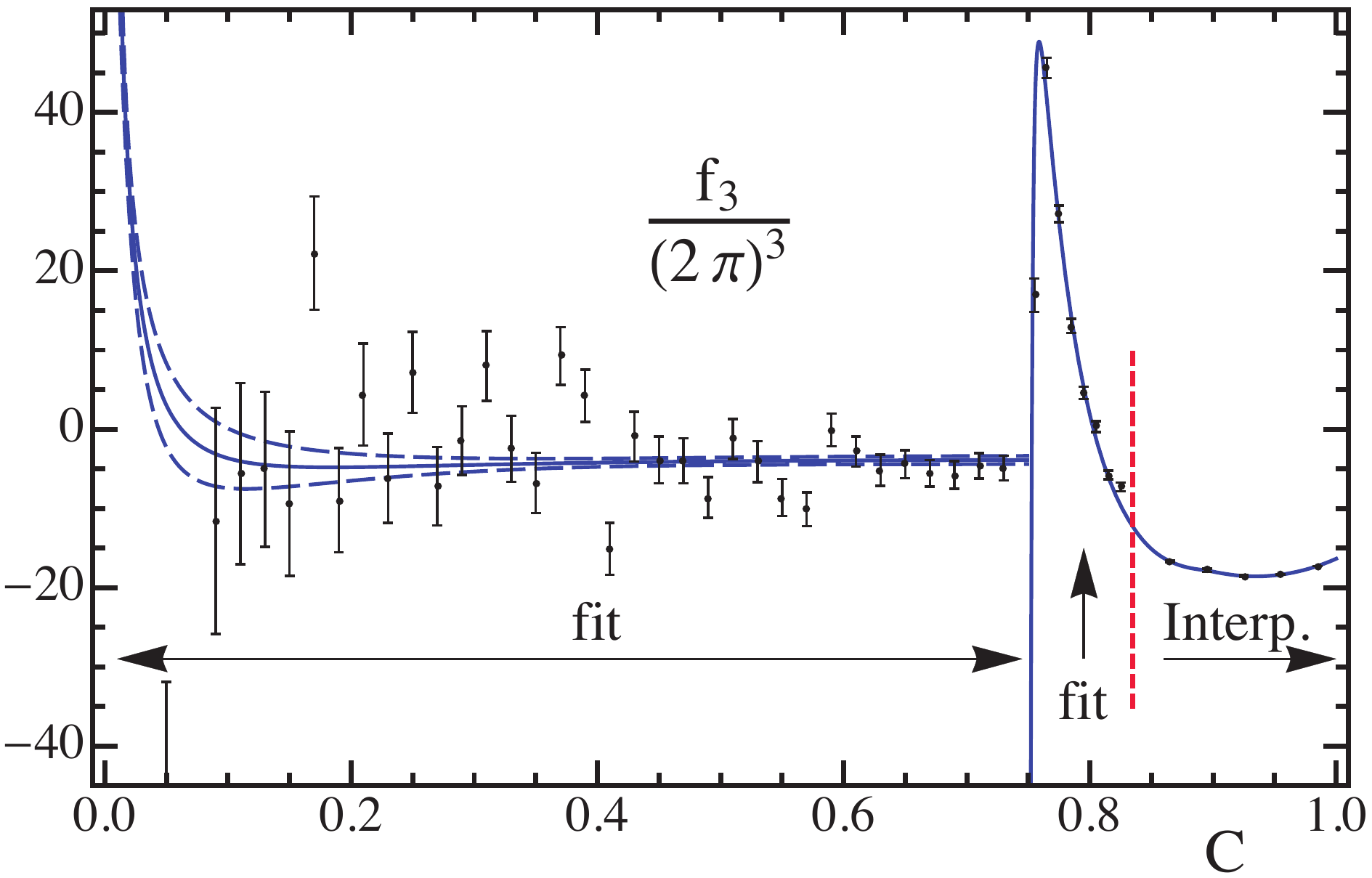}
\caption{$\ord {\alpha_s^3}$ nonsingular C-parameter distribution. The solid line shows our reconstruction, whereas dots with error bars correspond to the EERAD3 output with the singular terms subtracted. Our reconstruction consists of two fit functions, one for $C < 0.75$ and another one for $0.75 < C < 0.835$, and an interpolation for $C>0.835$ (to the right of the red dashed vertical line).}
  \label{fig:3-loopNS}
\end{center}
\end{figure}
At one loop, we can write down the exact form of the full distribution as a two-dimensional integral 
\cite{Ellis:1980nc}
\begin{align}\label{eq:1loop-FO}
&\frac{1}{\sigma_0} \bigg|\frac{\df \hat{\sigma}}{\df C}\bigg|^{\rm 1-loop} \,= \\
& \frac{\alpha_s}{2 \pi}\, C_F \!\!\int_0^1\! \df x_1\! \int_0^1\! \df x_2 \, \theta( x_1 + x_2 -1 ) 
\,\frac{x_1^2+x_2^2}{(1-x_1)(1-x_2)} \nonumber\\
&\times \delta\! \left( \!C - \frac{6\,(1-x_1)(1-x_2)(x_1+x_2-1)}{x_1 x_2 (2-x_1-x_2)}\right),\nonumber
\end{align}
which has support for $0<C<3/4$ and jumps to zero for $C> 3/4$. After resolving the delta function, it becomes a one-dimensional integral that can be easily evaluated numerically. After subtracting off the one-loop singular piece discussed in Sec.~\ref{sec:factorization-singular}, we obtain the result for $f_1$ shown in Fig.~\ref{fig:1-loopNS}. For $C>3/4$ the nonsingular distribution at this order is simply given by the negative of the singular, and for practical purposes one can find a parametrization for $f_1$ for $C < 3/4$, so we use
\begin{align}\label{eq:1NS}
& f_1(C<0.75) \,=\, -\,2.25168 + 0.506802\, C \,+\, 0.184585\, C^2 \nonumber\\
&+\, 0.121051\, C^3 \,+ (0.890476 - 0.544484 \,C \nonumber\\
&-\, 0.252937 \,C^2 - 0.0327797 \,C^3) \ln(C)\,,\nn\\
& f_1(C>0.75) \,=\, \frac{4}{3C}\bigg[3+4 \ln\bigg(\frac{C}{6}\bigg)\bigg].
\end{align}
For an average over $C$, this result for $f_1(C)$ is accurate to $10^{-7}$ and at worst for a particular $C$ is accurate at $10^{-5}$. An exact closed form in terms of elliptic functions for the integral in Eq.~(\ref{eq:1loop-FO}) has
been found in Ref.~\cite{Gardi:2003iv}.

The full $\ord{\alpha_s^2}$ and $\ord{\alpha_s^3}$ fixed-order distributions can be obtained numerically
from the Fortran programs EVENT2 \cite{Catani:1996jh, Catani:1996vz} and EERAD3
\cite{GehrmannDeRidder:2009dp,Ridder:2014wza}, respectively. At $\ord{\alpha_s^2}$ we use log-binning EVENT2 results
for $C<0.2$ and linear-binning (with bin size of 0.02) results for $0.2<C<0.75$. We then have additional
log binning from $0.75<C<0.775$ (using bins in $\ln[C-0.75]$) before returning to linear binning
for $C>0.775$. We used runs with a total of $3 \times 10^{11}$ events and an infrared cutoff $y_0=10^{-8}$. In the regions
of linear binning, the statistical uncertainties are quite low and we can use a numerical interpolation for $f_2(C)$.
For $C<0.15$ we use the ansatz, $f_2(C)=\sum_{i=0}^3 a_i \ln^i C + a_4\, C \ln C$  and fit the coefficients from EVENT2
output, including the constraint that the total fixed-order cross section gives the known $\ord{\alpha_s^2}$ coefficient for
the total cross section.  The resulting values for the $a_i$ are given as functions of $s_2^{\widetilde C}$,
the non-logarithmic coefficient in the partonic soft function. Details on the determination of this fit function and the determination of $s_2^{\widetilde C}$ can be found below in Sec.~\ref{sec:softtwoloop}. We  find 
\begin{align}
  s_2^{\widetilde C}=-\,43.2\,\pm\,1.0 \,, 
\end{align}
whose central value is used in Figs.~\ref{fig:2-loopNS}, \ref{fig:3-loopNS}, \ref{fig:component-plot}, and \ref{fig:component-plot-sum}, and whose uncertainty is included in our uncertainty analysis. For $0.75<C<0.8$ we employ another ansatz, \mbox{$f_2(C)\,=\,\sum_{i=0}^1 \sum_{k=0}^2 b_{i k}\, (C - 0.75)^i \ln^k [\,8\,(C-0.75)/3]$}. We use the values calculated in Ref.~\cite{Catani:1997xc}, \mbox{$b_{01}=59.8728$}, $b_{02}=43.3122$, and $b_{12} = -\,115.499$ and fit the rest of the coefficients to EVENT2 output. The final result for the two-loop nonsingular cross section coefficient then has the form
\begin{align}
f_2(C)\,+\,\epsilon^\text{low}_{2}\, \delta^\text{low}_{2}(C) \,
+\,\epsilon^\text{high}_{2}\, \delta^\text{high}_{2}(C) \,. 
\end{align}
Here $f_2(C)$ gives the best fit in all regions, and $\delta_2^\text{low}$ and  $\delta_2^\text{high}$ give the $1$-$\sigma$ error functions for the lower fit ($C<0.75$) and upper fit ($C>0.75$), respectively. The two variables $\epsilon^\text{low}_{2}$ and $\epsilon^\text{high}_{2}$ are varied during our theory scans in order to account for the error in the nonsingular function. In Fig.~\ref{fig:2-loopNS}, we show the EVENT2 data as dots and the best-fit nonsingular function as a solid blue line. The uncertainties are almost invisible on the scale of this plot.

In order to determine the $\ord{\alpha_s^3}$ nonsingular cross section $f_3(C)$, we follow a similar procedure. The EERAD3 numerical output is based on an infrared cutoff $y_0=10^{-5}$ and calculated with $6 \times 10^7$ events for the three leading color structures and $10^7$ events for the three subleading color structures. The results are linearly binned with a bin size of 0.02 for $C<0.835$ and a bin size of $0.01$ for $C>0.835$. As the three-loop numerical results have larger uncertainties than the two-loop results, we employ a fit for all $C<0.835$ and use interpolation only above that value. The fit is split into two parts for $C$ below and above $0.75$. For the lower fit, we use the ansatz $f_3(C)=\sum_{i=1}^5 a_i \ln^i (C)$. The results for the $a_i$ depend on the $\ord{\alpha_s^2}$ partonic soft function coefficient $s_2^{\widetilde C}$ and a combination of the three-loop coefficients in the partonic soft and jet function, $s_3^{\widetilde C}\,+\,2\, j_3$. Due to the amount of numerical uncertainty in the EERAD3 results, it is not feasible to fit for this combination, so each of these parameters is left as a variable that is separately varied in our theory scans. Above $C=0.75$ we carry out a second fit, using the fit form \mbox{$f_3(C)=\sum_{i=0}^4 b_i \ln^i(C-0.75)$}. We use the value \mbox{$b_4 = 122.718$} predicted by exponentiation in Ref.~\cite{Catani:1997xc}. The rest of the $b$'s depend on $s_2^{\widetilde C}$. The final result for the three-loop nonsingular cross section coefficient can once again be written in the form
\begin{align}
f_3(C)+\epsilon^\text{low}_{3} \delta_3^\text{low}(C) +\epsilon^\text{high}_{3} \delta^\text{high}_{3}(C) \,,
\end{align}
where $f_3(C)$ is the best-fit function and the $\delta_3$'s give the $1$-$\sigma$ error function for the low ($C<0.75$) and high ($C>0.75$) fits. Exactly like for the $\ord{\alpha_s^2}$ case, the $\epsilon_3$'s are varied in the final error analysis. In Fig.~\ref{fig:3-loopNS}, we plot the EERAD3 data as dots, the best-fit function $f_3$ as a solid line, and the nonsingular results with $\epsilon^\text{low}_{3}=\epsilon^\text{high}_{3}=\pm\,1$ as dashed lines. In this plot we take $s_2^{\widetilde C}$ to its best-fit value and $j_3=s_3^{\widetilde C}=0$. 

In the final error analysis, we vary the nonsingular parameters encoding the numerical extraction uncertainty $\epsilon_i$'s, as well as the profile parameter $\mu_{\rm ns}$. The uncertainties in our nonsingular fitting are obtained by taking $\epsilon^{\rm low}_2,\,\epsilon^\text{high}_{2},\,\epsilon^\text{low}_{3},$ and $\epsilon^\text{high}_3$ to be $-\,1$, $0$, and $1$ independently. The effects of $\epsilon^\text{low}_{2}$, $\epsilon^\text{high}_{2}$ and $\epsilon^\text{high}_{3}$ are essentially negligible in the tail region. Due to the high noise in the EERAD3 results, the variation of $\epsilon^\text{high}_{3}$ is not negligible in the tail region, but because it comes in at ${\cal O}(\alpha_s^3)$, it is still small. We vary the nonsingular renormalization scale $\mu_{\rm ns}$ in a way described in Sec. \ref{sec:profiles}.

In order to have an idea about the size of the nonsingular distributions with respect to the singular terms, we quote numbers
for the average value of the one-, two-, and three-loop distributions between $C = 0.2$ and $C = 0.6$:
$5, 21, 76$ (singular at one, two, and three loops); $-\,0.4,-\,1,-\,4$ (nonsingular at one, two, and three loops). Hence, in the region to be used for fitting $\alpha_s(m_Z)$, the singular distribution is $12$ (at 1-loop) to $20$ (at two and three loops) times larger than the nonsingular one and has the opposite sign. Plots comparing the singular and nonsingular cross sections for all $C$ values are given below in \Fig{fig:component-plot}.

\section{Determination of Two-Loop Soft Function Parameters}
\label{sec:softtwoloop}
In this section we will expand on the procedure used to extract the  $\ord{\alpha_s^2}$ non-logarithmic coefficient in the soft function using EVENT2. For a general event shape, we can separate the partonic cross section into a singular part  where the cross section involves $\delta(e)$ or $\ln^k(e)/e$,\,\footnote{The numerical outcome of a parton-level Monte Carlo such as EVENT2 contains only the power of logs, but the complete distributions can be obtained from knowledge of the SCET soft and jet functions.} and a nonsingular part with integrable functions, that diverge at most as $\ln^k(e)$. Of course, when these are added together and integrated over the whole spectrum of the event shape distribution, we get the correct fixed-order normalization:
\begin{equation} \label{eq:s2extractionintegral}
\hat\sigma_{\rm had} = \int_0^{e_{\rm max}} \!\df e \left( \frac{\df \hat\sigma_{\rm s}}{\df e} +
\frac{\df \hat\sigma_{\rm ns}}{\df e} \right).
\end{equation}
Here $e_\text{max}$ is the maximum value for the given event shape (for C-parameter, $e_\text{max}=1$).
Using SCET, we can calculate the singular cross section at $\mathcal{O}(\alpha_s^2)$, having the form
\begin{equation} \label{eq:ADs}
\dfrac{1}{\sigma_0} \frac{\df\hat\sigma_{\rm s}^{(2)}}{\df e}=A_{\delta}\,\delta(e)+\sum_{n=0}^3\,D_n\bigg[\frac{\ln^n e}{e}\bigg]_+\,.
\end{equation}
To define $\df\hat\sigma_{\rm s}^{(2)}/\df e$ we factor out $\alpha_s^2/16\pi^2$ and set $\mu=Q$. The only unknown term at ${\cal O} (\alpha_s^2)$ for the \mbox{C-parameter} distribution is the two-loop constant $s_{2}^{\widetilde C}$ in the soft function, which contributes to $A_\delta$. The explicit result for the terms in Eq.~(\ref{eq:ADs}) can be obtained from $H(Q,\mu)\,P(Q,QC/6,\mu)$ which are given in \App{ap:formulae}. This allows us to write the singular integral 
in (\ref{eq:s2extractionintegral}) as a function of $s_{2}^{\widetilde C}$ and known constants.

We extract the two-loop nonsingular portion of the cross section from EVENT2 data. Looking now at the specific case of
C-parameter, we use both log-binning (in the small $C$ region, which is then described with high accuracy)
and linear binning (for the rest). By default we use logarithmic binning for $C < C_{\rm fit} = 0.2$, but this boundary
is changed between $0.15$ and $0.25$ in order to estimate systematic uncertainties of our method. In the logarithmically
binned region we use a fit function to extrapolate for the full behavior of the nonsingular cross section.
In order to determine the coefficients of the fit function we use data between $C_{\rm cut}$ and $C_{\rm fit}$. By
default we take the value $C_{\rm cut} = 10^{-4}$, but we also explore different values between $5\times 10^{-5}$ and
$7.625\times 10^{-4}$ to estimate systematic uncertainties. We employ the following functional form, motivated by the expected nonsingular logarithms
\begin{align}
\dfrac{1}{\sigma_0}\frac{\df\hat\sigma^{\rm ns}_{\rm fit}}{\df C}=\sum_{i=0}^{3}\,a_i\,\ln^{i}C +
a_4\,C\ln^n C\,,
\end{align}
taking the value $n=1$ as default and exploring values \mbox{$0\le n \le 3$} as an additional source of systematic
uncertainty.

For the region with linear binning, we can simply calculate the relevant integrals by summing over the
bins. One can also sum bins that contain the shoulder region as its singular behavior is integrable.
These various pieces all combine into a final formula that can be used to extract the two-loop constant piece
of the soft function:
\begin{align} \label{eq:s2extractionfinalintegral}
\hat\sigma_{\rm had}^{(2)} = \!\int_0^{1}\! \df C\, \frac{\df \hat\sigma^{(2)}_{\rm s}}{\df C}
\,+ \!\int_0^{C_{\rm fit}}\!\! \df C \,\frac{\df \hat\sigma^{\rm ns}_{\rm fit}}{\df C} \,+
\!\int_{C_{\rm fit}}^{1}\!\! \df C\, \frac{\df \hat\sigma^{\text{ns}}_{\rm int}}{\df C}  .
\end{align}
Using \eq{s2extractionfinalintegral} one can extract $s_2^{\widetilde{C}}$, which can be decomposed into its various
color components as
\begin{align}
 s_2^{\widetilde{C}} = C_F^2\, s_{2}^{[C_F^2]} +  C_F\, C_A s_{2}^{[C_FC_A]} +  C_F\, n_f T_F s_{2}^{[n_f]} \,.
\end{align} 
The results of this extraction are
\begin{align} \label{s2results}
s_{2}^{[C_F^2]}  &= \,-\, \,0.46 \,\pm 0.75\,, \nn \\
s_{2}^{[C_FC_A]}  &= - \,29.08 \pm 0.13\,, \\
s_{2}^{[n_f]}  &= ~~\:21.87 \pm 0.03\,. \nn
\end{align}
The quoted uncertainties include a statistical component coming from the fitting procedure and a systematical component coming from the parameter variations explained above, added in quadrature. Note that the value for $s_{2}^{[C_F^2]}$ is  consistent with zero, as expected from exponentiation \cite{Hoang:2008fs}. For our analysis we will always take $s_{2}^{[C_F^2]}=0$. We have cross-checked that, when a similar extraction is repeated for the  case of thrust, the extracted values are consistent with those calculated analytically in Refs.~\cite{Kelley:2011ng,Monni:2011gb}. This indicates a high level of accuracy in the fitting procedure.  We have also confirmed that following the alternate fit procedure of Ref.~\cite{Hoang:2008fs} gives compatible results, as shown in App.~\ref{ap:MC-comparison}.

\section{Power Corrections and Renormalon-Free Scheme}
\label{sec:power}
The expressions for the theoretical prediction of the C-parameter distribution in the dijet region shown in  Eqs.~(\ref{eq:singular-nonperturbative}) and (\ref{eq:factorization-partonic-singular}) incorporate that the full soft function can be written as a convolution of the partonic soft function $\hat S_C$  and the nonperturbative shape function $F_C$\,\cite{Hoang:2007vb}\,\footnote{Here we use the relations $\hat S_C(\ell,\mu) = \hat S_{\widetilde C}(\ell/6,\mu)/6$ and $F_C(\ell) = F_{\widetilde C}(\ell/6)/6$. }:
\begin{align}\label{eq:soft-nonperturbative}
S_C(k,\mu) = \!\int \!\df k' \,\hat S_C(k-k',\mu) F_C(k',\Delta_{C})\,.
\end{align}
Here, the partonic soft function $\hat S_C$ is defined in fixed order in $\overline{\text{MS}}$. The shape function $F_C$ allows a smooth transition between the peak and tail regions, where different kinematic expansions are valid, and $\Delta_{C}$ is a parameter of the shape function that represents an offset from zero momentum and that will be discussed further below. By definition, the shape function satisfies the relations $F_C(k, \Delta_C) = F_C(k - 2\Delta_C)$ and $F_C(k <0) = 0$. In the tail region, where $QC/6 \gg \Lambda_{\text{QCD}}$, this soft function can be expanded to give
\begin{align}\label{eq:soft-OPE}
S_C(k,\mu) = \hat S_C(k) - \frac{\df \hat S_C(k)}{\df k}\, \overline{\Omega}_1^C
  + {\cal O}\Big( \frac{\alpha_s \Lambda_{\rm QCD}}{QC},\frac{\Lambda_{\rm QCD}^2}{Q^2 C^2}\Big) ,
\end{align}
where $ \overline{\Omega}_1^C$ is the leading nonperturbative power correction in $\overline{\rm MS}$ which effectively introduces a shift of the distribution in the tail region \cite{Abbate:2010xh}. The $\overline \Omega_1^C$ power correction 
\begin{align}
  \overline \Omega_1^C(\mu) & = \frac{1}{N_c} \big\langle 0 \big| {\rm tr}\: \overline Y_{\bar n}^T Y_{n}  (Q \widehat C) Y_n^\dagger  \overline Y_{\bar n}^*  \big| 0 \big\rangle \,,
\end{align}
is related to $\overline{\Omega}_1^\tau$, the first moment of the thrust shape function, as given in \Eq{eq:O1c}. In addition to the normalization difference that involves a factor of $3\pi/2$, their relation is further affected by hadron-mass effects which cause an additional deviation at the 2.5\% level (computed in \Sec{sec:hadronmass}). The dominant contributions of the ${\cal O}(\alpha_s \Lambda_{\rm QCD}/Q C)$ corrections indicated in \Eq{eq:soft-OPE} are log enhanced and will be captured once we include the $\mu$-anomalous dimension for $\Omega_1^C$ that is induced by hadron-mass effects~\cite{Mateu:2012nk}. There are additional ${\cal O}(\alpha_s \Lambda_{\rm QCD}/Q C)$ corrections, which we neglect, that do not induce a shift. We consider hadron-mass effects in detail in \Sec{sec:hadronmass}. 

From \Eq{eq:soft-nonperturbative} and the OPE of \Eq{eq:soft-OPE}, we can immediately read off the relations
\begin{align} \label{eq:gapmomentrelation}
&\!\int\! \df k' k' F_C(k',\Delta_{C}) = 2\,\Delta_{C} + \! \int \! \df k' k' F_C(k') = \overline{\Omega}_1^C\,,\nn\\
&\!\int\! \df k' F_C(k') = 1\,,
\end{align}
which state that the first moment of the shape function provides the leading power
correction and that the shape function is normalized. In the peak, it is no longer sufficient to keep only the first
moment, as there is no OPE when $QC/6 \sim \Lambda_{\text{QCD}}$ and we must keep the full dependence on the model
function in \Eq{eq:soft-nonperturbative}.

The partonic soft function in $\overline{\text{MS}}$ has an $\mathcal{O}(\Lambda_{\rm QCD})$ renormalon, an ambiguity which is related to a linear sensitivity in its perturbative series.
This renormalon ambiguity is in turn inherited to the numerical values for $\overline \Omega_1^C$ obtained in fits to the experimental data.
It is possible to avoid this renormalon issue by switching to
a different scheme for $\Omega_1^C$, which involves subtractions in the partonic soft function that remove this type of infrared sensitivity.
Following the results of \Ref{Hoang:2007vb}, we write  $\Delta_{C}$ as
\begin{align} \label{eq:deltasplitting}
\Delta_{C}  & =  \frac{3\pi}{2}[\,\bar{\Delta}(R,\mu) + \delta(R,\mu)\,]\,.
\end{align}
The term $\delta(R,\mu)$ is a perturbative series in $\alpha_s(\mu)$ which has the same
renormalon behavior as $\overline{\Omega}_1^C$. In the factorization formula, it is grouped into the partonic soft function $\hat S_C$ through the exponential factor involving $\delta(R,\mu)$ shown in \Eq{eq:singular-resummation}. 
Upon simultaneous perturbative expansion of the exponential together with $\hat S_C$, the $\mathcal{O}(\Lambda_{\rm QCD})$ renormalon is subtracted.
The term $\bar{\Delta}(R,\mu)$ then becomes a nonperturbative
parameter which is free of the $\mathcal{O}(\Lambda_{\rm QCD})$ renormalon. Its dependence on the subtraction scale $R$ and on $\mu$ is dictated by $\delta(R,\mu)$ since $\Delta_{C}$ is $R$ and $\mu$ independent.
The subtraction scale $R$ encodes the momentum scale associated with
the removal of the linearly infrared-sensitive fluctuations. 
The factor $3\pi/2$ is a normalization coefficient that relates the ${\cal O}(\Lambda_{\rm QCD})$ 
renormalon ambiguity of the $C$ soft function $S_{C}$ to the one for the thrust soft function.
Taking into account this normalization we can use for $\delta(R,\mu)$ the scheme for the thrust 
soft function already defined in \Ref{Abbate:2010xh},
\begin{equation} \label{eq:delta-scheme}
\!\!\!\!\delta(R,\mu) = \frac{R}{2} \, e^{\gamma_E} \frac{\df}{\df \ln(ix)}
\big[ \ln S^{\text{part}}_{\tau}(x,\mu) \big]_{x = (i R e^{\gamma_E})^{-1}},
\end{equation}
where $S^{\text{part}}_{\tau}(x,\mu)$ is the position-space thrust partonic soft function. 
From this, we find that the perturbative series
for the subtraction is
\begin{align} \label{eq:deltaseries}
\delta(R,\mu) =
R \,e^{\gamma_E} \!\sum_{i=1}^\infty \alpha_s^i(\mu)\, \delta^i(R,\mu)\,.
\end{align}
Here the $\delta_{i\ge 2}$ depend on both the adjoint Casimir
\mbox{$C_A=3$} and the number of light flavors in combinations that are unrelated to the QCD beta
function. Using five light flavors the first three coefficients have been calculated in \Ref{Hoang:2008fs} as
\begin{align} \label{eq:d123}
\frac{\pi}{2}\, \delta^1(R,\mu) &= -\,1.33333\, L_R \,, \nn \\
\frac{\pi}{2}\, \delta^2(R,\mu) &= -\,0.245482 - 0.732981\, L_R - 0.813459\, L_R^2 \,, \nn \\
\frac{\pi}{2}\, \delta^3(R,\mu) &= -\,0.868628\, - 0.977769\, L_R
  \nn\\&\quad -1.22085\, L_R^2 - 0.661715\, L_R^3 \,,
\end{align}
where $L_R = \ln(\mu/R)$. Using these $\delta$'s, we can make a scheme change on the first moment to what we
call the Rgap scheme:
\begin{align} \label{eq:omegaschemechange}
\Omega_1^{C}(R,\mu) &= \overline{\Omega}_1^C(\mu) - 3\pi\,\delta (R,\mu)\,.
\end{align}
In contrast to the $\overline{\text{MS}}$ scheme $\overline{\Omega}_1^C(\mu)$, the Rgap scheme $\Omega_1^{C}(R,\mu)$  
is free of the $\Lambda_{\rm QCD}$ renormalon. From \Eq{eq:gapmomentrelation} it is then easy to see that the first moment of the shape function becomes
\begin{equation}
\int \!\df k \,k\,F_C\big(k - 3\pi\, \overline\Delta(R,\mu)\big) = \Omega_1^C(R,\mu)  \,.
\end{equation}
The factorization in \Eq{eq:soft-nonperturbative} can now be written as
\begin{align} \label{eq:soft-nonperturbative-subtract}
\!\!\!S_C(k,\mu) & = \!\int \!\df k' \,e^{-\,3\pi\delta(R,\mu) \frac{\partial}{\partial k}}
\hat S_C(k-k',\mu) \nn\\
 & \times F_C\big(k' - 3\pi \, \overline{\Delta}(R,\mu)\big)\,.
\end{align}
The logs in \Eq{eq:d123} can become large when $\mu$ and $R$ are far apart. This imposes a constraint that $R \sim \mu$,
which will require the subtraction scale to depend on $C$ in a way similar to $\mu$. On the other hand, we also must consider the power
counting $\overline{\Omega}^C \sim \Lambda_\text{QCD}$, which leads us to desire using $R \simeq 1$\,GeV. In order to satisfy both of these constraints in the tail region, where $\mu \sim QC/6 \gg 1$\,GeV, we (i) employ $R \sim \mu$ for the subtractions in $\delta(R,\mu)$ that are part of the Rgap partonic soft function and (ii) use the \mbox{R-evolution} to relate the gap parameter $\bar\Delta(R,\mu)$ to the reference gap parameter $\bar\Delta(R_\Delta,\mu_\Delta)$ with $R_\Delta \sim \mu_\Delta \sim {\cal O}(1\,\text{GeV})$ where the $\Lambda_{\rm QCD}$ counting applies~\cite{Hoang:2008yj, Hoang:2009yr,Hoang:2008fs}. The formulae for the \mbox{$R$-RGE} and $\mu$-RGE are
\begin{align} \label{eq:RandmuRGE}
R \frac{\df}{\df R}  \bar{\Delta} (R,R) &= -\,R \sum_{n=0}^\infty \gamma_n^R \left( \frac{\alpha_s(R)}{4 \pi} \right)^{\!\!n+1}  , \nn \\
\mu \frac{\df}{\df\mu}  \bar{\Delta} (R,\mu) &= 2\, R\, e^{\gamma_E} \sum_{n=0}^\infty \Gamma_n^\text{cusp} \left( \frac{\alpha_s(\mu)}{4 \pi} \right)^{\!\!n+1}  ,
\end{align}
where for five flavors the $\Gamma_n^\text{cusp}$ is given in App.~\ref{ap:formulae} and the $\gamma^R$ coefficients are given by
\begin{align} \label{eq:gammaR}
\gamma_0^R &= 0, \; \gamma_1^R = -\,43.954260\,,
\;\gamma_2^R = -\,606.523329\,.
\end{align}
The solution to \Eq{eq:RandmuRGE} is given, at N$^k$LL, by
\begin{align} \label{eq:DeltaRevolution}
\bar{\Delta}(R,\mu) &= \bar{\Delta}(R_\Delta,\mu_\Delta) + R\, e^{\gamma_E} \omega\, [\, \Gamma^\text{cusp},\mu,R\,] \nn \\
&+ R_\Delta e^{\gamma_E} \omega\,[\,\Gamma^\text{cusp},R_\Delta,\mu_\Delta\,] \nn \\
&+ \Lambda_{\text{QCD}}^{(k)} \,\sum_{j=0}^{k} (-1)^j S_j e^{i \pi \hat{b}_1} \nn \\
&\times\big[\,\Gamma ( -\,\hat{b}_1 -j, t_1) - \Gamma(-\,\hat{b}_1  -j,t_0) \,\big]\nn\\[0.2cm]
&\equiv \bar{\Delta}(R_\Delta,\mu_\Delta) + \Delta^{\rm diff}(R_\Delta,R, \mu_\Delta,\mu)\,.
\end{align}
For the convenience of the reader, the definition for $\omega$ is provided in \Eq{eq:w}, and the values for $\hat{b}_1$ and the $S_j$ are given in \Eq{eq:Sjnor}. In order to satisfy the power counting criterion for $R$, we specify the parameter $\bar{\Delta} (R_\Delta,\mu_\Delta)$ at the low reference scales $R_\Delta = \mu_\Delta = 2 \,\text{GeV}$. We then use \Eq{eq:DeltaRevolution} to evolve this parameter up to a scale $R(C)$, which is given in \Sec{sec:profiles} and satisfies the condition $R(C) \sim \mu_S(C)$ in order to avoid large logs. This \mbox{R-evolution} equation yields a similar equation for the running of $\Omega_1^C(R,\mu_S)$, which is easily found from \Eqs{eq:deltasplitting}{eq:omegaschemechange}.
\begin{figure}[t!]
\begin{center}
\includegraphics[width=0.95\columnwidth]{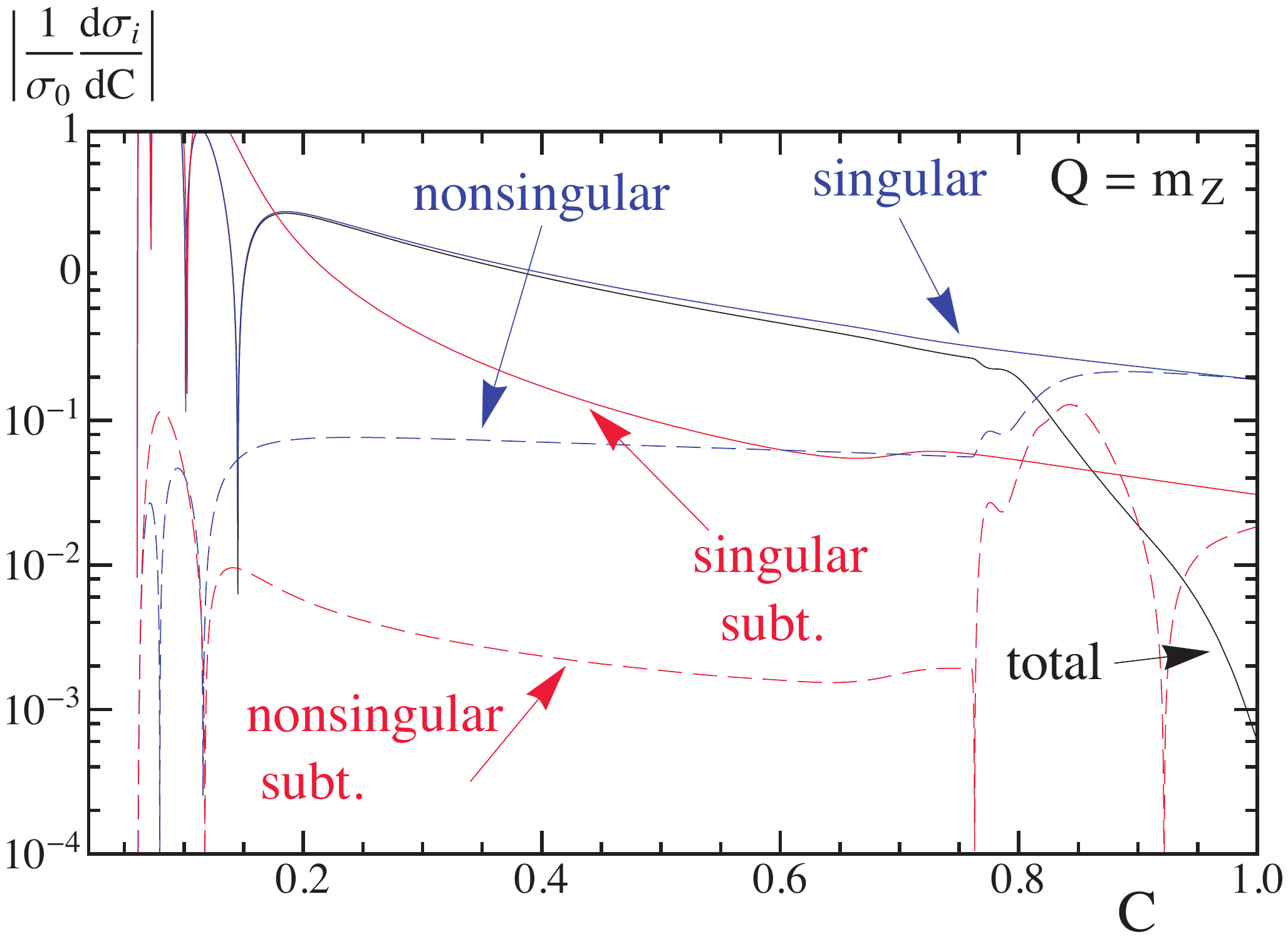}
\caption{Singular and nonsingular components of the fixed-order C-parameter cross section, including up to ${\cal O}(\alpha_s^3)$ terms, with $\Omega_1=0.25\,$GeV and $\alpha_s(m_Z) = 0.1141$.}
\label{fig:component-plot}
\end{center}
\end{figure}

We also apply the Rgap scheme in the nonsingular part of the cross section by using the convolution
\begin{align} \label{eq:nonsingular-shape-convolution}
& \!\!\!\int  \!\df k'\, e^{-\,3 \pi \frac{\delta(R,\mu_{\!s})}{Q}\frac{\partial}{\partial C}}
\frac{\df \hat{\sigma}_{\text{ns}}}{\df C} \Big( C - \frac{k'}{Q}, \frac{\mu_{\text{ns}}}{Q} \Big)
 \nn\\
 & \times F_C\big(k' - 3\pi \,\bar{\Delta}(R,\mu_S)\big) \,.
\end{align}
By employing the Rgap scheme for both the singular and nonsingular pieces,
the sum correctly recombines in a smooth manner to the fixed-order result in the far-tail region. 

Note that by using \Eq{eq:deltasplitting} we have defined the renormalon-free moment parameter $\Omega_1^C(R,\mu)$
in a scheme directly related to the one used for the thrust analyses in Refs.~\cite{Abbate:2010xh,Abbate:2012jh}.
This is convenient as it allows for a direct comparison to the $\Omega_1$ fit results we obtained in both these analyses. However, many other renormalon-free schemes can be devised, and all these schemes are perturbatively related to each other through their relation to the $\overline{\text{MS}}$ scheme $\Delta_C$. As an alternative, we could have defined a renormalon-free scheme for $\Omega_1^C$ by determining the subtraction $\delta$ directly from the $\widetilde C$ soft function $S_{\widetilde C}$ using the analog to \Eq{eq:delta-scheme}. For future reference we quote the results for the resulting subtraction function $\delta_{\widetilde C}$ in App.~\ref{ap:subtractionchoice}. 

\begin{figure}[t!]
\begin{center}
\includegraphics[width=0.95\columnwidth]{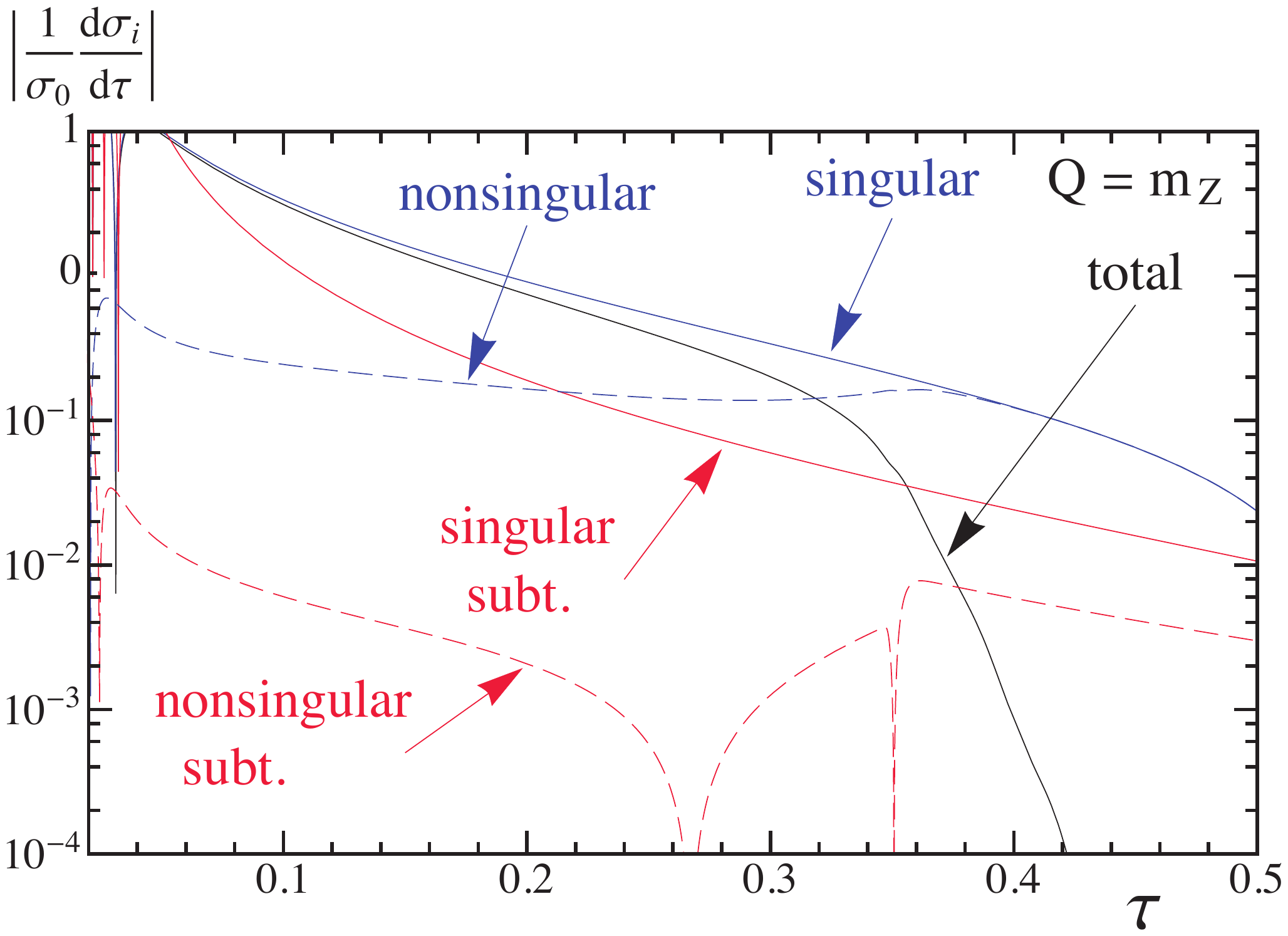}
\caption{Singular and nonsingular components of the fixed-order thrust cross section, including up to ${\cal O}(\alpha_s^3)$ terms, with \mbox{$\Omega_1=0.25\,$GeV} and $\alpha_s(m_Z) = 0.1141$.}
\label{fig:component-plot-tau}
\end{center}
\end{figure}
In close analogy to Ref.~\cite{Abbate:2010xh}, we parametrize the shape function $F_C$ in terms of the basis functions introduced in Ref.~\cite{Ligeti:2008ac}. In this expansion the shape function has the form
\begin{equation} \label{eq:shapebasis}
F_C(k,\lambda,\{c_i\}) = \frac{1}{\lambda} \left[ \sum_{n=0}^N c_n f_n\left(\frac{k}{\lambda}\right)\right]^2,
\end{equation}
where the $f_n$ are given by
\begin{align} \label{eq:basisfunctions}
f_n(z) = 8 \sqrt{\frac{2z^3(2n+1)}{3}} e^{-2z} P_n [\,g(z)\,], \nn \\
g(z)=\frac{2}{3} \left[3 - e^{-4z} (3 + 12 z + 24 z^2 + 32 z^3) \right] - 1\,,
\end{align}
and $P_n$ denote the Legendre polynomials. The additional parameter $\lambda$ is irrelevant when $N \to \infty$. For finite $N$ it is strongly correlated with the first moment $\Omega_1^C$ (and with $c_1$). The normalization of the shape function requires that $\sum_{n=0}^N c_n^2 = 1$. When plotting and fitting in the tail region, where the first moment of the shape
function $\Omega_1^C$ is the only important parameter, it suffices to take $c_0=1$ and all $c_{i>0}=0$. In this case
the parameter $\lambda$ directly specifies our $\Omega_1^C$ according to
\begin{align} \label{eq:omega-from-lambda}
\Omega_1^C(R_\Delta,\mu_\Delta) 
 = \lambda + 3 \pi\, \bar{\Delta}(R_\Delta,\mu_\Delta) 
 \,.
\end{align}
In the tail region where one fits for $\alpha_s(m_Z)$, there is not separate dependence on the nonperturbative parameters $\lambda$ and $\bar{\Delta}(R_\Delta,\mu_\Delta)$; they only appear together through the parameter $\Omega_1^C(R_\Delta,\mu_\Delta)$. In the peak region, one should keep more $c_i$'s in order to
correctly parametrize the nonperturbative behavior.

In Fig.~\ref{fig:component-plot} we plot the absolute value of the four components of the partonic fixed-order $C$ distribution at $\mathcal{O}(\alpha_s^3)$ in the Rgap scheme at $Q=m_Z$. Resummation has been turned off. The cross section components include the singular terms (solid blue), nonsingular terms (dashed blue), and separately the contributions from terms that involve the subtraction coefficients $\delta_i$, for both singular subtractions (solid red) and nonsingular subtractions (dashed red). The sum of these four components gives the total cross section (solid black line). One can observe that the nonsingular terms are significantly smaller than the singular ones in the tail region below the shoulder, i.e.\ for $C<0.7$. Hence the tail region is completely dominated by the part of the cross section described by the SCET factorization theorem, where resummation matters most. Above the shoulder the singular and nonsingular $C$ results have comparable sizes. An analogous plot for the thrust cross section is shown in Fig.~\ref{fig:component-plot-tau}.  We see that the portion of the C-parameter distribution where the logarithmic resummation in the singular terms is important, is substantially larger compared to the thrust distribution.

\subsection{Hadron Mass Effects}
\label{sec:hadronmass}

Following the analysis in \Ref{Mateu:2012nk}, we include the effects of hadron masses by including 
the dependence of  $\Omega_1^C$ on the distributions of transverse velocities,
\begin{equation} \label{eq:def-trans-vel}
r \equiv \frac{p_\perp}{\sqrt{p_\perp^2 + m_H^2}} \,,
\end{equation}
where $m_H$ is the nonzero hadron mass and $p_\perp$ is the transverse velocity with respect to the thrust axis.
For the massless case, one has $r=1$. However, when the hadron masses are nonzero, $r$ can take any value
in the range $0$ to $1$. The additional effects of the finite hadron masses cause non-trivial modifications in the form of the first moment of the shape function,
\begin{equation} \label{eq:omega1functionofg}
 \overline\Omega_1^e(\mu) \,=\, c_e\! \int_0^1\! \df r \, g_e(r)\, \overline\Omega_1(r,\mu)\,,
\end{equation}
where $e$ denotes the specific event shape that we are studying, $c_e$ is an event-shape-dependent constant, $g_e(r)$
is an event-shape-dependent function\,\footnote{As discussed in Ref.~\cite{Mateu:2012nk}, event shapes with
a common $g_e$ function belong to the same universality class. This means that their leading power corrections are
simply related by the $c_e$ factors. In this sense $g_e$ is universality-class dependent rather than event-shape
dependent.} that encodes the dependence on the hadron-mass effects and $\overline\Omega_1(r)$ is a
universal $r$-dependent generalization of the first moment, described by a matrix element of the transverse
velocity operator. $\overline\Omega_1(r,\mu)$ is universal for all recoil-insensitive event shapes. Note that once hadron masses are included there is no limit of the hadronic parameters that reduces to the case where hadron masses are not accounted for.

For the cases of thrust and C-parameter, we have
\begin{align} \label{eq:thrust-C-r-constants}
c_C &= 3 \pi\, , && g_C(r)  = \frac{2 r^2}{\pi} K(1-r^2)\,, \\
c_\tau &= 2\, , && g_\tau(r) \,= 1 - E(1-r^2) + r^2 K(1-r^2)\,, \nn
\end{align}
where $E(x)$ and $K(x)$ are complete elliptic integrals, whose definition can be found in~\Ref{Salam:2001bd,Mateu:2012nk}.
Notice that $g_C(r)$ and $g_\tau(r)$ are within a few percent of
each other over the entire $r$ range, so we expect the relation 
\begin{align}
 2\,\overline\Omega_1^C(\mu) = 3 \pi\,\overline\Omega_1^\tau(\mu)
   \big[ 1 +{\cal O}(2.5\%) \big]\,,
\end{align}
where the ${\cal O}(2.5\%)$ captures the breaking of universality due to the effects of hadron masses. We have determined the size of the breaking by 
\begin{align}
 \frac{\int_0^1 \df r  \big[\,g_C(r) - g_\tau(r) \big] }{\int_0^1 \df r \big[\,g_C(r) + g_\tau(r)\big]/2} = 0.025 .
\end{align}
All $r\in[\,0,1]$ contribute roughly an equal amount to this deviation, which is therefore well captured by 
this integral.

As indicated in \Eq{eq:omega1functionofg} the moment $\overline\Omega_1(r,\mu)$ in the $\overline{\text{MS}}$ scheme is renormalization-scale dependent and at LL satisfies the RGE of the form~\cite{Mateu:2012nk}
\begin{equation} \label{eq:omega1-r-murunning}
\overline \Omega_1(r,\mu) = \overline \Omega_1(r,\mu_0)\!
\left[ \frac{\alpha_s(\mu)}{\alpha_s(\mu_0)} \right]^{\hat{\gamma}_1(r)}.
\end{equation}
In App.~\ref{ap:hadronmassR} we show how to extend this running to the Rgap scheme in order to remove the ${\cal O}(\Lambda_{\rm QCD})$ renormalon.
The result in the Rgap scheme is
\begin{align}  \label{eq:omega1R-r-murunning}
&g_C(r)\,\Omega_1^C(R,\mu,r) = g_C(r)\!
\left[ \frac{\alpha_s(\mu)}{\alpha_s(\mu_\Delta)} \right]^{\hat{\gamma}_1(r)}\!
\Omega_1^C(R_\Delta, \mu_\Delta,r) \nn \\
&+ R\, e^{\gamma_E} \!\left(\frac{\alpha_s(\mu)}{\alpha_s(R)} \right)^{\!\!\hat{\gamma}_1(r)}\!
\omega\, [\,\Gamma^{\text{cusp}},\mu,R\,] \nn \\
&+ R_\Delta e^{\gamma_E} \!\left(\frac{\alpha_s(\mu)}{\alpha_s(R_\Delta)} \right)^{\!\!\hat{\gamma}_1(r)}\!
\omega\, [\,\Gamma^{\text{cusp}},R_\Delta,\mu_\Delta] \nn \\
&+ \Lambda_{\text{QCD}}^{(k)} \!\left( \frac{\beta_0 \alpha_s(\mu)}{2 \pi} \right)^{\!\!\hat{\gamma}_1(r)}\!
\sum_{j=0}^{k}S^r_j(r) (-1)^j e^{i \pi [\hat{b}_1 - \hat{\gamma}_1(r)]} \nn \\
&\ \ \times\big[\,\Gamma ( -\,\hat{b}_1 + \hat{\gamma}_1(r) -j, t_1) - 
\Gamma(-\,\hat{b}_1 + \hat{\gamma}_1(r) -j,t_0) \,\big] \nn\\
&\equiv g_C(r)\!\left[ \frac{\alpha_s(\mu)}{\alpha_s(\mu_\Delta)} \right]^{\hat{\gamma}_1(r)}\!
\Omega_1(R_\Delta, \mu_\Delta,r) \nn\\
&\ \ +  \Delta^{\rm diff}(R_\Delta,R,\mu_\Delta,\mu,r)\,.
\end{align}
Here the formula is resummed to N$^k$LL, and $\Lambda_{\text{QCD}}^{(k)}$ is the familiar N$^k$LO perturbative expression for $\Lambda_{\text{QCD}}$. We always use $R_\Delta = \mu_\Delta = 2\,$GeV to define the initial hadronic parameter. The values for $\hat{b}_1$, $\gamma_1(r)$, $t_1$, $t_0$,and  the $S_j$ can all be found in App.~\ref{ap:hadronmassR} and the resummed $\omega$ is given in ~\Eq{eq:w}.

In order to implement this running, we pick an ansatz for the form of the moment at the low scales, $R_\Delta$
and $\mu_\Delta$, given by
\begin{align} \label{eq:omega1-ansatz}
\Omega_1(R_\Delta,\mu_\Delta,r) &= \big[\, a(R_\Delta,\mu_\Delta) f_a(r) + b(R_\Delta,\mu_\Delta) f_b(r) \,\big]^2, \nn \\
f_a(r) &= 3.510 \, e^{-\frac{r^2}{1-r^2}}, \\
f_b(r) &= 13.585 \, e^{-\frac{2\,r^2}{1-r^2}} - 21.687 \,\, e^{-\frac{4\,r^2}{1-r^2}}. \nn
\end{align}
The form of $\Omega_1(R_\Delta,\mu_\Delta,r)$ was chosen to always be positive and to smoothly go to zero at the endpoint $r=1$. In the Rgap scheme, $\Omega_1$ can be interpreted in a Wilsonian manner as a physical hadronic average momentum parameter, and hence it is natural to impose positivity. As $r\to 1$ we are asking about the vacuum-fluctuation-induced distribution of hadrons with large $p_\perp$ which is anticipated to fall off rapidly.
We also check other ans\"atze that satisfied these conditions, but choosing different positive definite functions has a minimal effect on the distribution. The functions $f_a$ and $f_b$ were chosen to satisfy
$\int_0^1 \df r g_C(r) f_a(r)^2 = \int_0^1 \df r g_C(r) f_b(r)^2 = 1$ and $\int_0^1 \df r g_C(r) f_a(r)\, f_b(r) = 0$.
This allows us to write
\begin{align} \label{eq:omegafunctionofaandb}
\Omega_1^C(R_\Delta,\mu_\Delta) 
  &= c_C \!\!\int_0^1 \!\! \df r\:  g_C(r) \: \Omega_1(R_\Delta, \mu_\Delta ,r) 
  \nn\\
  &= 3\, \pi\, \big[\,a(R_\Delta,\mu_\Delta)^2 + b(R_\Delta,\mu_\Delta)^2\, \big]\,,
\end{align}
and to define an orthogonal variable,
\begin{equation} \label{eq:thetadef}
\theta(R_\Delta,\mu_\Delta) \equiv \arctan\left(\frac{b(R_\Delta,\mu_\Delta)}{a(R_\Delta,\mu_\Delta)}\right).
\end{equation}
The parameters $a$ and $b$ can therefore be swapped for $\Omega_1^C(R_\Delta,\mu_\Delta)$ and $\theta(R_\Delta,\mu_\Delta)$. This $\theta$ is defined as part of the model for the universal function $\Omega_1(R,\mu,r)$ and so should also exhibit universality between event shapes. In \Sec{subsec:hadronmass} below, we will demonstrate that $\theta$ has a small effect on the cross section for the C-parameter, and hence that $\Omega_1^C(R_\Delta,\mu_\Delta)$ is the most important hadronic parameter.

\section{Profile Functions}
\label{sec:profiles}

The ingredients required for cross section predictions at various resummed perturbative orders are given in \Tab{tab:ordercounting}. This includes the order for the cusp and non-cusp anomalous dimensions for $H$, $J_\tau$, and $S_C$; their perturbative matching order; the beta function $\beta[\alpha_s]$ for the running coupling: and the order for the nonsingular corrections discussed in \Sec{sec:nonsingular}.  It also includes the anomalous dimensions $\gamma_\Delta$ and subtractions $\delta$ discussed in this section. In our analysis we only use primed orders with the factorization theorem for the distribution. For the unprimed orders, only the formula for the cumulant cross section properly resums the logarithms, see Ref.~\cite{Almeida:2014uva}, but for the reasons discussed in Ref.~\cite{Abbate:2010xh}, we need to use the distribution cross section for our analysis. The primed order distribution factorization theorem properly resums the desired series of logarithms for $C$, and was also used in Refs.~\cite{Abbate:2010xh,Abbate:2012jh} to make predictions for thrust.

The factorization formula in Eq.~(\ref{eq:singular-resummation}) contains three characteristic renormalization scales, the hard scale $\mu_H$, the jet scale $\mu_J$, and the soft scale $\mu_S$. In order to avoid large logarithms, these scales
must satisfy certain constraints in the different $C$ regions:
\begin{align} \label{eq:profileconstraints}
& \text{1) nonperturbative:~} C \lesssim 3\pi\,\frac{\Lambda_\text{QCD}}{Q}\nn \\
&  \qquad  \mu_H  \sim Q,\  \mu_J \sim \sqrt{\Lambda_\text{QCD} Q},\  \mu_S \!\sim\! R \!\sim\!  \Lambda_\text{QCD} 
  \,, \nn \\[5pt]
& \text{2) resummation:~} 3\pi\,\frac{\Lambda_\text{QCD}}{Q} \ll C < 0.75
 \\
& \qquad  \mu_H \sim Q,\  \mu_J \sim Q \sqrt{\frac{C}{6}},\   \mu_S \! \sim \! R \!\sim\! \frac{QC}{6}  \gg \Lambda_{\rm QCD}
  \,, \nn \\
& \text{3) fixed-order:~} C > 0.75
   \nn \\
 & \qquad \mu_H   = \mu_J  = \mu_S = R \sim Q\gg \Lambda_{\rm QCD}\nn
  \,.\end{align}
In order to meet these constraints and have a continuous factorization formula, we make each scale a smooth function of $C$ using profile functions.

When one looks at the physical C-parameter cross-section, it is easy to identify the peak, tail, and far-tail as
distinct physical regions of the distribution. How much of the physical peak belongs to the nonperturbative vs
resummation region is in general a process-dependent statement, as is the location of the transition between the resummation
and fixed-order regions. For example, in $b\to s\,\gamma$ the entire peak is in the nonperturbative 
region~\cite{Ligeti:2008ac}, whereas for $pp\to H+1$ gluon initiated jet with $p_T\sim 400\,{\rm GeV}$, the entire peak
is in the resummation region~\cite{Jouttenus:2013hs}.  For thrust with $Q=m_Z$~\cite{Abbate:2010xh}, and similarly here
for C-parameter with $Q=m_Z$, the transition between the nonperturbative and resummation regions occurs near the maximum
of the physical peak.   Note that, despite the naming, in the nonperturbative region, where the full form of the shape function is needed, resummation is always important.
The tail for the thrust and C-parameter distributions is located in the resummation region, and the far-tail, which is dominated by events with three or more jets, exists in the fixed-order region.

For the renormalization scale in the hard function, we use
\begin{align}
\mu_H \,=\, e_H \, Q\,,
\end{align}
where $e_H$ is a parameter that we vary from 0.5 to 2.0 in order to account for theory uncertainties.
\begin{table}[t!]
\begin{tabular}{c|ccccccc}
 &  cusp &  non-cusp  &  matching  &  $\beta[\alpha_s]$  &  nonsingular  &  $\gamma_\Delta^{\mu,R,r}$  &  $\delta$  \\
\hline
LL  &  1 &  -  &  tree  &  1  &  -  &  -  &  - \\
NLL  &  2 &  1  &  tree  &  2  &  -  &  1  &  - \\
N${}^2$LL  &  3 &  2  &  1  &  3  &  1  &  2  &  1 \\
N${}^3$LL  &  4${}^\text{pade}$ &  3  &  2  &  4  &  2  &  3  &  2 \\
\hline
NLL$^\prime$  &  2 &  1  &  1  &  2  &  1  &  1  &  1 \\
N${}^2$LL$^\prime$  &  3 &  2  &  2  &  3  &  2  &  2  &  2 \\
N${}^3$LL$^\prime$  &  4${}^\text{pade}$ &  3  &  3  &  4  &  3  &  3  &  3 \\
\end{tabular}
\caption{Loop corrections required for specified orders.}
\label{tab:ordercounting}
\end{table}
\begin{figure}[t!]
\begin{center}
\includegraphics[width=0.95\columnwidth]{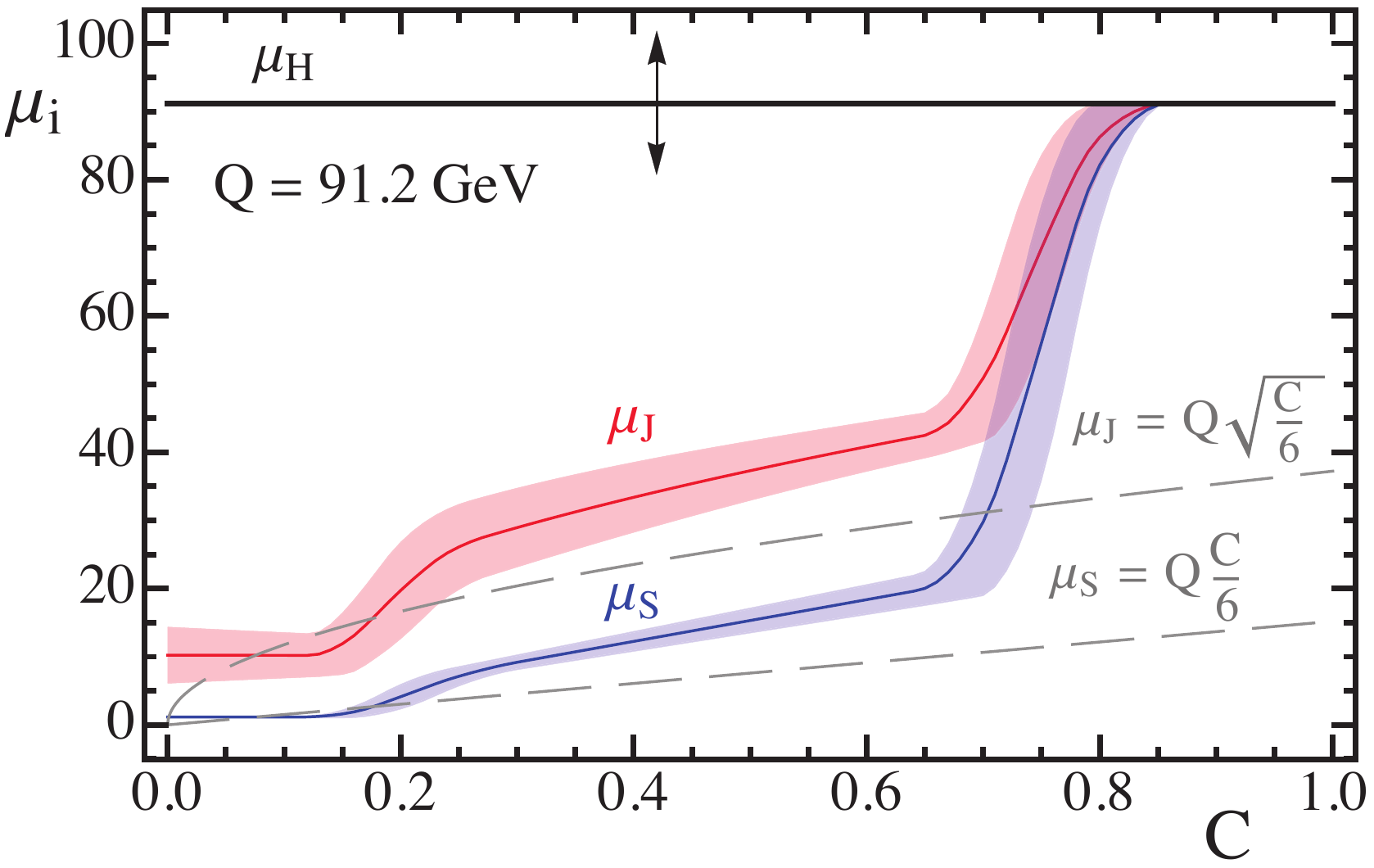}
\caption{
Solid lines are the central results for the profile functions for the renormalization scales $\mu_H$, $\mu_J(C)$, $\mu_S(C)$ at $Q=m_Z$. The bands and up-down arrow indicate the results of varying the profile parameters. The result for $R(C)$ is identical to $\mu_S(C)$ at the resolution of this figure, differing only at small $C$. Above $C=t_s\simeq 0.8$ all the scales merge, $\mu_H=\mu_J=\mu_S=R$.   }
\label{fig:profile-variation}
\end{center}
\end{figure}

The profile function for the soft scale is more complicated, and we adopt the following form:
\begin{equation} \label{eq:muSprofile}
 \!\mu_{S} = \left\{ \begin{tabular}{p{.5\columnwidth} l}
 $\mu_0$                                             & $0 \le C < t_0$ \\
 $\zeta(\mu_0,\,0,\,0,\,\frac{r_s\,\mu_H}{6},\,t_0,\,t_1,\,C)$  & $t_0 \le C < t_1$ \\
 $r_s \,\mu_H \frac{C}{6}$         & $t_1 \le C < t_2$ \\
 $\zeta(0,\,\frac{r_s\,\mu_H}{6},\,\mu_H,0,\,t_2,\,t_s,\,C)$ & $t_2 \le C < t_s$ \\
 $\mu_H$     &           $t_s \le C < 1$
 \end{tabular}
 \right.\!.
\end{equation}
Here the 1st, 3rd, and 5th lines satisfy the three constraints in \Eq{eq:profileconstraints}. 
In particular, $\mu_0$ controls the intercept of the soft scale at $C=0$. The term $t_0$ controls the boundary of the purely nonperturbative region and the start of the transition to the resummation region, and $t_1$ represents the end of this 
transition. As the border between the nonperturbative and perturbative regions is $Q$ dependent, we actually use  $n_0 \equiv t_0 (Q/1$\,{\rm GeV}) and $n_1 \equiv t_1 (Q/1$\,{\rm GeV}) as the profile parameters. In the resummation region $t_1< C<t_2$, the parameter $r_s$ determines the linear slope with which $\mu_S$ rises.
The parameter $t_2$ controls the border and transition between the resummation and fixed-order
regions. Finally, the $t_s$ parameter sets the value of $C$ where the renormalization
scales all join. We require both $\mu_{S}$ and its first derivative to be continuous, and to
this end we have defined the function $\zeta(a_1,b_1,a_2,b_2,t_1,t_2,\,t)$ with $t_1 < t_2$,
which smoothly connects two straight lines of the form $l_1(t) = a_1 \,+\, b_1\,t$ for $t < t_1$
and $l_2(t) = a_2 \,+\, b_2\,t$ for $t > t_2$ at the meeting points $t_1$ and $t_2$. We find that
a convenient form for $\zeta$ is a piecewise function made out of two quadratic functions patched
together in a smooth way.
These two second-order polynomials join at the middle point $t_m=(t_1 + t_2)/2$:
\begin{align}
& \!\!\!\!\!\!\!
 \zeta(a_1,b_1,a_2,b_2,t_1,t_2,t) 
\nn\\*
& = \left\{ \!\begin{tabular}{p{.5\columnwidth} l}
 $\hat a_1 + b_1(t - t_1) + e_1(t - t_1)^2$  &  $~t_1 \le t \le t_m$  \nn \\
 $\hat a_2 + b_2(t - t_2) + e_2(t - t_2)^2$  &  $~t_m \le t \le t_2$
 \end{tabular}
 \right.\!,\\[0.1cm]
 \hat a_1 & = a_1 + b_1\,t_1\,,\qquad \hat a_2 = a_2 + b_2\,t_2\,,\nn\\
e_1 &=\frac{4\,(\hat a_2-\hat a_1)-(3\,b_1 + b_2)\,(t_2-t_1)}{2\,(t_2-t_1)^2}\,,\nn\\[0.1cm]
e_2 &=\frac{4\,(\hat a_1-\hat a_2)+(3\,b_2 + b_1)\,(t_2-t_1)}{2\,(t_2-t_1)^2}\,.
\end{align}
The soft scale profile in \Eq{eq:muSprofile} was also used in Ref.~\cite{Stewart:2014nna} for jet-mass distributions in $pp\to Z+1$-jet.
\begin{table}[t!]
\begin{tabular}{ccc}
parameter\ & \ default value\ & \ range of values \ \\
\hline
$\mu_0$ & $1.1$\,GeV & - \\
$R_0$ & $0.7$\,GeV &  - \\
$n_0$ & $12$ & $10$ to $16$\\
$n_1$ & $25$ & $22$ to $28$\\
$t_2$ & $0.67$ & $0.64$ to $0.7$\\
$t_s$ & $0.83$ & $0.8$ to $0.86$\\
$r_s$ & $2$ & $1.78$ to $2.26$\\
$e_J$ & $0$ & $-\,0.5$ to $0.5$\\
$e_H$ & $1$ & $0.5$ to $2.0$\\
$n_s$ & $0$ & $-\,1$, $0$, $1$\\
\hline
$\Gamma^{\rm cusp}_3$ & $1553.06$ & $-\,1553.06$ to $+\,4659.18$ \\
$s_2^{\widetilde C}$ & $-\,43.2$ & $-\,44.2$ to $-\,42.2$ \\
$j_3$ & $0$ & $-\,3000$ to $+\,3000$ \\
$s_3^{\widetilde C}$ & $0$ & $-\,500$ to $+\,500$ \\
\hline
$\epsilon^\text{low}_{2}$ & $0$ & $-\,1$, $0$, $1$ \\
$\epsilon^\text{high}_{2}$ & $0$ & $-\,1$, $0$, $1$ \\
$\epsilon^\text{low}_{3}$ & $0$ & $-\,1$, $0$, $1$ \\
$\epsilon^\text{high}_{3}$ & $0$ & $-\,1$, $0$, $1$ \\
\end{tabular}
\caption{C-parameter theory parameters relevant for estimating the theory uncertainty, their
default values, and range of values used for the scan for theory uncertainties.}
\label{tab:theoryerr}
\end{table} 
\begin{table}[tbh!]
\begin{tabular}{ccc}
parameter\ & \ default value\ & \ range of values \ \\
\hline
$\mu_0$ & $1.1$\,GeV & -\\
$R_0$ & $0.7$\,GeV & -\\
$n_0$ & $2$ & $1.5$ to $2.5$ \\
$n_1$ & $10$ & $8.5$ to $11.5$\\
$t_2$ & $0.25$ & $0.225$ to $0.275$\\
$t_s$ & $0.4$ & $0.375$ to $0.425$\\
$r_s$ & $2$ & $1.77$ to $2.26$\\
$e_J$ & $0$ & $-\,1.5$ to $1.5$\\
$e_H$ & $1$ & $0.5$ to $2.0$\\
$n_s$ & $0$ & $-\,1$, $0$, $1$\\
\hline
$j_3$ & $0$ & $-\,3000$ to $+\,3000$ \\
$s_3^{\tau}$ & $0$ & $-\,500$ to $+\,500$ \\
\hline
$\epsilon_{2}$ & $0$ & $-\,1$, $0$, $1$ \\
$\epsilon_{3}$ & $0$ & $-\,1$, $0$, $1$ \\
\end{tabular}
\caption{Thrust theory parameters relevant for estimating the theory uncertainty, their
default values, and range of values used for the scan for theory uncertainties.}
\label{tab:theoryerrthrust}
\end{table}

In Ref.~\cite{Abbate:2010xh} slightly different profiles were used. For instance there was no
region of constant soft scale. This can be reproduced from our new profiles by choosing
\mbox{$t_0 = 0$}. Moreover, in Ref.~\cite{Abbate:2010xh} there was only one quadratic form after
the linear term, and the slope was completely determined by other parameters. These new profiles have several
advantages. The most obvious is a variable slope, which allows us to balance the introduction of logs and the
smoothness of the profiles. Additionally, in the new set up, the parameters for different regions are more
independent. For example, the $n_0$ parameter will only affect the nonperturbative region in the new profiles,
while in the old profiles, changing $n_0$ would have an impact on the resummation region. This independence
makes analyzing the different regions more transparent.

For the jet scale, we introduce a ``trumpeting'' factor that modifies the natural relation to the hard and soft
scales in the following way:
\begin{equation} \label{eq:muJprofile}
 \!\!\!\!\!\mu_J(C) = \left\{\!\! \begin{array}{lr}
 \big[\,1 + e_J (C-t_s)^2\,\big] \sqrt{ \mu_H\, \mu_{S} (C)} & C \le t_s\\
 \,\mu_H & C > t_s
 \end{array}
 \right.\!.
\end{equation}
The parameter $e_J$ is varied in our theory scans.

The subtraction scale $R(C)$ can be chosen to be the same as $\mu_S(C)$ in the resummation
region to avoid large logarithms in the subtractions for the soft function. In the
nonperturbative region we do not want the $\mathcal{O}(\alpha_s)$ subtraction piece
to vanish, see \Eq{eq:d123}, so we choose the form
\begin{equation} \label{eq:muRprofile}
\!\!\!\!\!R(C) = \left\{\!\! \begin{array}{ll}
R_0                             & 0 \le C < t_0 \\
\zeta(R_0,\,0,\,0,\,\frac{r_s\,\mu_H}{6},\,t_0,\,t_1,\,C) & t_0 \le C < t_1 \\
\mu_S(C)                        & t_1 \le C \le 1
\end{array}
\right.\!\!.
\end{equation}
The only free parameter in this equation, $R_0$, simply sets the value of $R$ at $C=0$.
The requirement of continuity at $t_1$ in both $R(C)$ and its first derivative are again
ensured by the $\zeta$ function.

In order to account for resummation effects in the nonsingular partonic cross section,
which we cannot treat coherently, we vary $\mu_{\rm ns}$. We use three possibilities:
\begin{equation} \label{eq:muNSprofile}
 \mu_{\rm ns}(C) = \left\{\! \begin{array}{ll}
 \frac{1}{2} \big[\,\mu_H(C) + \mu_J (C)\,\big] &~ n_s \,= ~~\,1 \\
 \mu_H  & ~n_s \,= ~~\,0 \\
 \frac{1}{2} \big[\,3\,\mu_H(C) - \mu_J (C)\,\big] &~ n_s \,= -\,1
 \end{array}
 \right.\!\!.
\end{equation}
Using these variations, as opposed to those in Ref.~\cite{Abbate:2010xh}, gives more symmetric uncertainty
bands for the nonsingular distribution.

The plot in Fig.~\ref{fig:profile-variation} shows the scales for the default parameters for the case $Q=m_Z$ (thick lines). Also shown (gray dashed lines) are plots of $QC/6$ and $Q\sqrt{C/6}$.  In the resummation region, these correspond fairly well with the profile functions, indicating that in this region our analysis will avoid large logarithms. Note that the soft and jet scales in the plot would exactly match the gray dashed lines in the region $0.25<C<0.67$ if we took $r_s=1$ as our default. For reasons discussed in \Sec{subsec:slope-C} we use $r_s=2$ as our default value. We also set as default values $\mu_0 = 1.1\, {\rm GeV}$, $R_0 = 0.7\, {\rm GeV}$, $e_H=1$, $e_J=0$, and $n_s=0$. Default central values for other profile parameters for $C$ are listed in \Tab{tab:theoryerr}.

Perturbative uncertainties are obtained by varying the profile parameters. We hold $\mu_0$ and $R_0$ fixed, which are the parameters relevant in the region impacted by the entire nonperturbative shape function. They influence the meaning of  the nonperturbative soft function parameters in $F_C$.  The difference of the two parameters is important for renormalon subtractions and hence should not be varied ($\mu_0-R_0 = 0.4\, {\rm GeV}$) to avoid changing the meaning of $F_C$.  Varying $\mu_0$ and $R_0$ keeping the difference fixed has a very small impact compared to variations from $F_C$ parameters, as well as other profile parameters, and hence is also kept constant.  We are then left with eight profile parameters to vary during the theory scan, whose central values and variation ranges used in our analysis are $r_s = 2\times 1.13^{\pm1}$, $n_0=12\,\pm\,2$, $n_1=25\,\pm\,3$, $t_2 = 0.67\,\pm\,0.03$, $t_s=0.83\,\pm\,0.03$, $e_J = 0\,\pm\, 0.5$,  $e_H = 2^{\pm1}$, and $n_s = 0\,\pm\,1$.  The resulting ranges are also listed in \Tab{tab:theoryerr}, and the effect of these variations on the scales is plotted in \Fig{fig:profile-variation}. Since  we have so many events in our EVENT2 runs, the effect of $\epsilon_2^{\rm low}$ is completely negligible in the theory uncertainty scan. Likewise, the effect of $\epsilon_2^{\rm high}$ is also tiny above the shoulder
region.

Due to the advantages of the new profile functions, we have implemented them for the thrust predictions from Refs.~\cite{Abbate:2010xh} as well. For thrust we redefine $r_s \to 6\, r_s$, which eliminates all four appearances of the factor of 1/6 in \Eqs{eq:muSprofile}{eq:muRprofile}. After making this substitution, we can specify the theory parameters for thrust, which are summarized in \Tab{tab:theoryerrthrust}.These choices create profiles and profile variations that are very similar to those used in Ref.~\cite{Abbate:2010xh}. The only noticeable difference is the flat $\mu_S$ in the nonperturbative region (which is relevant for a fit to the full shape function but is irrelevant for the $\alpha_s$ tail fit).

\begin{figure*}[t!]
\subfigure[]{
\includegraphics[width=0.3\textwidth]{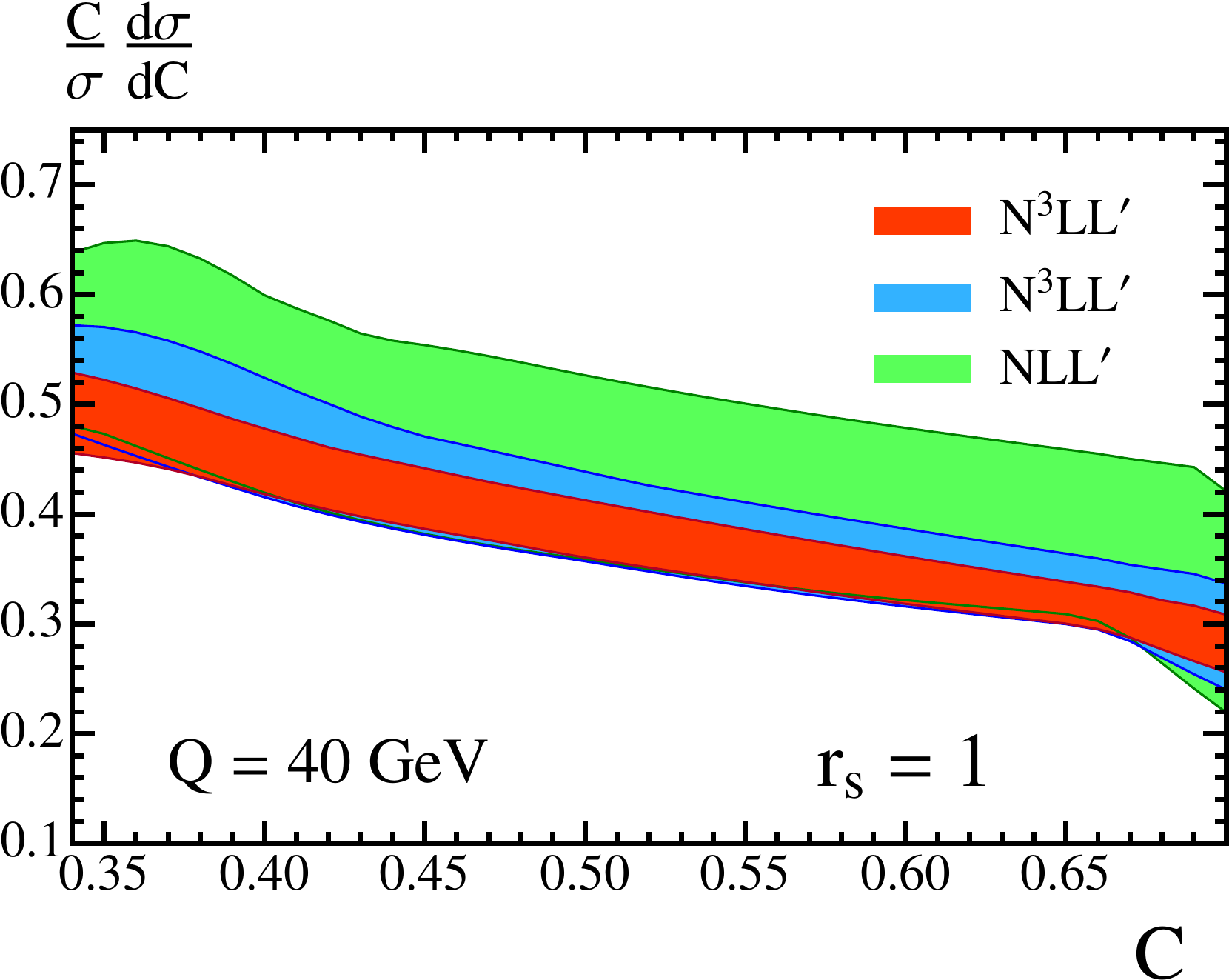}
\label{fig:slope1-40}
}
\subfigure[]
{
\includegraphics[width=0.3\textwidth]{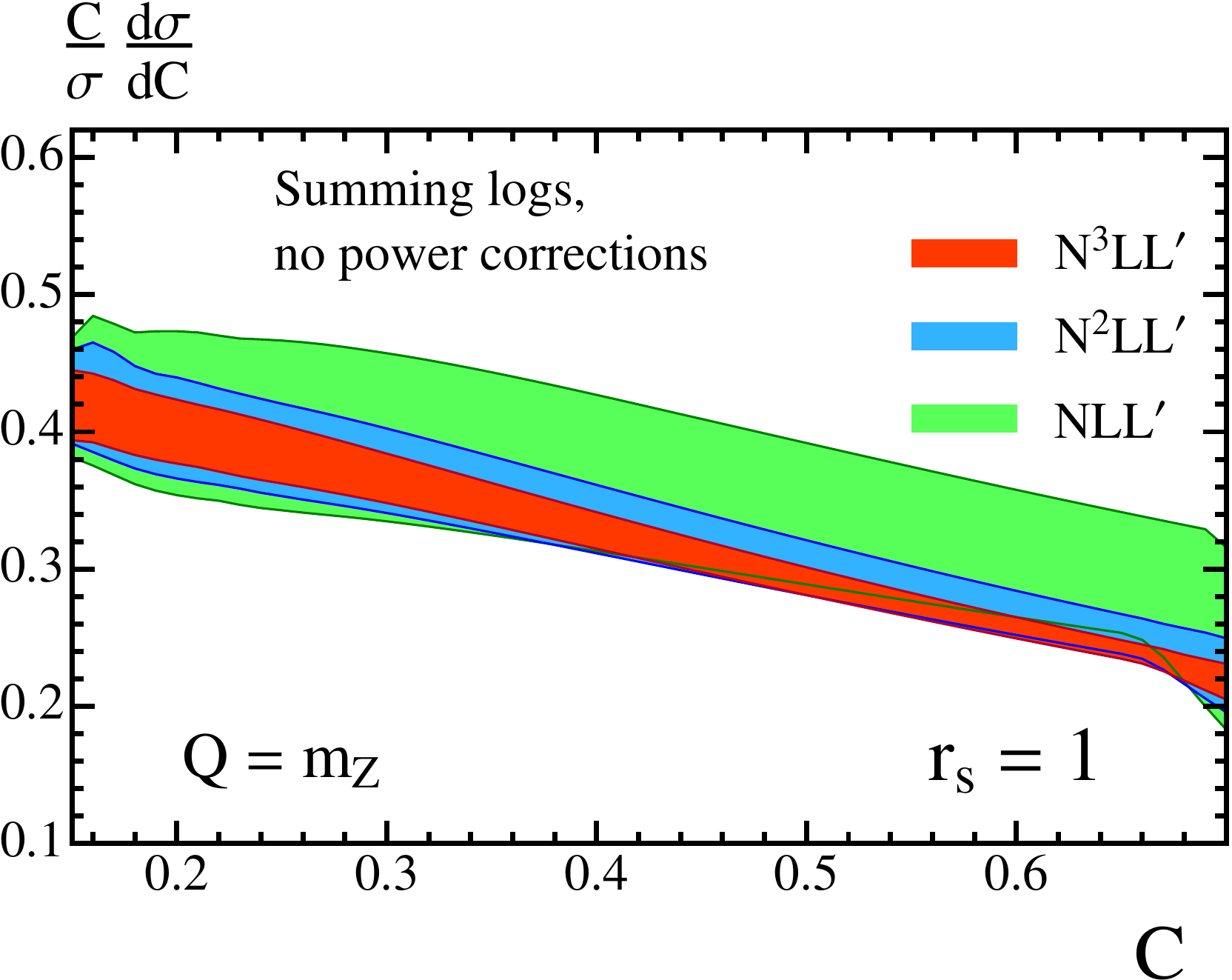}
\label{fig:slope1-Mz}
}
\subfigure[]{
\includegraphics[width=0.3\textwidth]{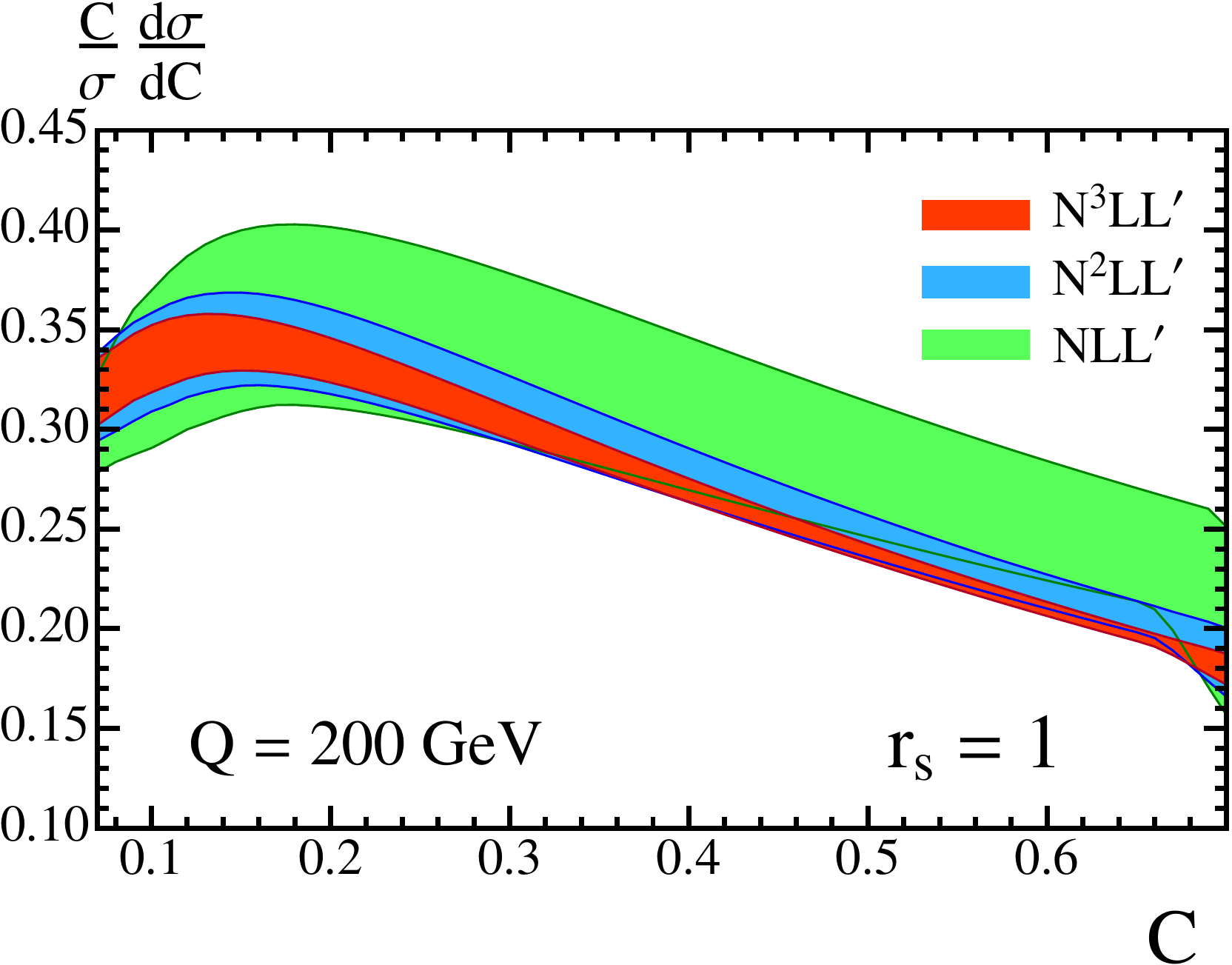}
\label{fig:slope1-200}
}
\subfigure[]{
\includegraphics[width=0.3\textwidth]{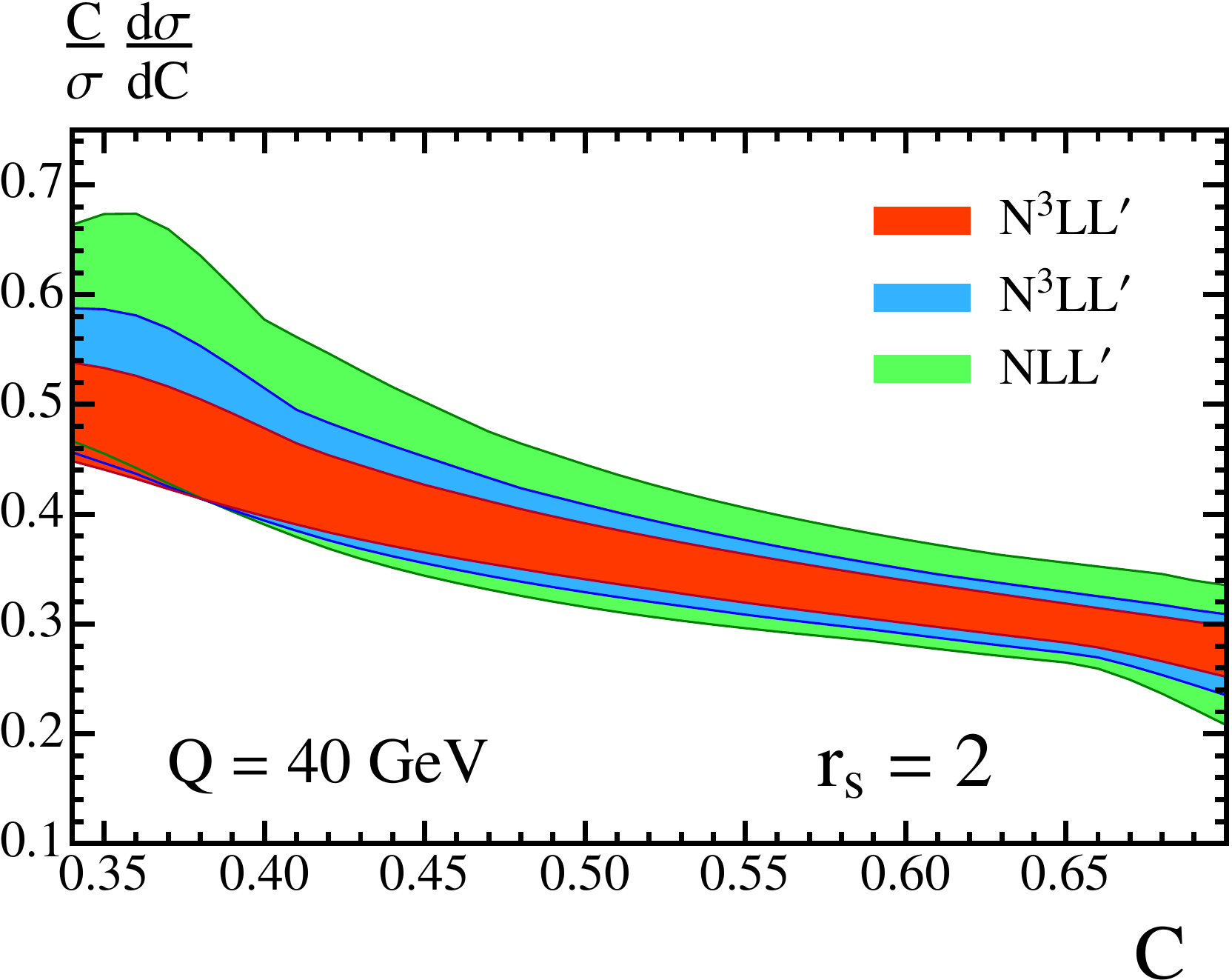}
\label{fig:slope2-40}
}
\subfigure[]{
\includegraphics[width=0.3\textwidth]{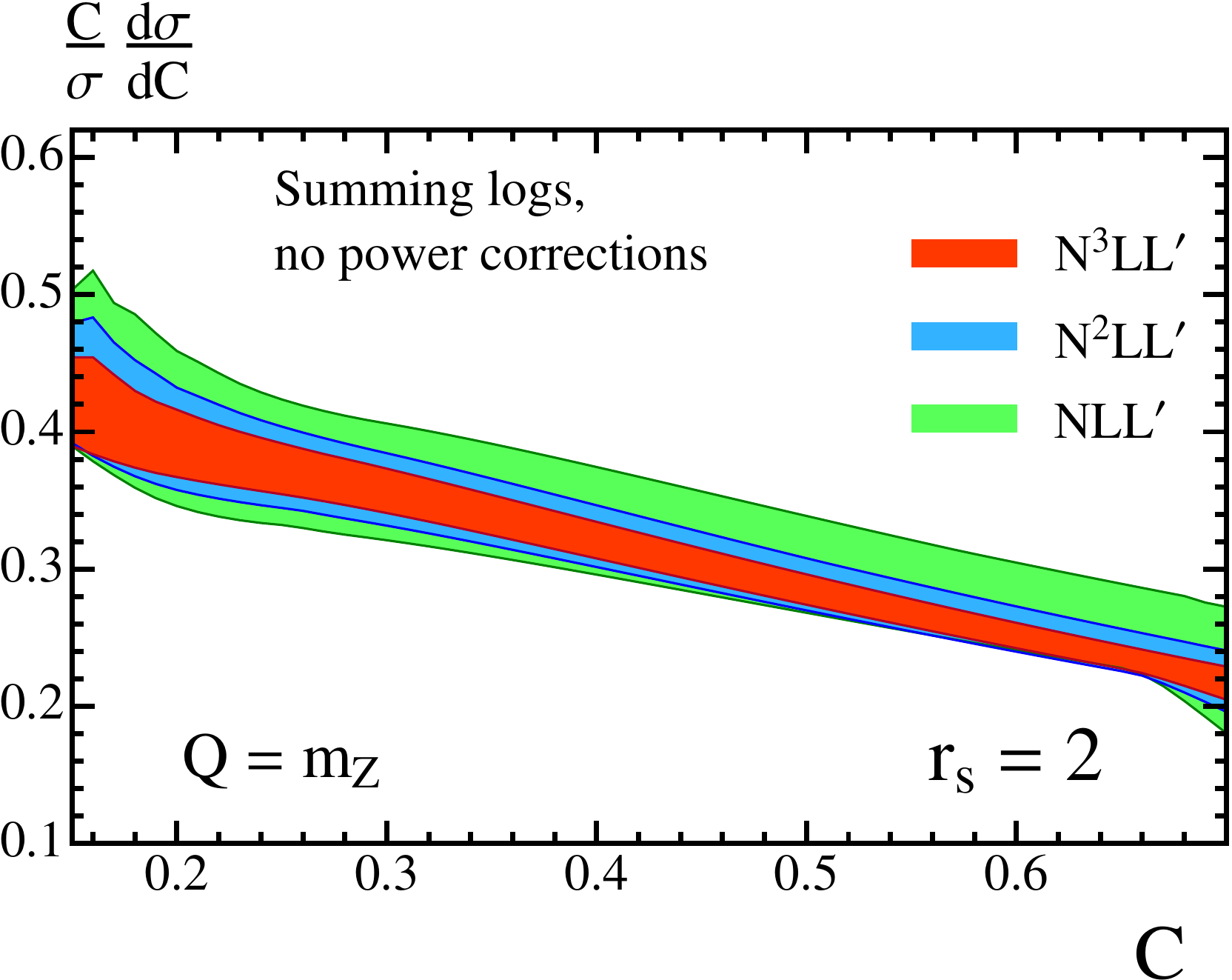}
\label{fig:slope2-Mz}
}
\subfigure[]{
\includegraphics[width=0.3\textwidth]{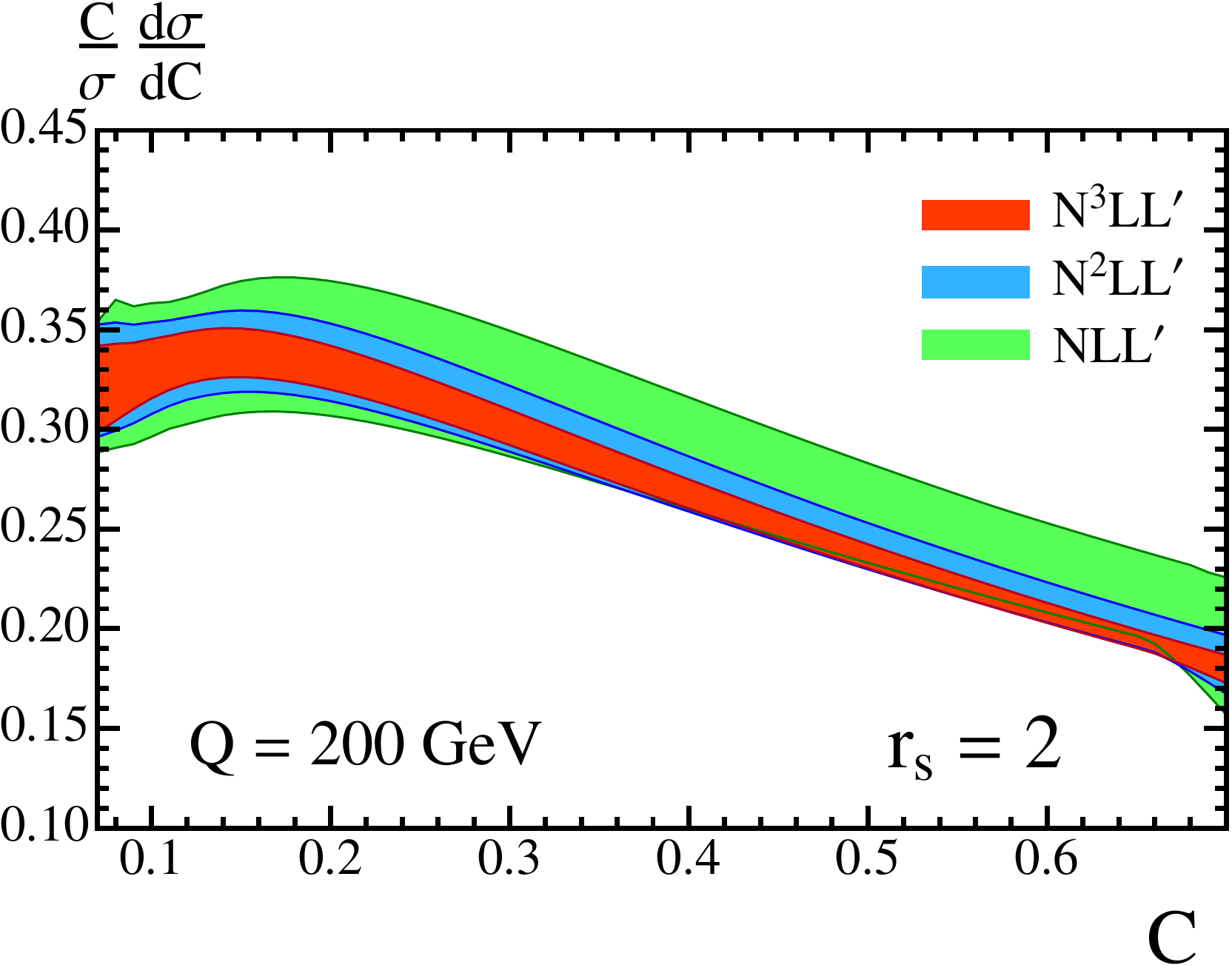}
\label{fig:slope2-200}
}
\subfigure[]
{
\includegraphics[width=0.3\textwidth]{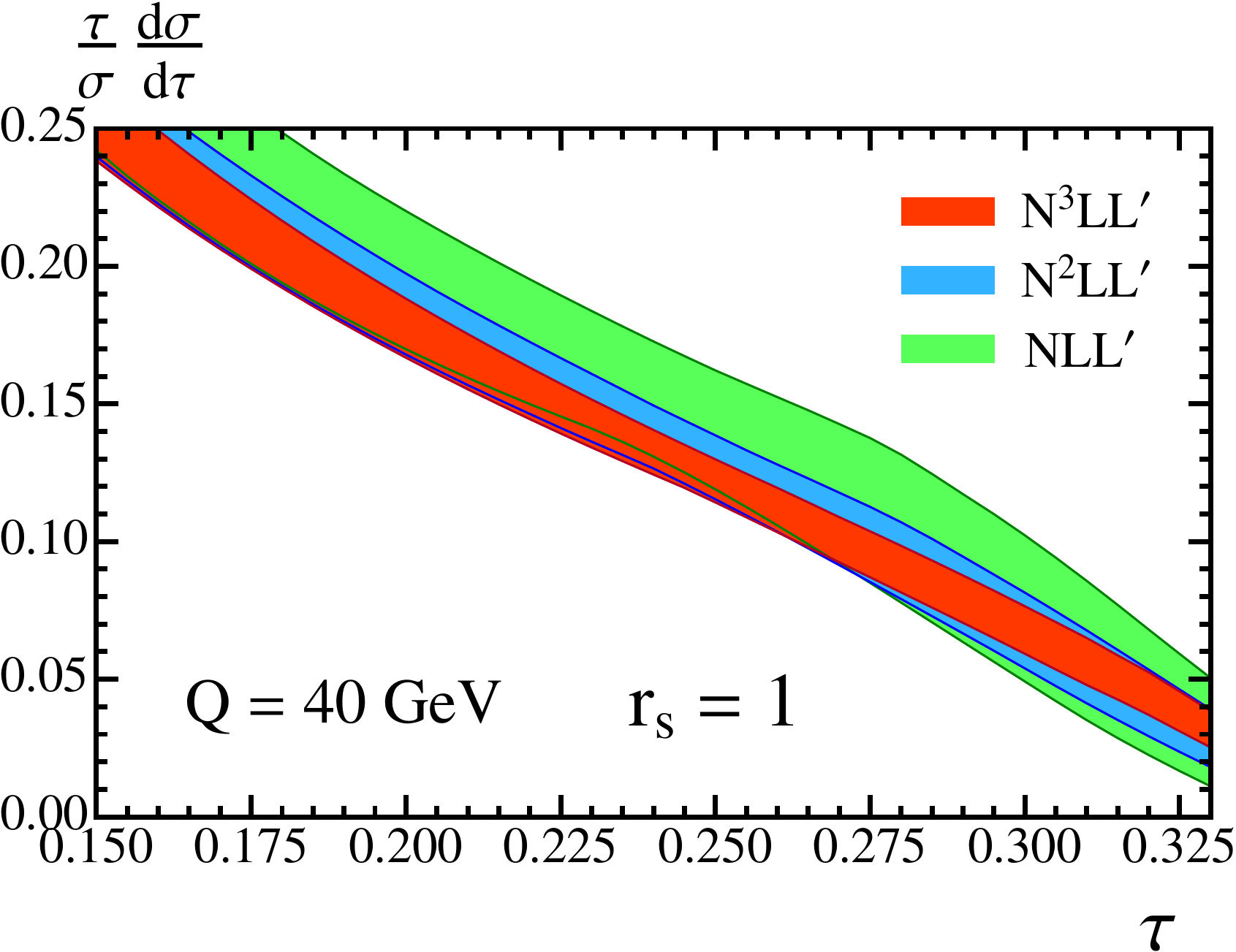}
\label{fig:slope1-40-T}
}
\subfigure[]{
\includegraphics[width=0.3\textwidth]{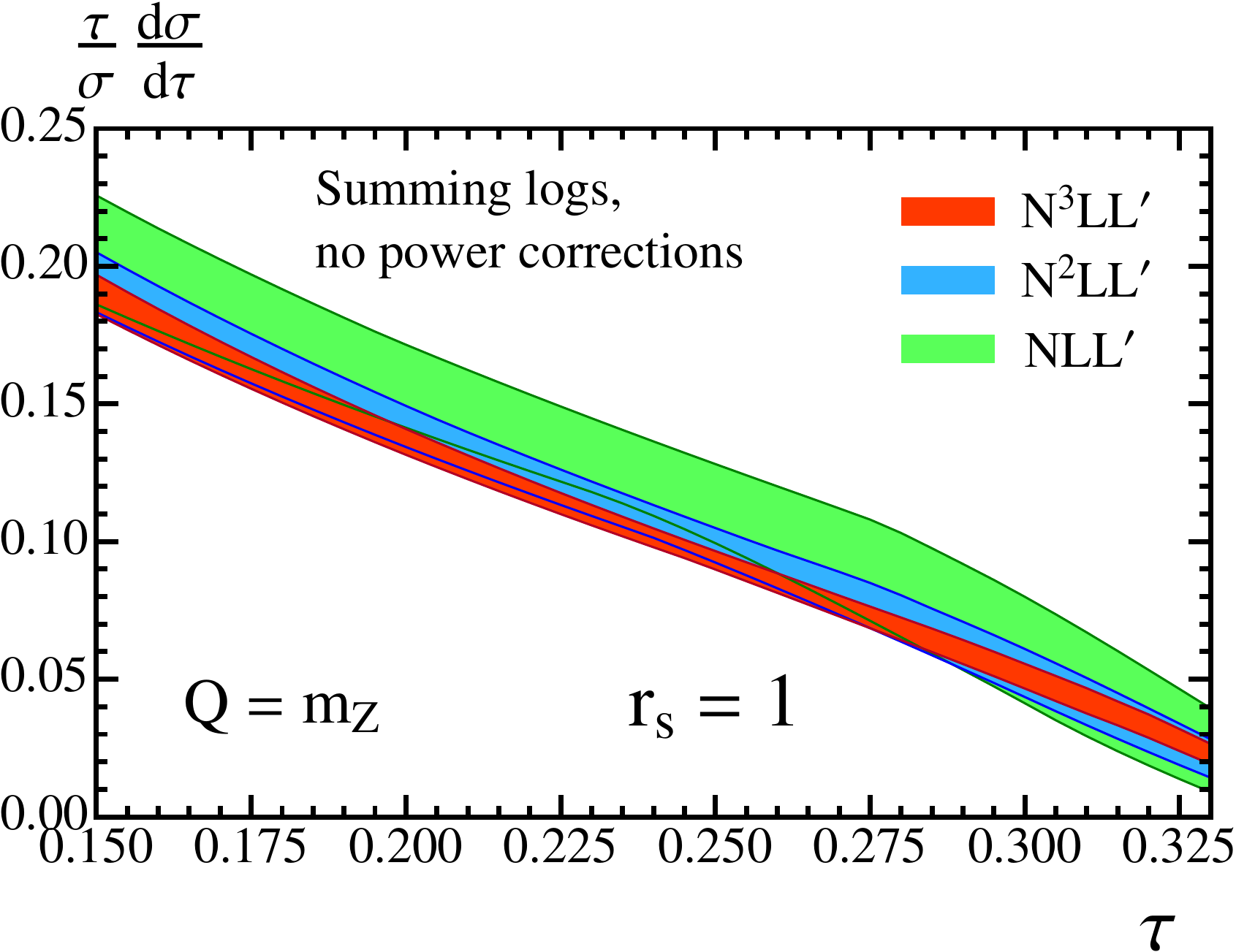}
\label{fig:slope1-Mz-T}
}
\subfigure[]{
\includegraphics[width=0.3\textwidth]{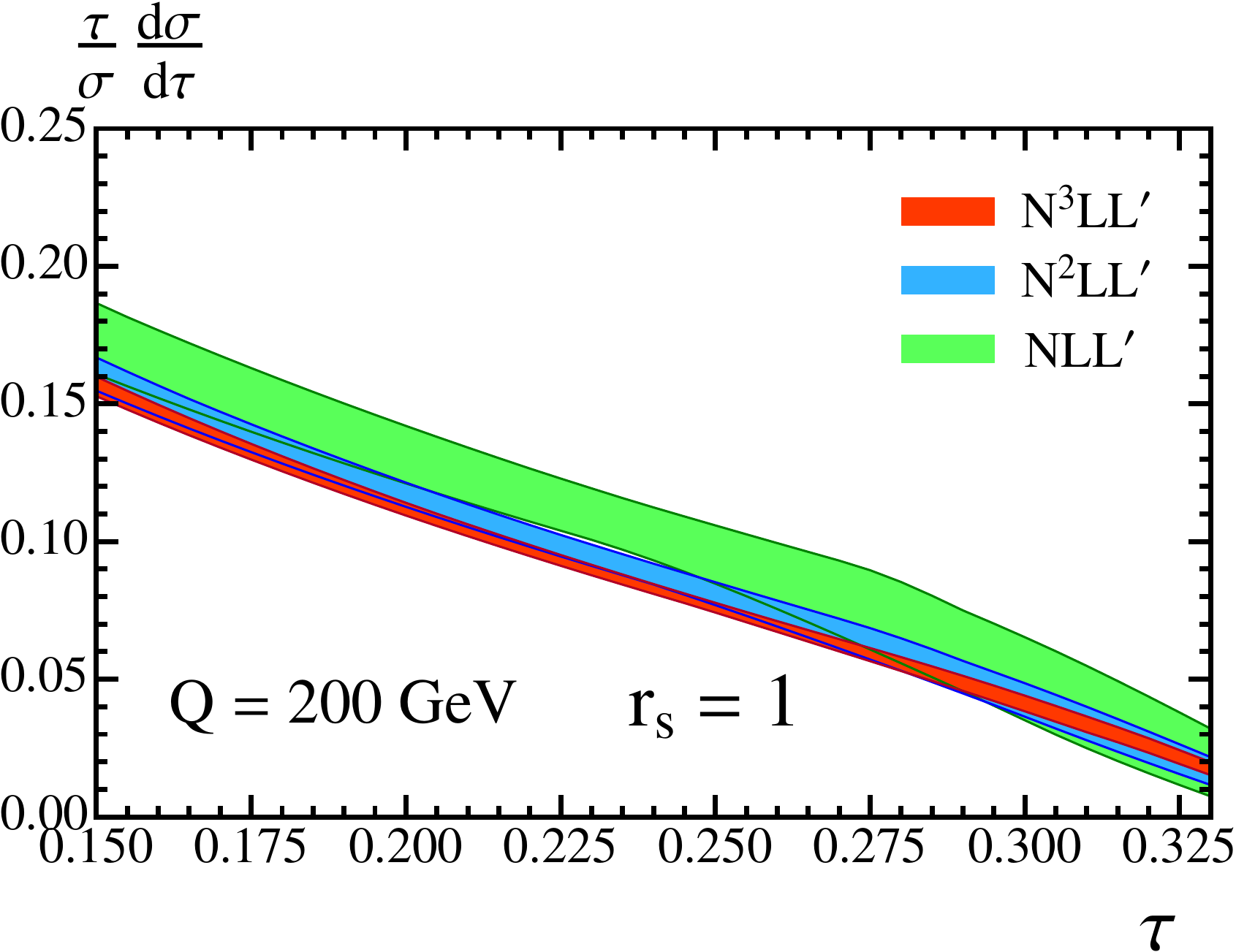}
\label{fig:slope1-200-T}
}
\subfigure[]{
\includegraphics[width=0.3\textwidth]{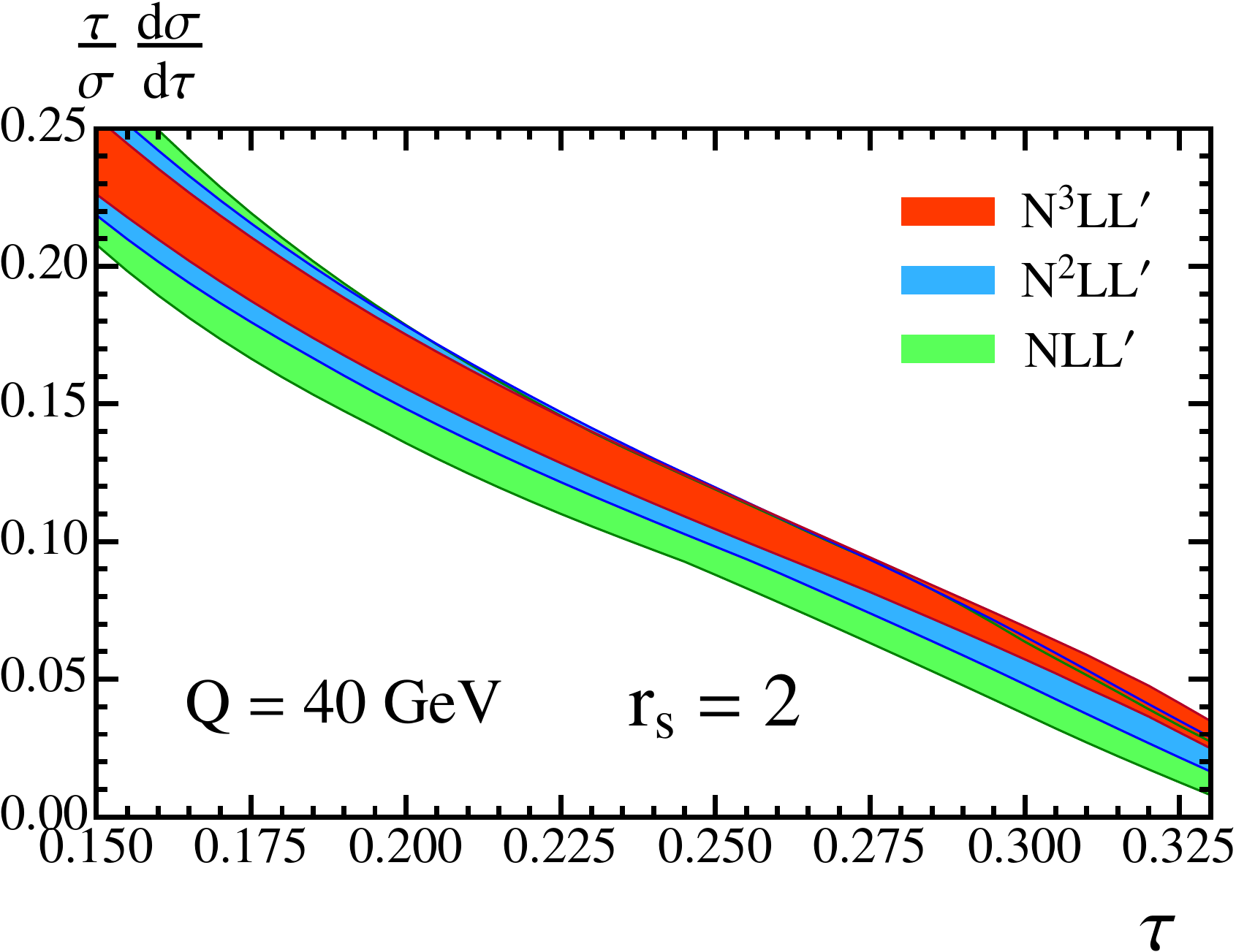}
\label{fig:slope2-40-T}
}
\subfigure[]{
\includegraphics[width=0.3\textwidth]{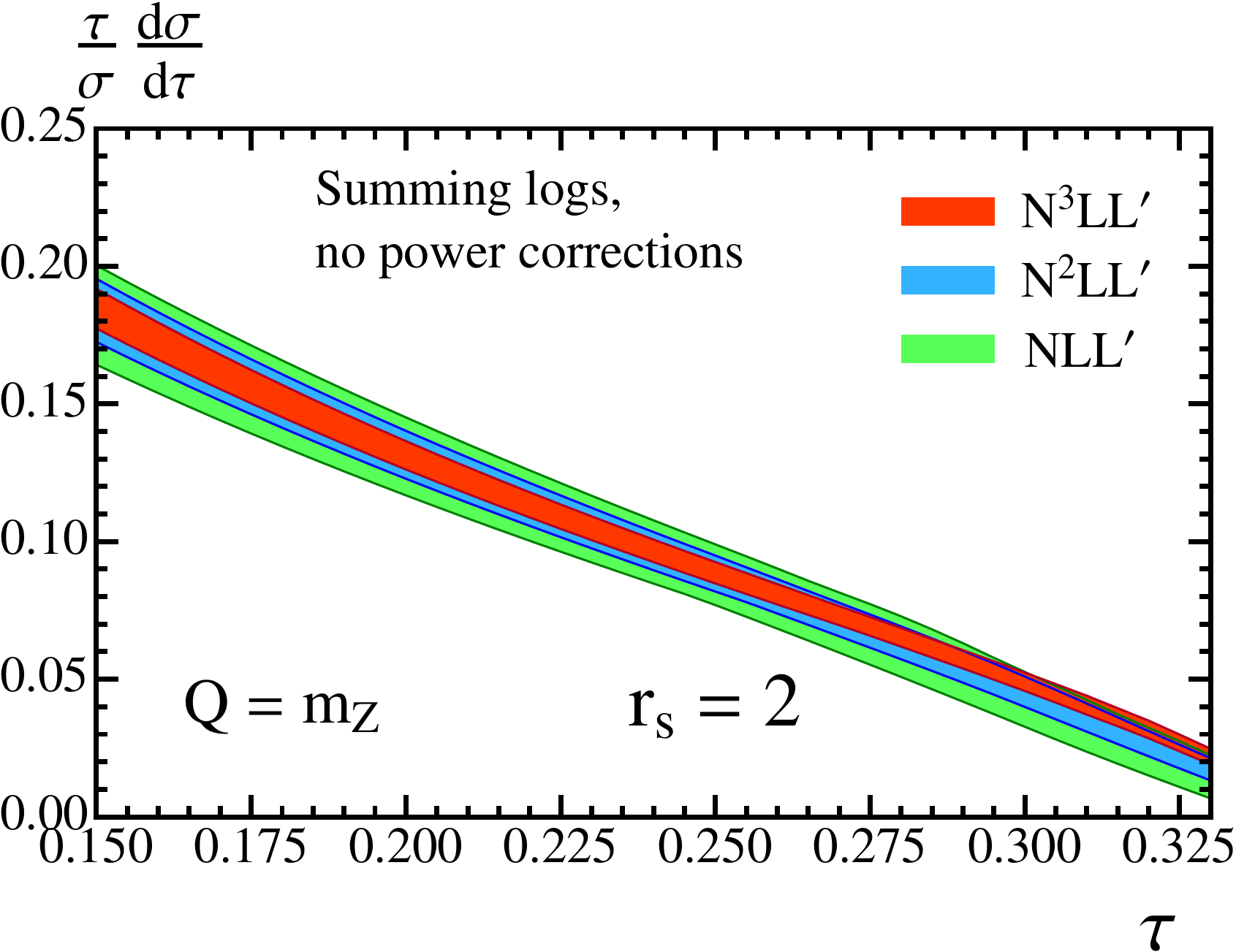}
\label{fig:slope2-Mz-T}
}
\subfigure[]{
\includegraphics[width=0.3\textwidth]{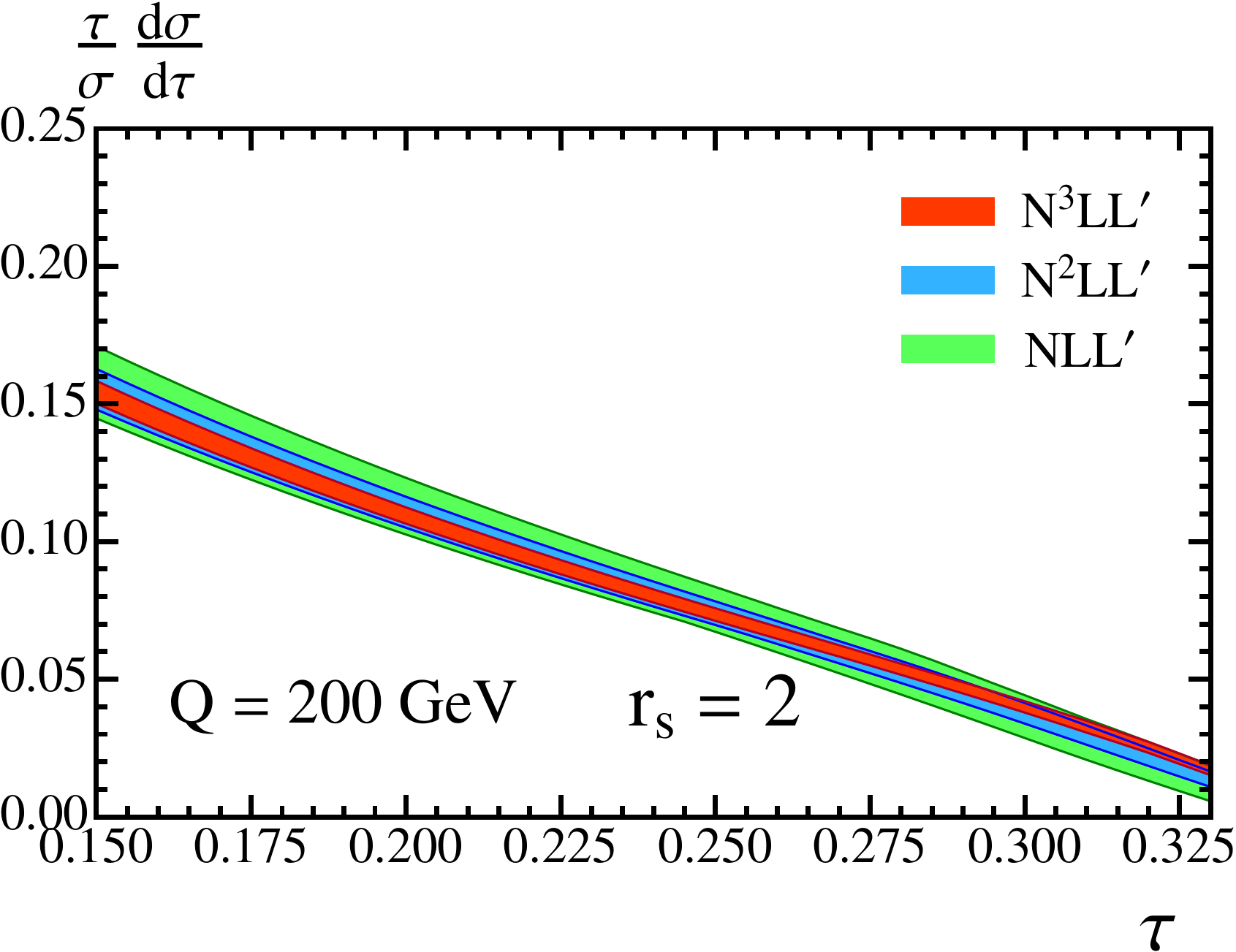}
\label{fig:slope2-200-T}
}
\caption{Theory scan for C-parameter [\,panels (a) to (f)\,] and Thrust [\,panels (g) to (l)\,] cross section uncertainties at
various orders, for few center-of-mass energies $Q$ and slopes $r_s$. The theoretical predictions include log resummation and are purely perturbative. The respective upper and lower rows use $r_s = 1$ and $r_s = 2$, respectively. The left, center, and right columns correspond to $Q = 40$\,GeV, $91.2$\,GeV, and $200$\,GeV, respectively. Here we use $\alpha_s(m_Z)=0.1141$. }
\label{fig:12-plots}%
\end{figure*}

\section{Results}
\label{sec:results}
In this section we present our final results for the \mbox{C-parameter} cross section, comparing  to the thrust cross section  when appropriate.  We will use different levels of theoretical accuracy for these analyses. When indicating the perturbative precision, and whether or not the power correction $\Omega_1$ is included and at what level of precision, we follow the notation
\begin{align}
 & {\cal O}(\alpha_s^k) 
   & \phantom{x} & \text{fixed order up to ${\cal O}(\alpha_s^k)$} 
  \nn\\ 
 & \text{N}^k\text{LL}^\prime \!+\! {\cal O}(\alpha_s^k) 
   & \phantom{x} & \text{perturbative resummation} 
  \nn\\ 
 & \text{N}^k\text{LL}^\prime \!+\! {\cal O}(\alpha_s^k)  \!+\! {\overline \Omega}_1
   & \phantom{x} & \text{$\overline{\rm MS}$ scheme for $\Omega_1$} 
  \nn\\ 
 & \text{N}^k\text{LL}^\prime \!+\! {\cal O}(\alpha_s^k) 
   \!+\!  {\Omega}_1(R,\mu)
   & \phantom{x} & \text{Rgap scheme for $\Omega_1$} 
  \nn\\ 
 & \text{N}^k\text{LL}^\prime \!+\! 
   {\cal O}(\alpha_s^k) \!+\! {\Omega}_1(R,\mu,r)
   & \phantom{x} & \text{Rgap scheme with }
  \nn\\
 & & \phantom{x} & \  \text{hadron masses for $\Omega_1$} 
  \,. \nn
\end{align}
In the first three subsections, we discuss the determination of higher-order perturbative coefficients in the cross section, the impact of the slope parameter $r_s$, and the order-by-order convergence and uncertainties. 
Since the effect of the additional running induced by the presence of hadron masses is a relatively small effect on the cross section, we leave their discussion to the final fourth subsection.

\begin{table}[t!]
\begin{tabular}{c|c}
Resummation Order&  Calculable $G_{ij}$'s and $B_i$'s \\
\hline
LL  &  $G_{i,\,i+1}$ \\
NLL$^\prime$  &  $G_{i,\,i}$ and $B_1$\\
N${}^2$LL$^\prime$  &  $G_{i,i-1}$ and $B_2$\\ 
N${}^3$LL$^\prime$  &  $G_{i,i-2}$ and $B_3$\\ 
\end{tabular}
\caption{By doing resummation to the given order, we can access results for the entire hierarchy of $G_{ij}$'s listed.}
\label{tab:Gijorders}
\end{table}

\begin{figure*}[t!]
\includegraphics[width=0.95\columnwidth]{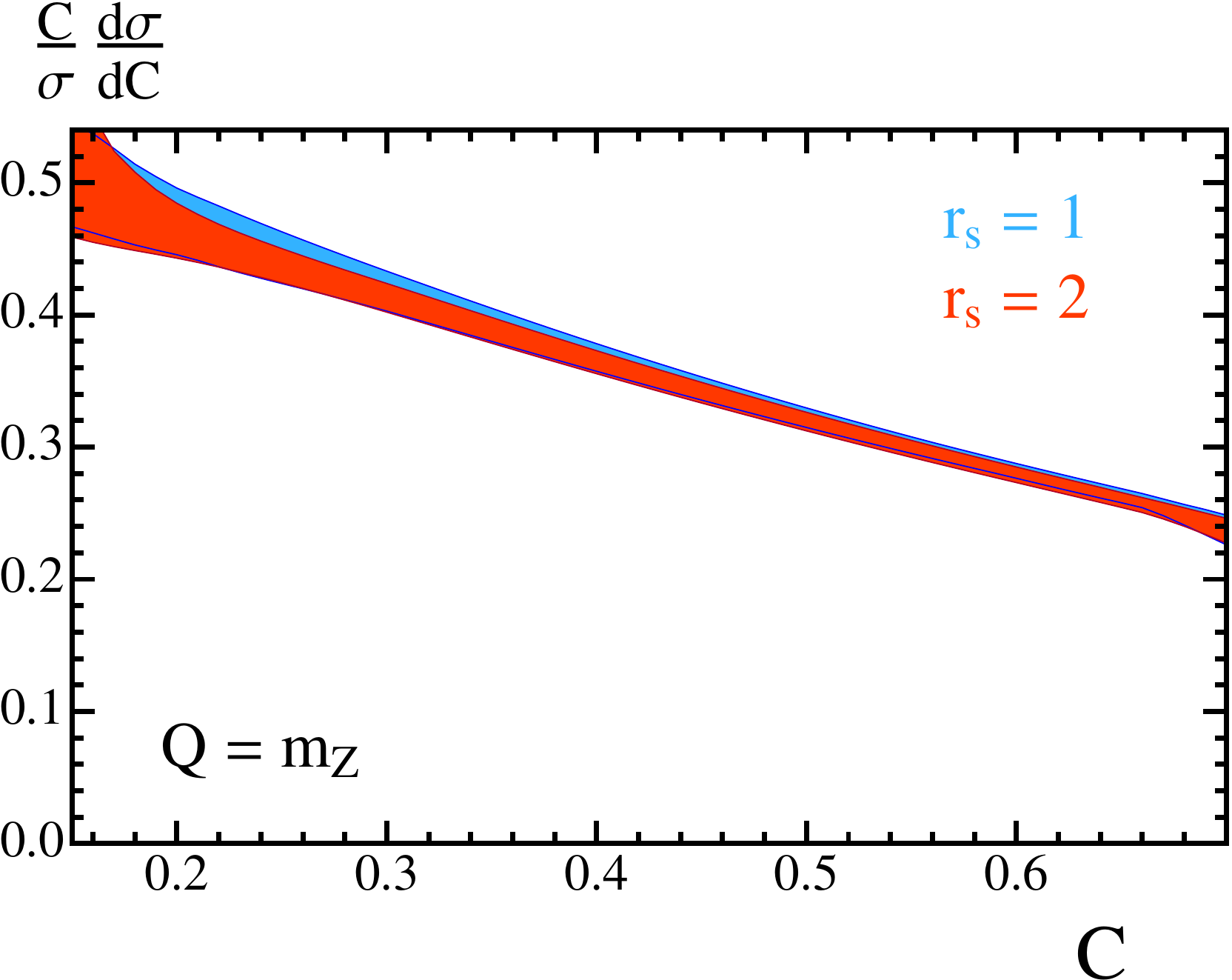}
\includegraphics[width=0.95\columnwidth]{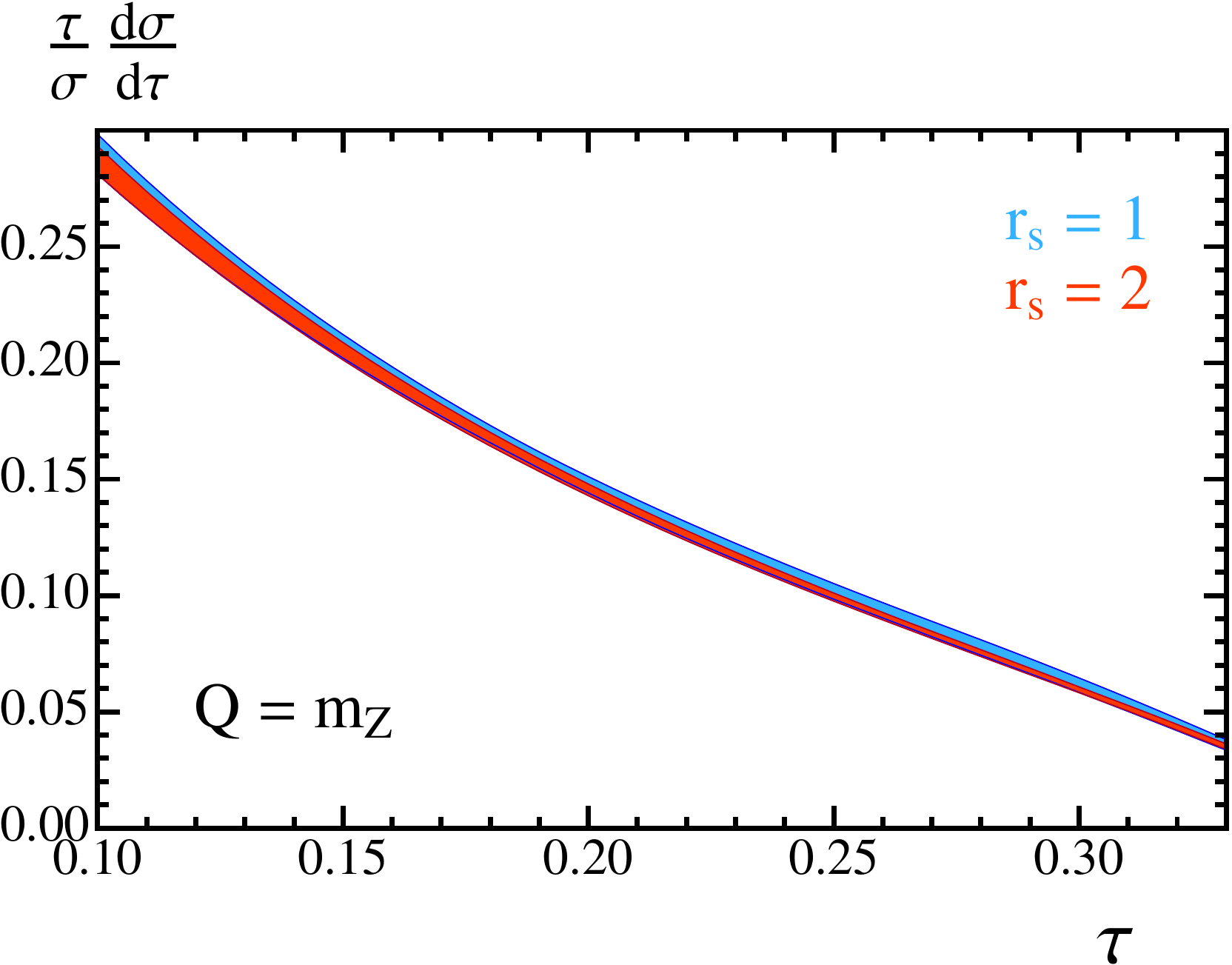}
\caption{C-parameter (left panel) and thrust (right panel) cross section predictions at N$^3$LL$^\prime+{\cal O}(\alpha_s^3)$ + $\Omega_1(R,\mu)$ for $r_s = 1$ (blue) and $r_s = 2$ (red).
For these plots we use our most complete setup, with power corrections in the renormalon-free Rgap scheme. Here we use $\alpha_s(m_Z)=0.1141$ and $\Omega_1(R_\Delta,\mu_\Delta)=0.33\,{\rm GeV}$.}
\label{fig:highest-slope}
\end{figure*}
\subsection{$\mathbf{G_{ij}}$ expansion}
\label{subsec:Gij}
Our N$^3$LL$^\prime$ resummed predictions can be used to compute various coefficients of the most singular terms in the cross section. In this determination only perturbative results are used.  In order to exhibit the terms that are determined by the logarithmic resummation, one can take $\mu=Q$ and write the cumulant function as:
\begin{align}\label{eq:BG1}
\Sigma_0 (C) = & \,\dfrac{1}{\sigma_0}\!\int^C_0\!\!\! \df C^\prime\, \dfrac{\df \sigma}{\df C^\prime} =
\left(\! 1 + \sum_{m=1}^{\infty} B_{m}^{[0]} \!\left( \dfrac{\alpha_s(Q)}{2 \pi} \right)^{\!\!m} \right) \nn \\
&\times \exp\! \left( \sum_{i=1}^{\infty} \sum_{j=1}^{i+1} G_{i j}\! \left( \dfrac{\alpha_s(Q)}{2 \pi} \right)^{\!\!i}
\!\ln^j\!\!\left( \dfrac{6}{C} \right) \!\right)\!,
\end{align}
where $\sigma_0$ is the tree-level total cross section.
A different normalization with respect to the total cross section including all QCD corrections is also used in the literature and reads
\begin{align}\label{eq:BG2}
\Sigma (C) = &\, \dfrac{1}{\sigma_\text{had}}\!\int^C_0 \!\df C^\prime \dfrac{\df \sigma}{\df C^\prime} =
\left(\! 1 + \sum_{m=1}^{\infty} B_m\! \left( \dfrac{\alpha_s(Q)}{2 \pi} \right)^{\!\!m} \right) \nn \\
& \times\exp\! \left( \sum_{i=1}^{\infty} \sum_{j=1}^{i+1} G_{i j}\! \left( \dfrac{\alpha_s(Q)}{2 \pi} \right)^{\!\!i}
\!\ln^j\!\! \left( \dfrac{6}{C} \right)\! \right)\!.
\end{align}
Notice that the different normalizations do not affect the $G_{ij}$'s and only change the non-logarithmic pieces. 
Our $\text{N}^3 \text{LL}^\prime$ result allows us to calculate the $B_{m}^{[0]}$'s and $B_{m}$'s to third order and 
entire hierarchies of the $G_{ij}$ coefficients as illustrated in \Tab{tab:Gijorders}.

The results for these coefficients through $G_{34}$ are collected in App.~\ref{ap:Gijcoefficients}. Note that, due to the equivalence of the thrust and C-parameter distributions at NLL (when using $\widetilde{C}$), we know that the $G_{i,\,i+1}$ and $G_{i,\,i}$ series are equal for these two event shapes~\cite{Catani:1998sf}. From our higher-order resummation analysis, we find that the $G_{i,i-1}$ and $G_{i,i-2}$ coefficients differ for the C-parameter and thrust event shapes because the fixed-order $s_1^{\widetilde C,\tau}$ and $s_2^{\widetilde C,\tau}$ constants differ and enter into the resummed results at N$^2$LL$^\prime$ and N$^3$LL$^\prime$ respectively. The values of all the $B_i$ coefficients are also different. (If we had instead used the unprimed counting for logarithms, N$^k$LL, there is less precision obtained at a given order, and each index of a $B_i$ in \Tab{tab:Gijorders} would be lowered by $1$.)

The $G_{ij}$ and $B_m$ serve to illustrate the type of terms that are included by having resummation and fixed-order terms at a given order. They are not used explicitly for the resummed analyses in the following sections, which instead exploit the full resummed factorization theorem in \Eq{eq:singular-resummation}.

\begin{figure*}[t!]
\includegraphics[width=1\columnwidth]{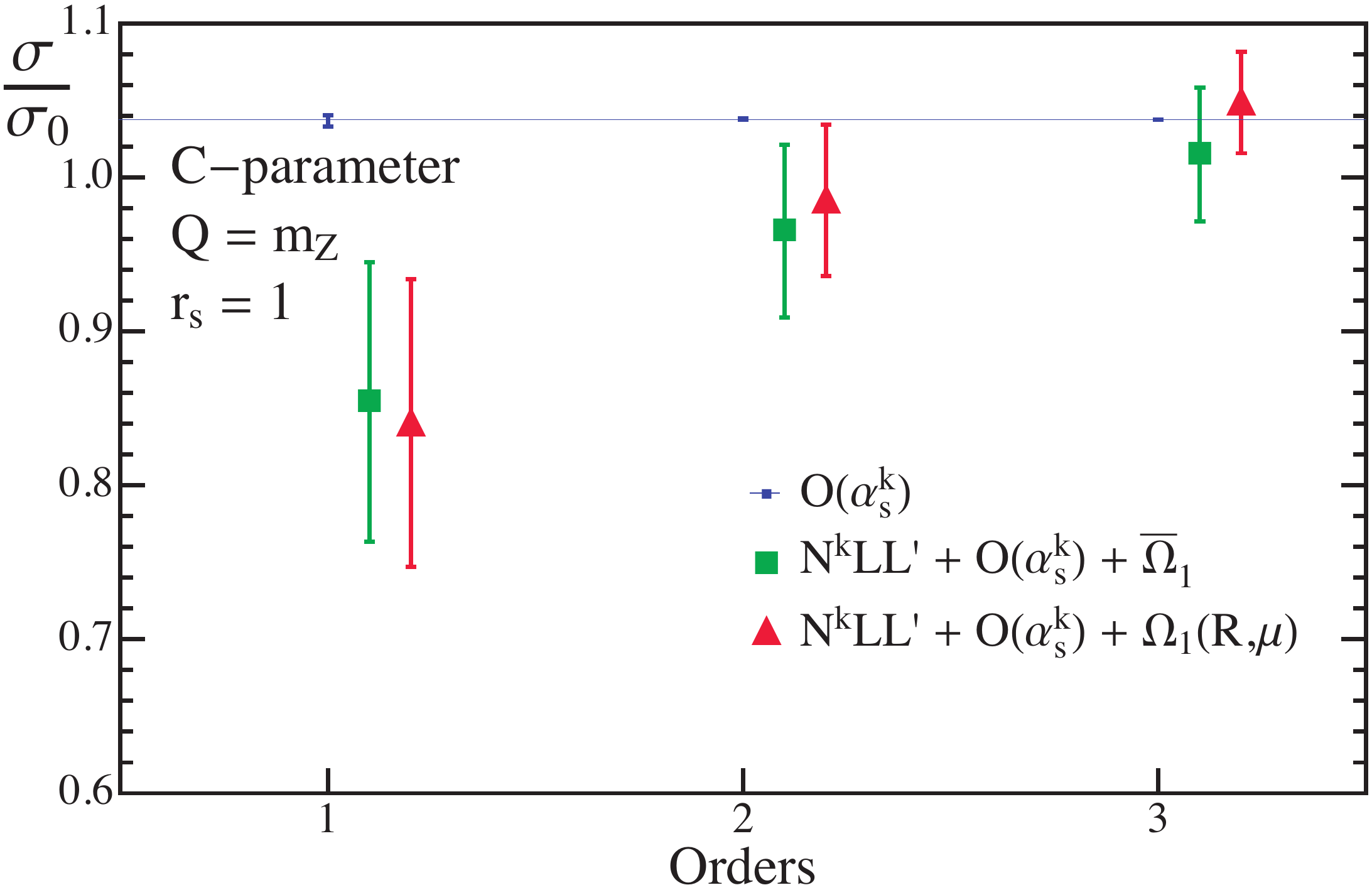}
\includegraphics[width=1\columnwidth]{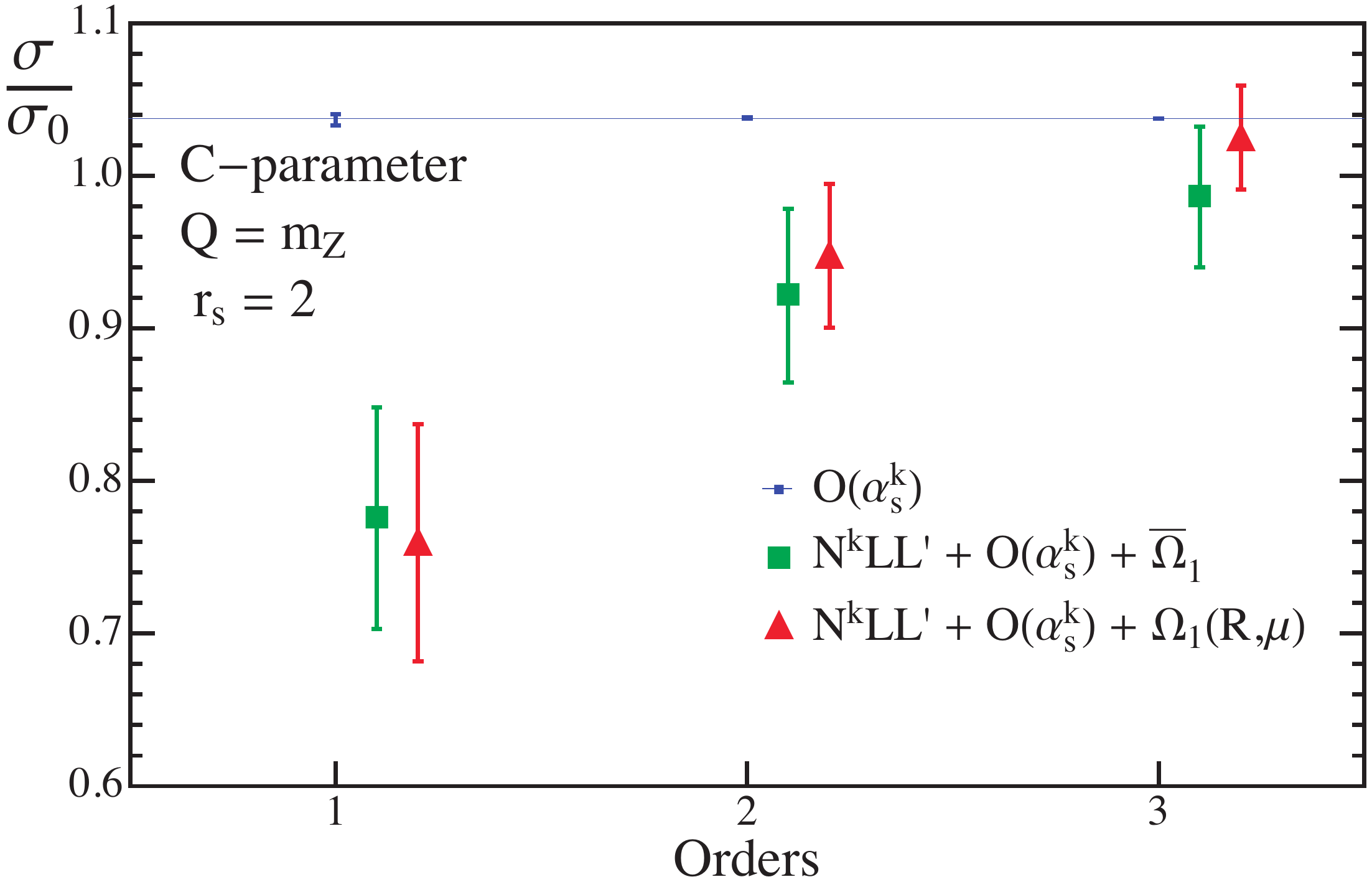}
\includegraphics[width=1\columnwidth]{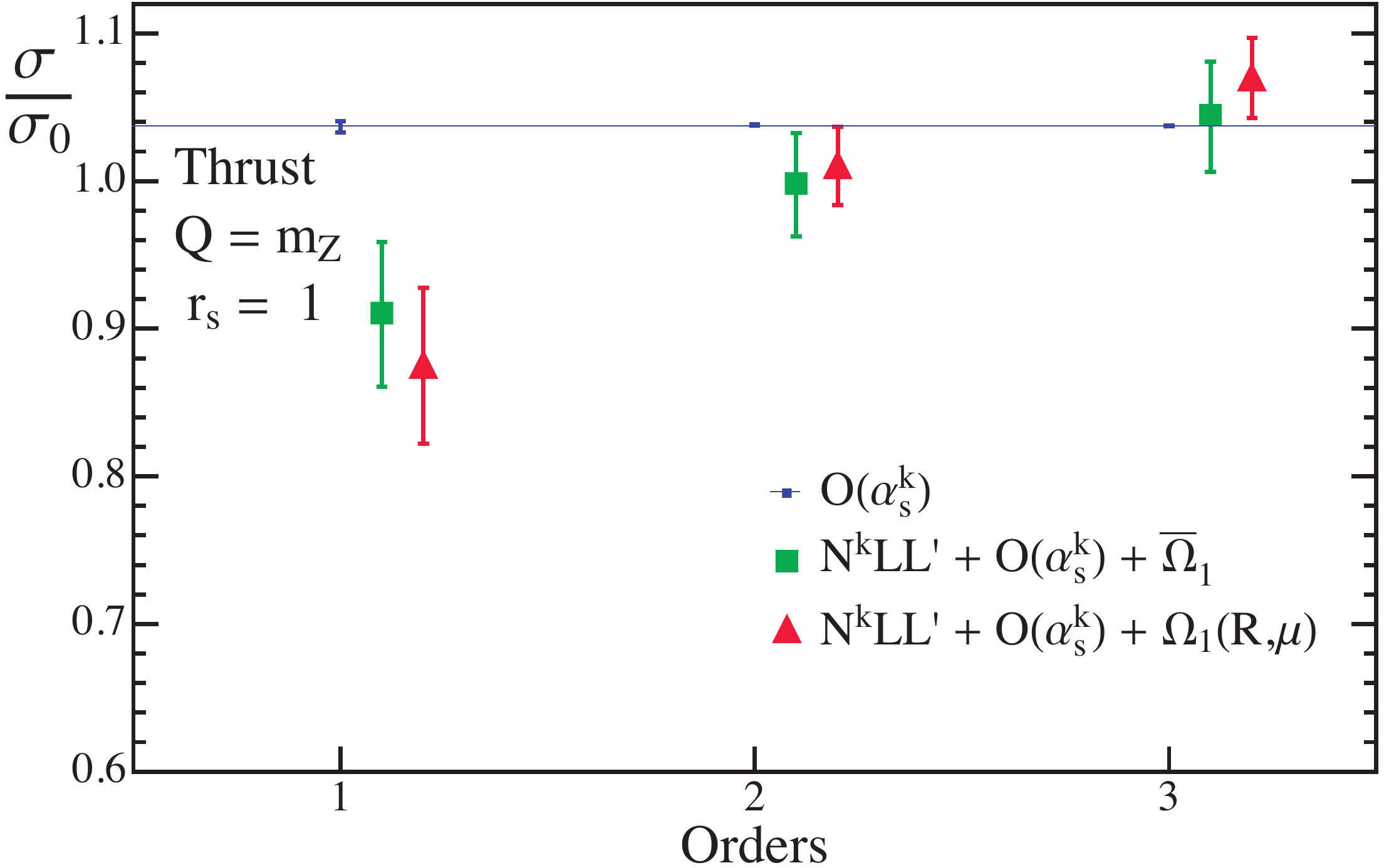}
\includegraphics[width=1\columnwidth]{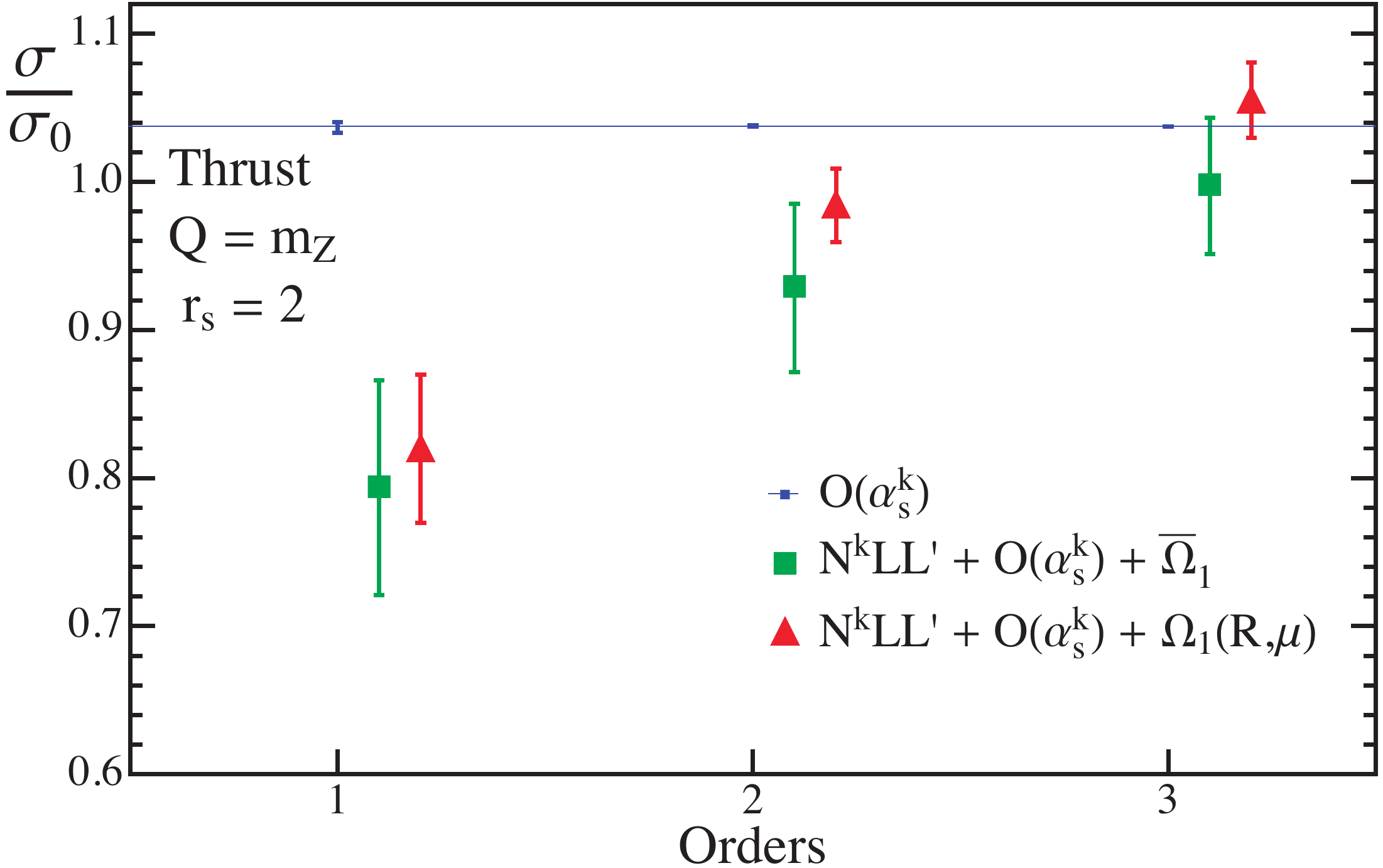}
\caption{\label{fig:norm-C}Total hadronic cross section obtained from integrating the resummed cross section. The top two panels show the prediction for $r_s = 1$ and $r_s = 2$ for C-parameter, respectively. Likewise, the bottom two panels  show the thrust results. Green squares correspond to the prediction with log resummation and the power correction in the $\msbar$ scheme, whereas red triangles have log resummation and the power correction in  the Rgap scheme. The blue points correspond to the fixed-order prediction, and the blue line shows the highest-order FO prediction. }
\end{figure*}
\begin{figure}[t!]
\begin{center}
\includegraphics[width=0.95\columnwidth]{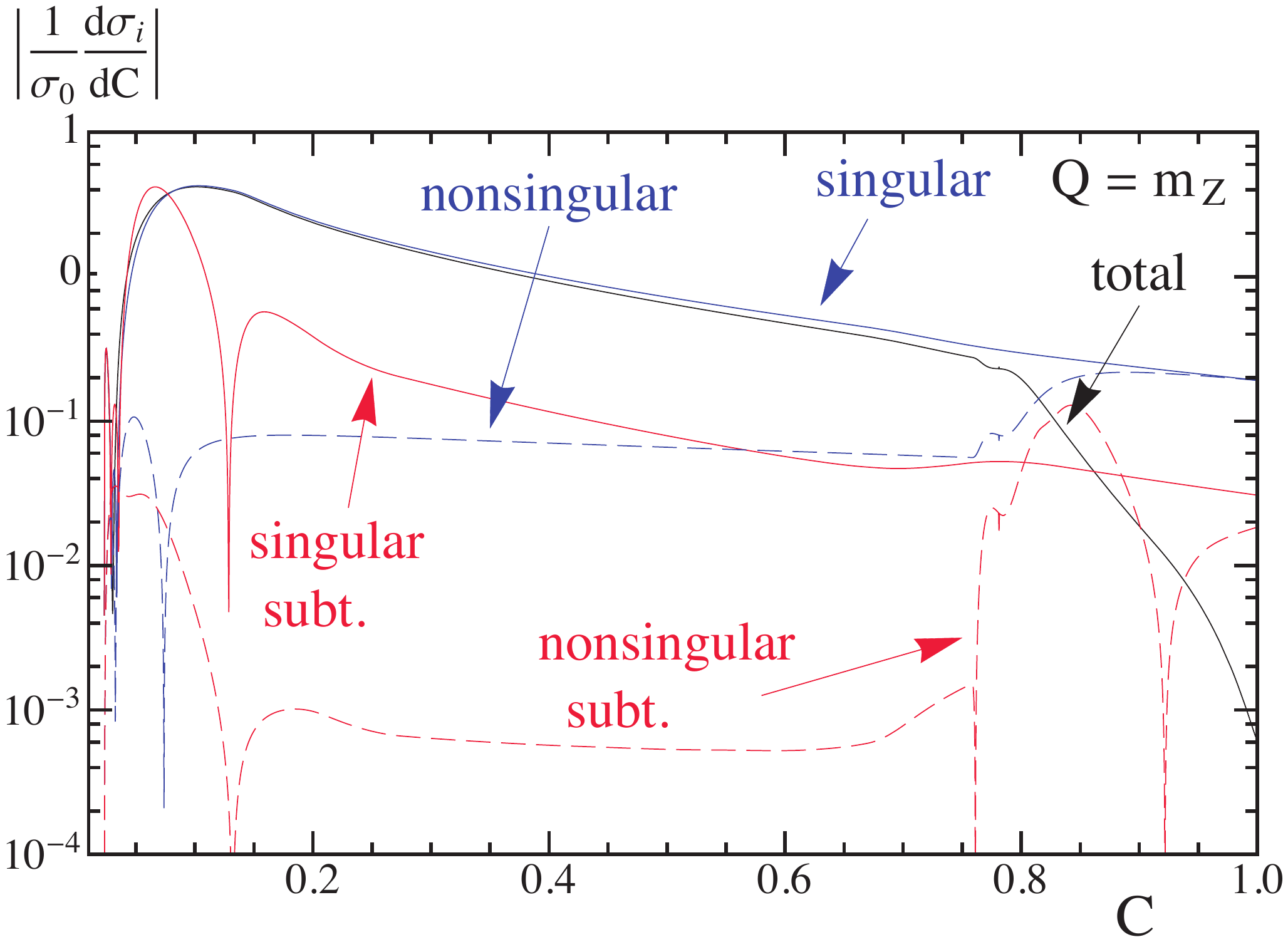}
\caption{Components of the C-parameter cross section with resummation  at N$^3$LL$^\prime+{\cal O}(\alpha_s^3)+\Omega_1(R,\mu)$ with $\Omega_1(R_\Delta,\mu_\Delta)=0.33\,$GeV
and $\alpha_s(m_Z) = 0.1141$.}
  \label{fig:component-plot-sum}
\end{center}
\end{figure}

\subsection{The slope $\mathbf{r_s}$ for C-parameter and Thrust}
\label{subsec:slope-C}
In the profiles of \Sec{sec:profiles}, the parameter $r_s$ was defined as the dimensionless slope of $\mu_S$ in the resummation region. It would seem natural to pick $r_s=1$ to eliminate the powers of $\ln(6\, \mu_S/(QC))$ and $\ln(\mu_S/(Q\tau))$ that appear in the cross section formula for $C$ and $\tau$. However, having a slightly steeper rise may also yield benefits by smoothing out the profile. Using an $r_s$ that is larger than 1, such as $r_s=2$, will only shift small $\ln(r_s)$ factors between different orders of the resummed cross section. Just like other profile parameters the dependence on $r_s$ will decrease as we go to higher orders in perturbation theory, but the central value choice may improve the accuracy of lower-order predictions.

In order to determine whether $r_s=1$ or $r_s=2$ is a better choice for the slope parameter, we examine the convergence of the cross section between different orders of resummation. For this analysis we will compare the three perturbative orders N$^3$LL$^\prime +{\cal O}(\alpha_s^3)$, N$^2$LL$^\prime +{\cal O}(\alpha_s^2)$, and NLL$^\prime +{\cal O}(\alpha_s)$. We also fix $\alpha_s(m_Z)=0.1141$, which is the value favored by the QCD only thrust fits~\cite{Abbate:2010xh}. [Use of larger values of $\alpha_s(m_Z)$ leads to the same conclusions that we draw below.]  In \Fig{fig:12-plots} we show the perturbative C-parameter cross section (upper two rows) and thrust cross section (lower two rows) with a scan over theory parameters (without including $\Omega_1$ or the shape function) for both $r_s=1$ (first respective row) and $r_s=2$ (second respective row). Additionally, we plot with different values of $Q$, using $Q=40$\,GeV in the first column, $Q=91.2$\,GeV  in the second column, and $Q=200$\,GeV in the third column.  The bands here correspond to a theory parameter scan with $500$ random points taken from Tabs.~\ref{tab:theoryerr} and \ref{tab:theoryerrthrust}. We conclude from these plots that $r_s=2$ has better convergence between different orders than $r_s=1$. For all of the values of $Q$, we can see that in the slope 1 case the N${}^2$LL$^\prime$ band lies near the outside of the edge of the NLL$^\prime$ band, while in the slope 2 plots, the scan for N${}^2$LL$^\prime$ is entirely contained within the scan for NLL$^\prime$. A similar picture can be seen for the transition from N${}^2$LL$^\prime$ to N${}^3$LL$^\prime$. This leads us to the conclusion that the resummed cross section prefers $r_s=2$ profiles which we choose as our central value for the remainder of the analysis. (In the thrust analysis of \Ref{Abbate:2010xh}, the profiles did not include an independent slope parameter, but in the resummation region, their profiles are closer to taking $r_s = 2$ than $r_s=1$.)
\begin{figure*}[t!]
\subfigure[]
{
\includegraphics[width=0.48\textwidth]{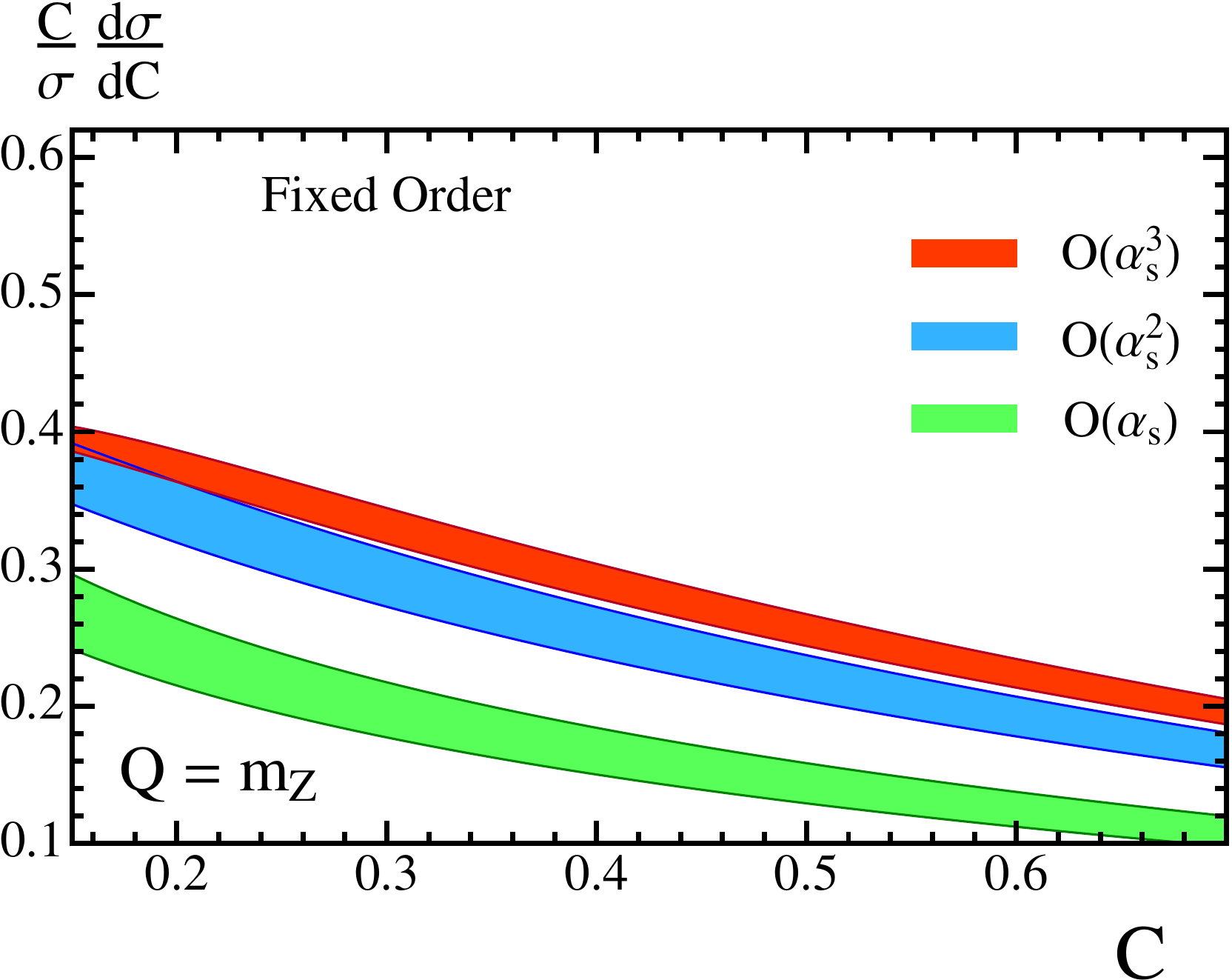}
\label{fig:tailbandFO}
}
\subfigure[]{
\includegraphics[width=0.48\textwidth]{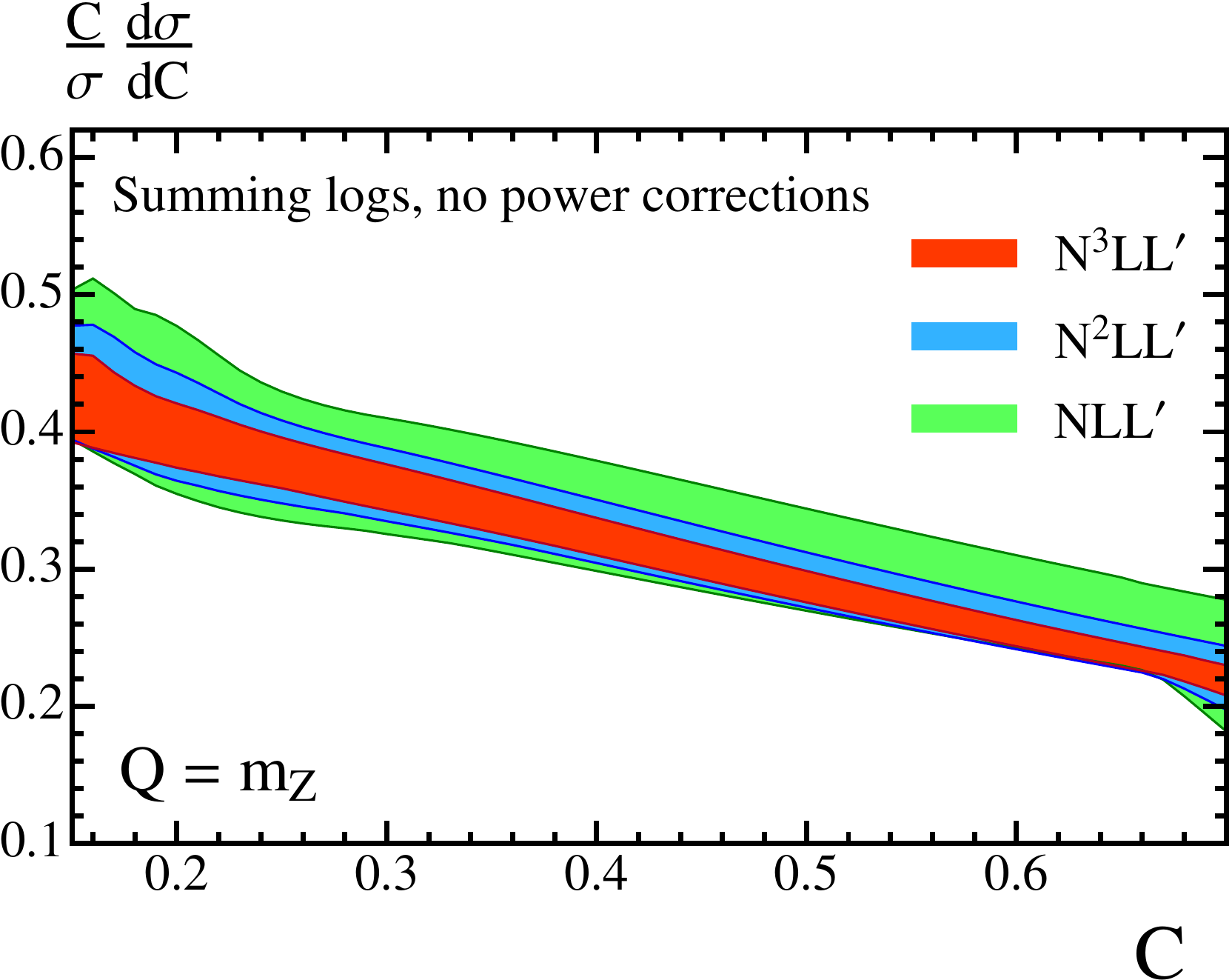}
\label{fig:tailbandnoF}
}
\subfigure[]{
\includegraphics[width=0.48\textwidth]{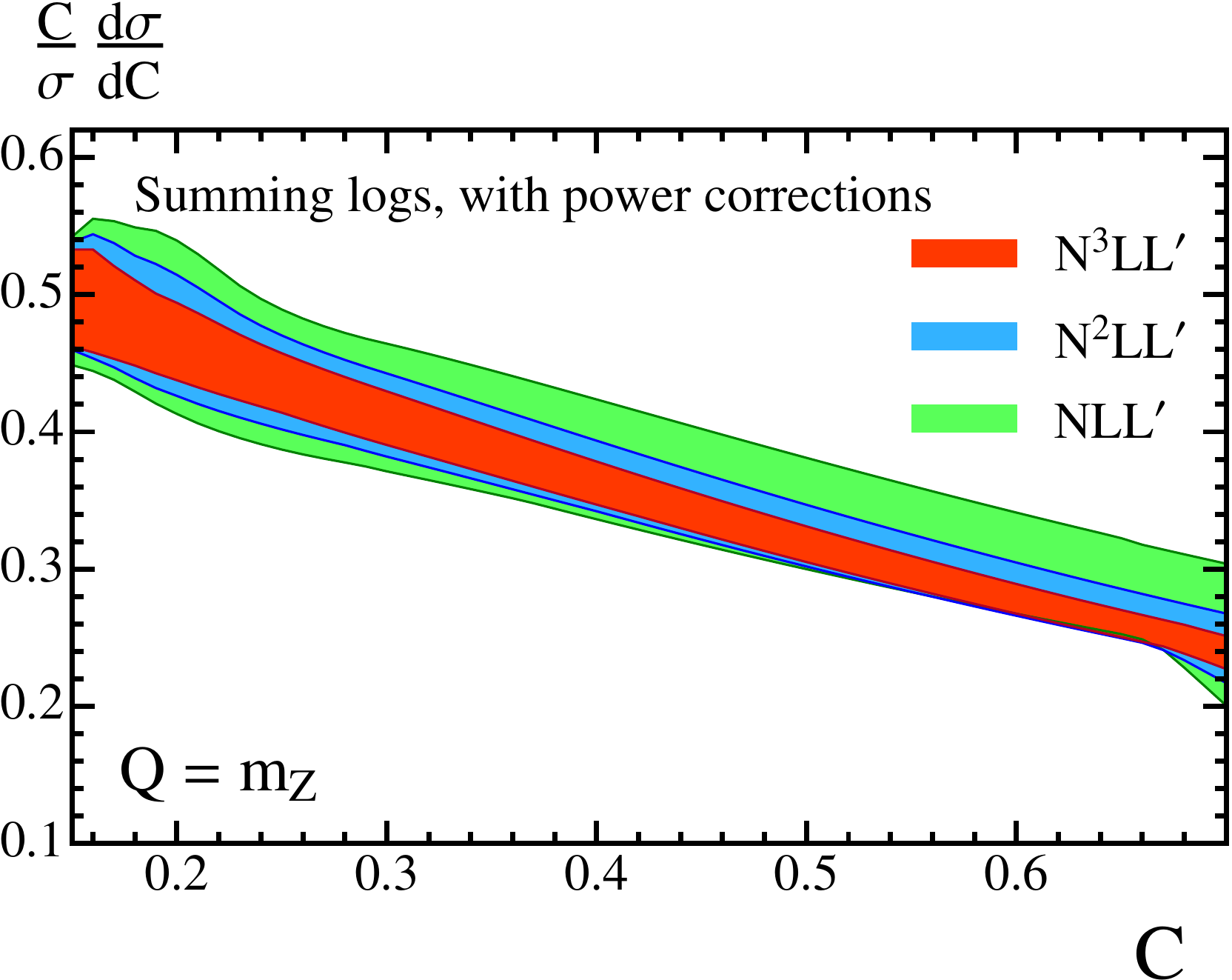}
\label{fig:tailbandnogap}
}
\subfigure[]{
\includegraphics[width=0.48\textwidth]{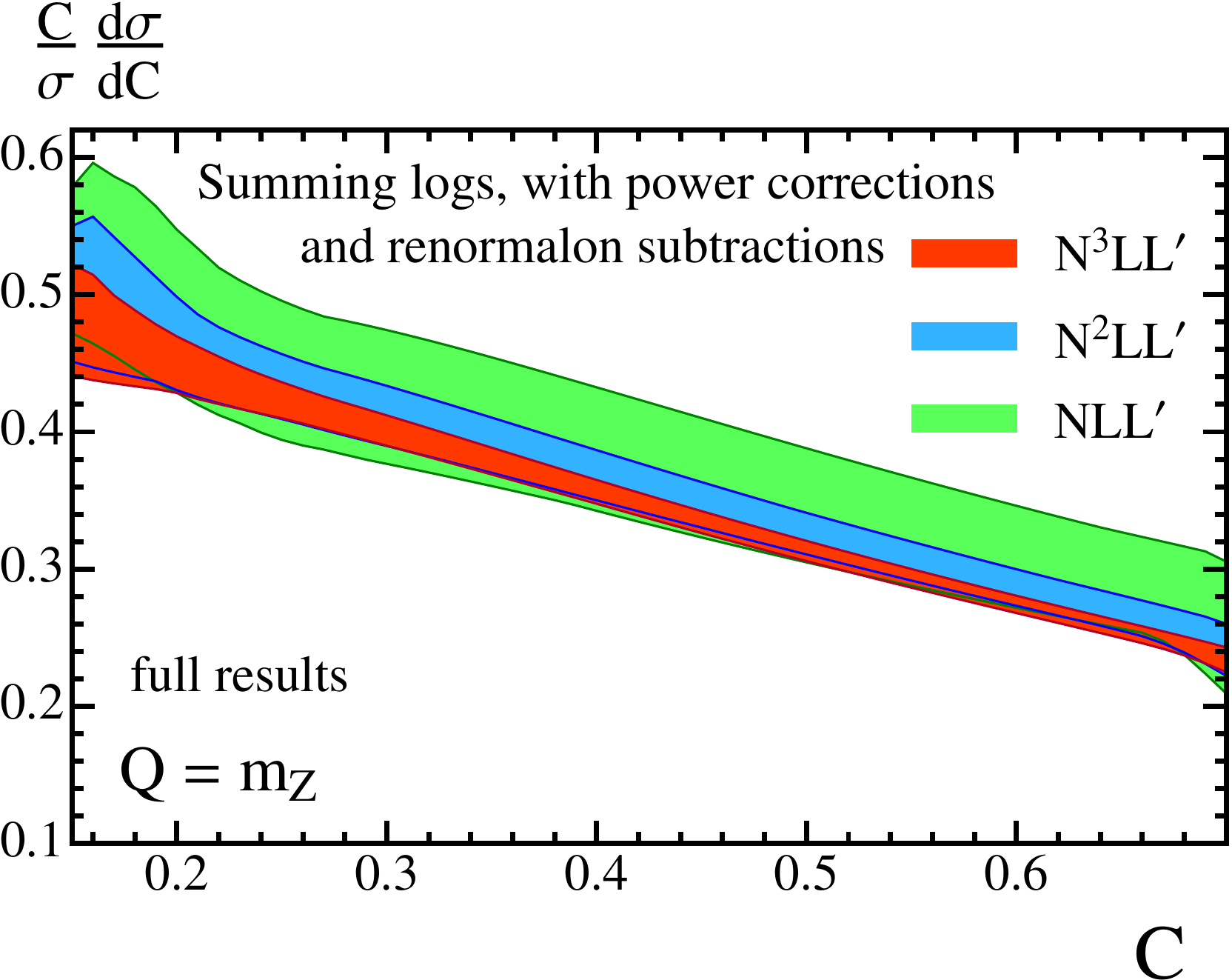}
\label{fig:tailbandwithgap}
}
\caption{Theory scan for cross section uncertainties in C-parameter. The panels are (a) fixed order, (b) resummation with no nonperturbative function, (c) resummation with a nonperturbative function using the $\msbar$ scheme for $\overline\Omega_1^{C}$ without renormalon subtraction, and (d) resummation with a nonperturbative function using the Rgap scheme for $\Omega_1^{C}$ with renormalon subtraction.}
\label{fig:4-plots}%
\end{figure*}

At the highest order, N${}^3$LL$^\prime+{\cal O}(\alpha_s^3)$, the choice of $r_s=1$ or $r_s=2$ has very little impact on the resulting cross section, both for C-parameter and thrust. Indeed, the difference between these two choices is smaller than the remaining (small) perturbative uncertainty at this order. This is illustrated in Fig.~\ref{fig:highest-slope}, which shows the complete N$^3$LL$^\prime+{\cal O}(\alpha_s^3)+\Omega_1(R,\mu)$ distributions for $r_s = 1$ (blue) and $r_s = 2$ (red) for C-parameter in the left panel, and for thrust in the right panel. Here we use $\Omega_1(R_\Delta,\mu_\Delta)=0.33\,{\rm GeV}$. Thus, the choice of $r_s$ is essentially irrelevant for our highest-order predictions, but has a bit of impact on conclusions drawn about the convergence of the lower to highest orders. Another thing that is clear from this figure is that the perturbative uncertainties for the \mbox{C-parameter} cross section at $Q=m_Z$, which are on average $\pm \,2.5\%$ in the region $0.25<C<0.65$, are a bit larger than those for thrust where we have on average $\pm\, 1.8\%$ in the region $0.1< \tau< 0.3$. This $\pm\, 1.8\%$, obtained with the profile and variations discussed here, agrees well with the $\pm\, 1.7\%$ quoted in Ref.~\cite{Abbate:2010xh}.

One can also look at the effect that the choice of $r_s$ has on the total integral over the C-parameter and thrust distributions at N$^k$LL$^\prime+{\cal O}(\alpha_s^k)+\Omega_1(R,\mu)$, which should reproduce the total hadronic cross section. For C-parameter, the outcome is shown in the first row of Fig.~\ref{fig:norm-C}, where green squares and red triangles represent resummed cross sections with power corrections in the $\msbar$ and Rgap schemes, respectively. In blue we display the fixed-order prediction. Comparing the predictions for $r_s = 1$ (left panel) and $r_s = 2$ (right panel), we observe that the former achieves a better description of the fixed-order prediction at N$^2$LL$^\prime$, in agreement with observations made in Ref.~\cite{Alioli:2012fc}.  For the case of thrust (second row of Fig.~\ref{fig:norm-C}), similar conclusions as for C-parameter can be drawn by observing the behavior of the total cross section. Again, at the highest order the result is  independent of the choice of the slope within uncertainties for both $C$ and thrust. Since our desired fit to determine $\alpha_s(m_Z)$ and $\Omega_1^C$ requires the best predictions for the shape of the normalized cross section, we do not use the better convergence for the normalization as a criteria for using $r_s=1$. Our results for cross section shapes are self-normalized using the central profile result.

In Fig.~\ref{fig:component-plot-sum} we present a plot analogous to Fig.~\ref{fig:component-plot} but including resummation at N$^3$LL$^\prime+\,{\cal O}(\alpha_s^3)\,+\,\Omega_1(R,\mu)$ with $r_s = 2$  (a similar plot for thrust can be found in Ref.~\cite{Abbate:2010xh}). The suppression of the dashed blue nonsingular curve relative to the solid upper blue singular curve is essentially the same as observed earlier in  Fig.~\ref{fig:component-plot}. The subtraction components are a small part of the cross section in the resummation region but have an impact at the level of precision obtained in our computation. Above the shoulder region, the singular and nonsingular terms appear with opposite signs and largely cancel. This is
clear from the figure where the individual singular and nonsingular lines are
larger than the total cross section in this region. The same cancellation
occurs for the singular subtraction and nonsingular subtraction terms.
The black curve labeled total in Fig.~\ref{fig:component-plot-sum} shows the central value for our full prediction. Note that the small dip in this black curve, visible at $C\simeq 0.75$, is what survives for the log-singular terms in the shoulder after the convolution with $F_C$.

\subsection{Convergence and Uncertainties: Impact of Resummation and Renormalon Subtractions}
\label{subsec:impact}

Results for the C-parameter cross sections at $Q=m_Z$ are shown at various levels of theoretical sophistication in Fig.~\ref{fig:4-plots}. The simplest setup is the purely perturbative fixed-order ${\cal O}(\alpha_s^k)$ QCD prediction (i.e.\ no resummation and no power corrections), shown in panel (a), which does not make use of the new perturbative results in this paper. Not unexpectedly, the most salient feature at fixed order is the lack of overlap between the $\mathcal{O}(\alpha_s)$ (green), $\mathcal{O}(\alpha_s^2)$ (blue), and $\mathcal{O}(\alpha_s^3)$ (red) results. This problem is cured once the perturbative resummation is included with N$^k$LL$^\prime+{\cal O}(\alpha_s^k)$ predictions shown in panel (b): the NLL$^\prime$ (green), N$^2$LL$^\prime$ (blue), and N$^3$LL$^\prime$ (red) bands now nicely overlap. To achieve this convergence and overlap with our setup, it is important to normalize the cross section bands with the integrated norm at a given order using the default profiles. (The convergence for the normalization in \Fig{fig:norm-C} was slower. Further discussion of this can be found in Ref.~\cite{Abbate:2010xh}.)

The panel (b) results neglect power corrections. Including them in the $\msbar$ scheme, N$^k$LL$^\prime+{\cal O}(\alpha_s^k)+{\overline\Omega}_1$, as shown in panel (c), does not affect the convergence of the series but rather simply shifts the bands toward larger $C$ values. This was mentioned above in \Eq{eq:shift}.

\begin{figure*}[t!]
\begin{center}
\includegraphics[width=0.95\columnwidth]{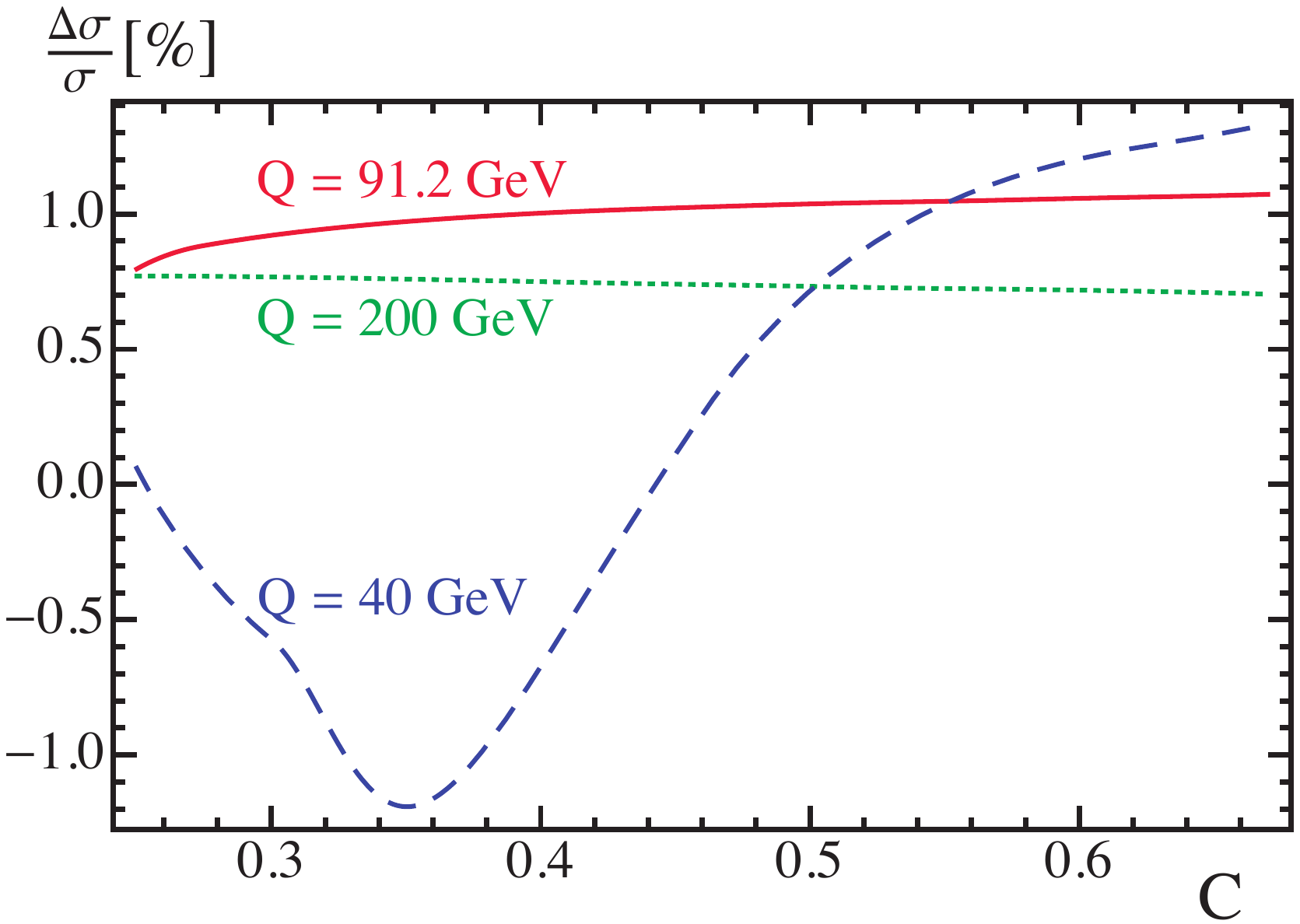}~~~~~
\includegraphics[width=0.95\columnwidth]{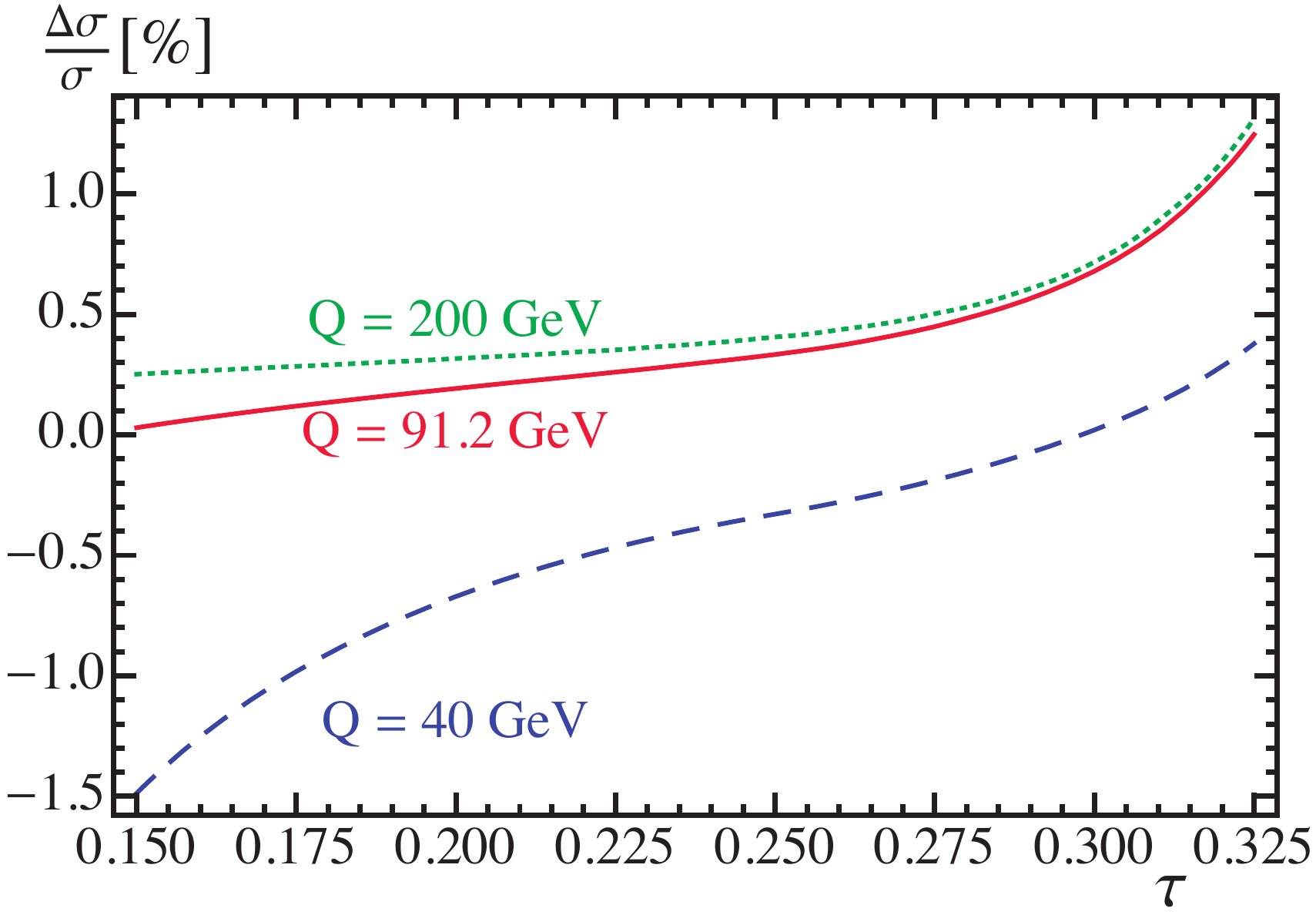}
\caption{Effects of hadron masses on the differential cross section for C-parameter (left) and thrust (right). Curves correspond to the percent difference between the cross section with and without hadron-mass effects, for three center-of-mass energies: $Q = 91.2$\,GeV (solid red), $200$\,GeV (dashed blue), and $40$\,GeV (dotted green). The cross section with hadron-mass effects uses $\theta(R_\Delta,\mu_\Delta) = 0$ and $\Omega_1(R_\Delta,\mu_\Delta) = 0.32\,(0.30)\,$GeV for C-parameter\,(thrust), while the cross section without hadron-mass effects has $\Omega_1(R_\Delta,\mu_\Delta) = 0.33\,$GeV. We set $\alpha_s(m_Z) = 0.1141$.}
  \label{fig:hadron-masses}
\end{center}
\end{figure*}

In panel (d) we show our results, which use the Rgap scheme for the power correction, at N$^k$LL$^\prime+{\cal O}(\alpha_s^k)+{\Omega}_1(R,\mu)$. In this scheme a perturbative series is subtracted from the partonic soft function to remove its ${\cal O}(\Lambda_{\rm QCD})$ renormalon. This subtraction entails a corresponding scheme change for the parameter $\Omega_1$ which becomes a subtraction-scale-dependent quantity. In general the use of renormalon-free schemes stabilizes the perturbative behavior of cross sections. The main feature visible in panel (d) is the noticeable reduction of the perturbative uncertainty band at the two highest orders, with the bands still essentially contained inside lower-order ones.

We can see the improvement in convergence numerically by comparing the average percent uncertainty between different orders at $Q=m_Z$. If we first look at the results without the renormalon subtraction, at N$^k$LL$^\prime+{\cal O}(\alpha_s^k)+{\overline\Omega}_1$, we see that in the region of interest for $\alpha_s(m_Z)$ fits ($0.25<C<0.65$) the NLL$^\prime$ distribution has an average percent error of $\pm\,11.7\%$, the N$^2$LL$^\prime$ distribution has an average percent error of $\pm\,7.0\%$, and the highest-order N$^3$LL$^\prime$ distribution has an average percent error of only $\pm\,4.3\%$. Once we implement the Rgap scheme to remove the renormalon, giving N$^k$LL$^\prime+{\cal O}(\alpha_s^k)+{\Omega}_1(R,\mu)$, we see that the NLL$^\prime$ distribution has an average percent error of $\pm\,11.8\%$, the N$^2$LL$^\prime$ distribution has an average percent error of $\pm\,4.9\%$, and the most precise N$^3$LL$^\prime$ distribution has an average percent error of only $\pm\,2.5\%$. Although the renormalon subtractions for C-parameter induce a trend toward the lower edge of the perturbative band of the predictions at one lower order, the improved convergence of the perturbative series makes the use of these more accurate predictions desirable.

\begin{figure*}[t!]
\begin{center}
\includegraphics[width=0.95\columnwidth]{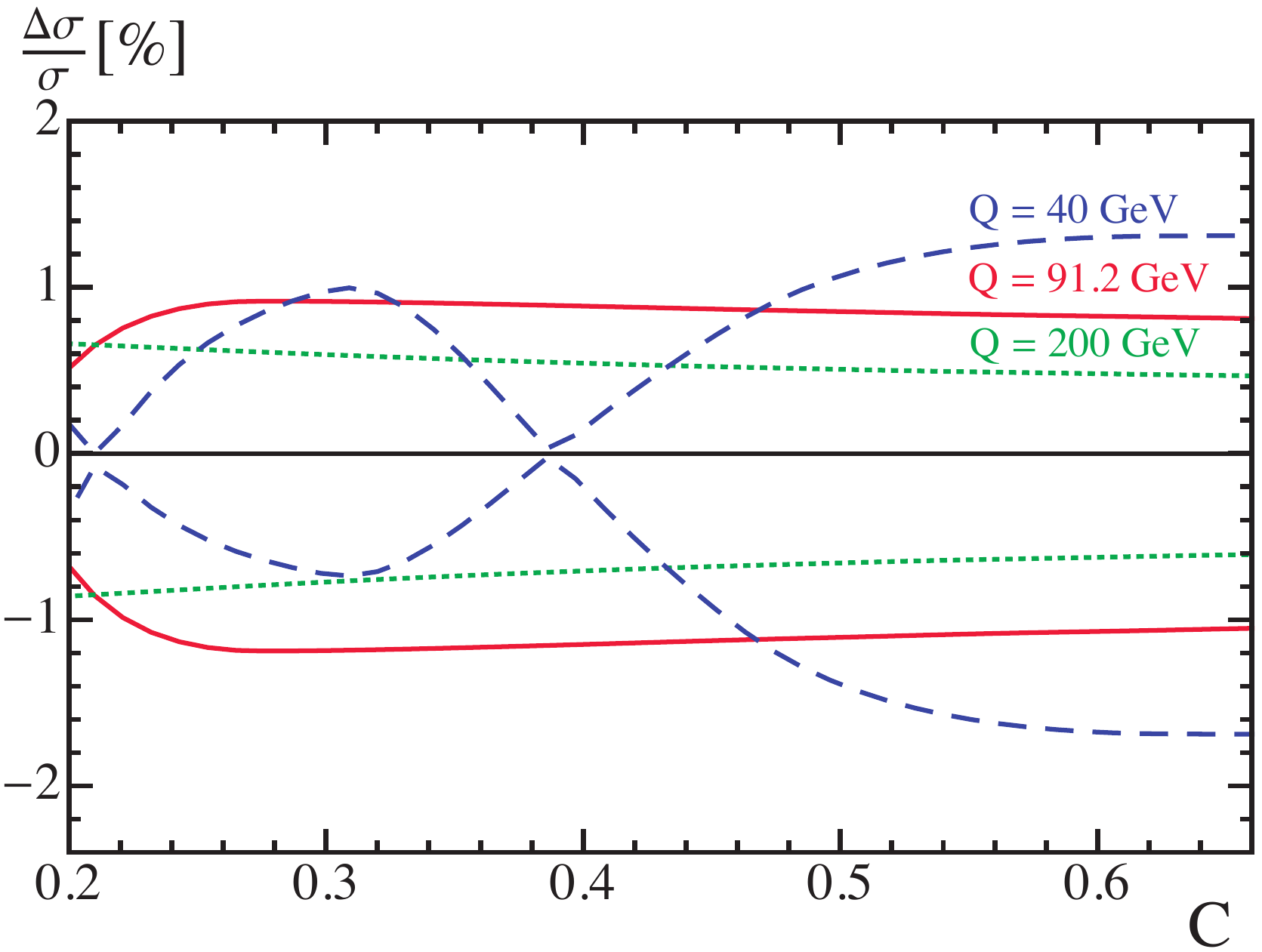}~~~~~
\includegraphics[width=0.98\columnwidth]{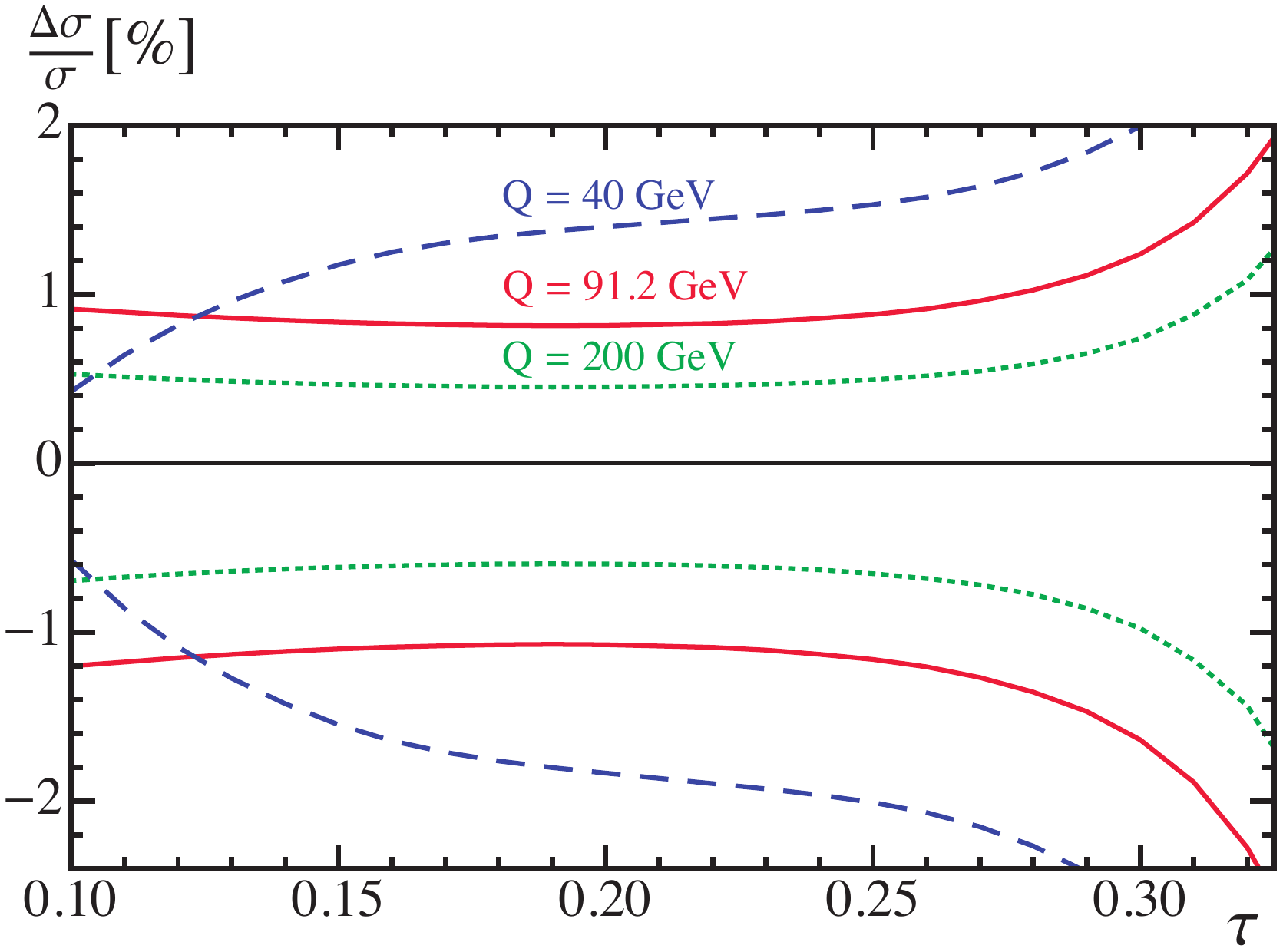}
\caption{Effects of $\theta(R_\Delta,\mu_\Delta)$ on the cross section for C-parameter (left) and thrust (right). The lines correspond to the largest variation achievable by varying $\theta(R_\Delta,\mu_\Delta)$ in both directions (which happen for $\theta = 0.23\,\pi$ and $\theta = -\,0.27\,\pi$), with respect to the cross section with $\theta(R_\Delta,\mu_\Delta) = 0$. The solid (red), dashed (blue), and dotted (green) lines correspond to $Q = 91.2$, $200$, and $40$\,GeV, respectively. We use $\Omega_1(R_\Delta,\mu_\Delta) = 0.32\,(0.30)\,$GeV for C-parameter\,(thrust), and $\alpha_s(m_Z) = 0.1141$.}
  \label{fig:theta-effect}
\end{center}
\end{figure*}
\subsection{Impact of Hadron Mass Effects}
\label{subsec:hadronmass}

In this section we discuss the impact of adding hadron-mass effects, which gives the orders denoted \mbox{N$^k$LL$^\prime+{\cal O}(\alpha_s^k)+{\Omega}_1(R,\mu,r)$}. Hadron masses induce an additional anomalous dimension for $\Omega_1$ and an associated series of logarithms of the form $\ln(Q C / \Lambda_{\rm QCD})$ that need to be resummed. They also impact the definition of $\Omega_1$ and its RGE equations in the $\overline{\rm MS}$ and Rgap schemes.  Since the overall effect of hadron masses on C-parameter and thrust are rather small, they do not change the perturbative convergence discussed in the previous section. Therefore, we study these effects here making use of only the highest-order perturbative results at N$^3$LL$^\prime+{\cal O}(\alpha_s^3)$.

In Fig.~\ref{fig:hadron-masses} we show the effect of hadron-mass running on the cross section. We compare differential cross sections with and without hadron masses, at the same center-of-mass energies. When the running effects from hadron masses are turned off, the value of $\Omega_1(R_\Delta,\mu_\Delta)$ preferred by the experimental data will attempt to average away these effects by absorbing them into the value of the initial parameter. (The running of $\Omega_1$ with and without hadron masses is shown below in Fig.~\ref{fig:Omega-run}.) Therefore, the specific values used in Fig.~\ref{fig:hadron-masses} are obtained by fixing $\alpha_s(m_Z)$ and then fitting for $\Omega_1(R_\Delta,\mu_\Delta)$ to minimize the difference between the cross section with and without hadron masses in the tail region. As the values for $\alpha_s(m_Z)$ and $\Omega_1$ in the case of no hadron-mass effects come from a fit to data (from Ref.~\cite{Abbate:2010xh}), choosing $\Omega_1(R_\Delta,\mu_\Delta)$ with the outlined procedure is similar to a full fit to data and will give results that allow comparison between the two cases. With hadron masses on, we fix $\theta(R_\Delta,\mu_\Delta)=0$, so the effects observed in Fig.~\ref{fig:hadron-masses} are related to the additional log resummation for $\Omega_1$. The effect is largest at $Q = 40\,$GeV where it varies between a $-\,1.25\%$ and $+\,1.25\%$ shift for C-parameter and between a $-\,1.5\%$ and $+\,0.5\%$ shift for thrust. For $Q=m_Z$ it amounts to a $1.0\%$ shift for C-parameter and a shift of $0\%$ to $1.3\%$ for thrust.

With hadron masses the additional hadronic parameter $\theta(R_\Delta,\mu_\Delta)$ encodes the fact that the extra resummation takes place in $r$ space, and therefore induces some dependence on the shape of the $\Omega_1(R,\mu,r)$ parameter. In contrast the dominant hadronic parameter $\Omega_1(R_\Delta,\mu_\Delta)$ is the normalization of this $r$-space hadronic function. In Fig.~\ref{fig:theta-effect} we show the effect that varying \mbox{$-\,\pi/2< \theta(R_\Delta,\mu_\Delta)<\pi/2$} has on the cross section for three center-of-mass energies, fixing \mbox{$\Omega_1(R_\Delta,\mu_\Delta) = 320$\,MeV} for C-parameter and $300$\,MeV for thrust.  The sensitivity of $\Delta\sigma$ here is proportional to $\Omega_1(R_\Delta,\mu_\Delta)$.  In these plots we pick the value of $\theta$ that gives the largest deviation from the $\theta = 0$ cross section (these values are listed in the figure caption). This maximum deviation is only $1.5\%$  for C-parameter and occurs at $Q = 40$\,GeV at larger $C$. For thrust, the largest deviation also occurs for $Q=40$\,GeV and for higher values of $\tau$ and is $\lesssim 2.0\%$ for $\tau\le 0.3$. For $Q=m_Z$ the effect is roughly $\pm 1.0\%$ for C-parameter and around $1.0\%$ for thrust when $\tau < 0.25$, growing to $2\%$ by $\tau=0.33$.

We conclude that the effect of hadron masses on log resummation should be included if one wishes to avoid an additional $\sim 1.5\%$ uncertainty on the cross section. Furthermore, one should consider fitting $\theta(R_\Delta,\mu_\Delta)$ as an additional parameter if one wants to avoid another $\sim 1\%$ uncertainty in the cross section that it induces. (Recall that Fig.~\ref{fig:theta-effect} shows the worst-case scenario.)

\section{Conclusions}
\label{sec:conclusions}

We have provided a factorization formula for the \mbox{C-parameter} distribution in $e^+e^-$ annihilation and analyzed its utility in making precision cross section predictions, extending earlier work on thrust \cite{Abbate:2010xh}. We have determined or computed the ingredients needed to achieve a summation of large logarithms at N${}^3$LL order for the singular terms in the dijet limit where $C$ is small, pushing beyond the classical resummed computations in Ref.~\cite{Catani:1998sf} by two additional orders. To achieve this goal, we demonstrated that the anomalous dimensions, as well as the hard and jet functions, are identical in the \mbox{C-parameter} and thrust distributions to all orders in perturbation theory. We then computed the only distinct piece, the soft function, to one-loop order analytically, two-loop order numerically, and with logarithmic accuracy at three loops using its anomalous dimension. (We also present a master formula for the one-loop soft function, valid for any recoil-free dijet event shape, in \App{ap:softoneloop}.) Our factorization formula also incorporates the previously known ${\cal O}(\alpha_s^2)$ and ${\cal O}(\alpha_s^3)$ perturbative QCD corrections in fixed order~\cite{GehrmannDeRidder:2007bj, GehrmannDeRidder:2007hr, GehrmannDeRidder:2009dp,Ridder:2014wza, Weinzierl:2008iv,Weinzierl:2009ms}.

Using our factorization theorem and computation for the soft function, we were able to analytically predict the fixed-order log-singular terms of C-parameter up to ${\cal O}(\alpha_s^3)$. These results were used to consistently incorporate fixed-order results at the same order by subtracting them from the numerical ${\cal O}(\alpha_s^3)$ results, to obtain the nonsingular terms plotted in Figs.~\ref{fig:1-loopNS}, \ref{fig:2-loopNS}, and \ref{fig:3-loopNS} and compared to the singular cross section in \Fig{fig:component-plot}.

The factorization formula we used incorporates a systematic description of nonperturbative power corrections using a shape function $F_C$, whose moments are given by quantum field theory matrix elements. This nonperturative shape function describes the dynamics of soft particle radiation at large angles. Although the shape function of thrust and C-parameter are {\it a priori} unrelated, one can show that, up to small deviations caused by hadron masses, their respective first moments are proportional to each other involving a simple calculable coefficient, as shown in Eq.~(\ref{eq:O1c}). Tests of this universality are very important, since the first moment of the shape function constitutes the most important nonperturative power correction in the tail of the distribution, where one can write down an OPE for the event-shape distributions. The results in this paper allow for universality to be tested with data accounting for a high degree of perturbative precision and incorporating hadron-mass effects.

In order to reduce the sensitivity to an $\mathcal{O}(\Lambda_{\rm QCD})$ renormalon present in the soft function, we switched to a short-distance scheme for the leading power correction, which translates into a more stable perturbative behavior. Using universality relations, it is possible (and desirable) to use thrust gap subtractions in the C-parameter distribution, with the appropriate prefactor. Furthermore, we introduced hadron-mass running effects into this formalism, and showed how one can consistently account for Rgap and hadron-mass running simultaneously.

We have used profile functions, which are $C$-dependent renormalization scales, to both properly implement the required conditions for the scales in different regions in
\Eq{eq:profileconstraints}, including the resummation of large logarithms in the nonperturbative and resummation regions, and to smoothly carry out the transition between regions. There are two significant differences for the C-parameter relative to thrust: the nonperturbative region is enlarged by a factor $3\pi/2$, and the slope of $\mu_S$ in the resummation region is reduced by a factor of $6$. In \Sec{subsec:slope-C} we made a numerical investigation of possible choices for the slope of the soft function in the resummation region and concluded both for thrust and C-parameter that having a slope that is twice the canonical value, $r_s=2$ rather than $r_s=1$, is advantageous  to have a better order-by-order convergence. At the highest order we achieves, which is N$^3$LL$^\prime$, we also observe that the choice of $r_s=1$ versus $r_s=2$ becomes irrelevant.

Our most accurate theoretical description has a very good perturbative convergence, and achieves a perturbative uncertainty of on average $2.5\%$ in the tail of the distribution for $Q=m_Z$; see \Fig{fig:tailbandwithgap}. We have also shown that the effect of hadron-mass corrections is at the $<2\%$ level (depending on $Q$, and differing somewhat between C-parameter and thrust); see \Figs{fig:hadron-masses}{fig:theta-effect}. The theory developments made here are necessary for a high-precision determination of the strong coupling constant $\alpha_s(m_Z)$ by fits to experimental data. Moreover, by simultaneously fitting to the leading power correction $\Omega_1$, our formalism allows us to check for universality between thrust and C-parameter, exploiting the high level of perturbative precision and also accounting for hadron-mass effects. These results will be presented in
a separate paper~\cite{Hoang:2015hka}.

\begin{acknowledgments}
This work was supported by the offices of Nuclear Physics of the
U.S. Department of Energy (DOE) under under Contract No.\ DE-SC0011090 and the
European Community's Marie-Curie Research Networks under contract PITN-GA-2010-264564
(LHCphenOnet). IS was also supported in part by the Simons Foundation Investigator Grant No.\ 327942.
VM was supported by a Marie Curie Fellowship under Contract No.\ PIOF-GA-2009-251174 while part of
this work was completed. This work was also supported in part by MIT MISTI global seed funds
and the ESI summer program on ``Jets and Quantum Fields for the LHC and Future Colliders''.
VM thanks Thomas Hahn for computing support. We thank Thomas Gehrmann for providing us with numerical results from the program EERAD3.
\end{acknowledgments}

\appendix

\section{Formulae} \label{ap:formulae}
In this appendix we collect all the remaining formulae used in our analysis for the case of massless quarks. Since we want to compare to experimental data, which are normalized to the total number of events, we need to calculate $(1/\sigma) \df \sigma / \df C$, our normalized cross section. To do this we can either self-normalize our results by integrating over $C$ or use the fixed-order result for the total hadronic cross section, which, at three loops for massless quarks at \mbox{$\mu=Q$}, is (see Ref.~\cite{Chetyrkin:1996ia} for further discussion)
 \begin{align} \label{eq:Rhad}
 R_{\rm had}\,&=\,1 + 0.3183099\,\alpha_s(Q) + 0.1427849\,\alpha_s^2(Q)
 \nn\\
 &\ -\,0.411757\,\alpha_s^3(Q)\,.
 \end{align}

Throughout our analysis, we use $m_{Z}=91.187\,\mathrm{GeV}$, and all numerical results quoted below are for SU(3) color with $n_f=5$ active light flavors, for simplicity.

\head{Singular Cross Section Formula}\vspace*{1pt}

\noindent
For the singular part of the differential cross section given in \Eq{eq:singular-resummation}, we simplify the numerical evaluation using our freedom to take $\mu=\mu_J$. This means
$U_J^\tau(s-s',\mu_J,\mu_J)=\delta(s-s')$, and we can write
\begin{align} \label{eq:dswithP}
&\frac{1}{\sigma_0}\!\int\! {\rm d}k\, \frac{{\rm d}\hat\sigma_s}{ {\rm d}C}\Big(C-\frac{k}{Q}\Big) 
F_{C} \Big(k- 3 \pi \bar\Delta(R,\mu_S)\Big) \nn\\
&= \frac{Q}{6}
H(Q,\mi_H)\,U_H\big(Q,\mi_H,\mi_J\big)\,\nonumber
\\&\times
\int {\rm d}k\,P\big(Q,Q C/6-k/6,\mi_J\big)\,\nn\\
&\times
e^{-{3 \pi\, \delta(R,\mu_s)}\frac{{\rm d}}{{\rm d}k}}\, F_{C} \Big(k-3 \pi\, \bar\Delta(R,\mu_S)\Big),
\end{align}
Here $\sigma_0$ is the tree-level (Born) cross section for \mbox{$e^+e^-\to q\bar q$}. 
Here we have combined the perturbative corrections from the partonic soft function, jet
function, and soft evolution factor into a single function, 
\begin{align}
P(Q,k,\mu_J) & =\int \!{\rm d}s\!\int\!
{\rm d}k' J_\tau(s,\mu_J)\, U_S^{\tau}(k',\mu_J,\mu_S)
  \nn \\
&\times\hat{S}_{\widetilde{C}}(k-k'-s/Q,\mu_S) \
  \,.
\end{align}
The large logarithms of $C/6$ are summed up in the evolution factors $U_H$ and $U_S^{\tau}$. We can carry out the integrals in $P$ exactly, and the results are enumerated below. The shape function \mbox{$F_{C}(k-3\pi\,\bar{\Delta})$} is discussed in Sec.~\ref{sec:power}, and we have used integration by parts in \Eq{eq:dswithP} to have the derivative in the exponential act on $F_C$, which is simpler than acting the derivative on the perturbative soft function. For our numerical calculation, we expand $H$, $J_{\tau}$, and $\hat{S}_{\widetilde{C}}$ order by order as a series in $\alpha_s(\mu_H)$, $\alpha_s(\mu_J)$, and $\alpha_s(\mu_S)$ respectively, with no large logs. Additionally, we expand $\exp(-\,3 \pi\, \delta(R,\mu_S) {\rm d}/{\rm d}k)$ [see \eq{deltaseries}] as a series in $\alpha_s(\mu_S)$, which must be done consistently to cancel the renormalon present in $\hat{S}_{\widetilde{C}}$.

The hard function to ${\cal O}(\alpha_s^3)$ with $n_f=5$ is~\cite{Matsuura:1987wt,Matsuura:1988sm,Gehrmann:2005pd,Moch:2005id,Lee:2010cg,Baikov:2009bg,Gehrmann:2010ue,Abbate:2010xh}
\begin{align} \label{eq:Hardnumeric}
& H(Q,\mu_H) \nn\\
& =
1+\alpha_{s}(\mu_H)\Big(\!0.745808\!-\!1.27324 L_{Q}\!-\!0.848826L_{Q}^{2}\Big)
\notag\\&
+\alpha_{s}^{2}(\mu_H)\Big(2.27587- 0.0251035\, L_{Q}- 1.06592\, L_{Q}^{2}
\notag\\&
+0.735517L_{Q}^{3}+0.360253L_{Q}^{4}\Big)
\notag\\&
+\alpha_{s}^{3}(\mu_H)\Big(0.00050393 \,h_{3}+2.78092 L_{Q}-2.85654 L_{Q}^{2}
\notag\\&
-0.147051 L_{Q}^{3}+0.865045L_{Q}^{4}-0.165638 L_{Q}^{5}
\notag\\&
-0.101931\, L_{Q}^{6}\Big)\,,
\end{align}
where $L_Q=\ln\frac{\mi_H}{Q}$ and $h_3=8998.080$ from Ref.~\cite{Baikov:2009bg}.

The resummation of large logs between $\mu_H$ and $\mu_J$ is given by
$U_H(Q,\mi_H,\mi_J)$, the solution of the RGE for the
hard function,  which can be written as \cite{Bauer:2000yr}
\begin{align} \label{appendix:U_H}
U_H(Q,\mi_H,\mi)=
e^{2 K(\Gamma_H,\gamma_H,\mi,\mi_H)}
\bigg(\frac{\mi_H^2}{Q^2}\bigg)^{\!\!\omega(\Gamma_H\!,\,\mi,\mi_H)},
\end{align}
where $\omega$ and $K$ are given in \eqs{w}{K} below.

In momentum space, we can use the results from Ref.~\cite{Ligeti:2008ac} to calculate the convolution of the plus-functions in $P$ to give the form
\begin{align} \label{appendix:P}
&P\big(Q,k,\mi_J\big)=
\frac{1}{\xi}\,E_S\big(\xi,\mi_J,\mi_S\big)\nonumber\\
&\times\sum_{\substack{n,m,k,l=-1\\m+n+1\geq k\\k+1\geq l}}^{\infty}
V_{k}^{mn}\,\,J_m\Big[\as(\mi_J),\frac{\xi\, Q}{\mi^2_{J}}\Big]
\,S_{n}^{\widetilde C}\Big[\as(\mi_S),\frac{\xi}{\mi_S}\Big]\,
\notag\\
&\times
V_l^k\big[-2\,\omega(\Gamma_S,\mi_J,\mi_S)\big]\,\mathcal
L_l^{-2\omega(\Gamma_S,\mi_J,\mi_S)}\Big(\frac{k}{\xi}\Big) \,.
\end{align}
Here $\xi$ is a dummy variable that does not affect the value of the result. \footnote{When convolved
with $F_C$ we use our freedom in choosing $\xi$ to simplify
the final numerical integration, picking \mbox{$\xi = QC/6 - 3 \pi\, \bar{\Delta}(R,\mu_S)$}.} $E_S(\xi,\mu_J,\mu_S)$ is given by~\cite{Balzereit:1998yf,Neubert:2004dd}
\begin{align} \label{appendix:E}
&E_{S}(\xi,\mi_J,\mi_S)=
\exp\!\big[2 K(\Gamma_S,\gamma_S,\mi_J,\mi_S)\big]
\\&\times
\Big(\frac{\xi}{\mi_S}\Big)^{-2\omega(\Gamma_S,\mi_J,\mi_S)}\ 
\frac{\exp\!\big[2\gamma_E\, \omega(\Gamma_S,\mi_J,\mi_S) \big]}
{\Gamma\big[1-2\,\omega(\Gamma_S,\mi_J,\mi_S)\big]} \,,
\notag
\end{align}
and encodes part of the running between $\mu_S$ and $\mu_J$. The rest of the running is included in the $V$ coefficients and the plus-functions, $\mathcal L_l$.

The $J_m$ and $S_n$ in Eq.~(\ref{appendix:P}) are the coefficients of the momentum-space soft and jet functions, given by
\begin{align}
J_\tau(p^-k,\mu_J) &= \frac{1}{p^-\xi} \sum_{m=-1}^\infty
J_m\Big[\alpha_s(\mu_J),\frac{p^-\xi}{\mu_J^2}\Big] {\cal L}_m\Big(\frac{k}{\xi}\Big) , \nn\\
\hat S_{\widetilde C}(k,\mu_S) &= \frac{1}{\xi} \sum_{n=-1}^\infty
S_n^{\widetilde C}\Big[\alpha_s(\mu_S),\frac{\xi}{\mu_S}\Big] {\cal
L}_n\Big(\frac{k}{\xi}\Big) .
\end{align}
Here the C-parameter soft function coefficients are
\begin{align}
S_{-1}^{\widetilde C}[\alpha_s,x] 
&= S_{-1}^{\widetilde C}(\alpha_s) + \sum_{n=0}^\infty S_n^{\widetilde C}(\alpha_s)
\frac{\ln^{n+1} x}{n+1} \,,\nn \\
S_n^{\widetilde C}[\alpha_s,x] 
&= \sum_{k=0}^\infty \frac{(n+k)!}{n!\, k!} S_{n+k}^{\widetilde C}(\alpha_s) \ln^k x
\,,
\end{align}
which for $n_f=5$ can be written as
\begin{align} \label{eq:Sncoeff}
S_{-1}^{\widetilde C}(\alpha_s) &= 1 + 1.0472 \alpha _s +
(1.75598 + 0.012666\, s_2^{\widetilde{C}}) \alpha _s^2 
\notag\\
&+ \left (
2.59883 + 0.0132629 \, s_2^{\widetilde{C}} 
+ 0.00100786\, s_3^{\widetilde C} \right) \alpha_s^3 
\nn \,,\\
S_0^{\widetilde C}(\alpha_s) &= 1.22136 \alpha _s^2 + 
(2.63481- 0.0309077\, s_2^{\widetilde{C}}) \alpha _s^3 
\nn\,,\\
S_1^{\widetilde C}(\alpha_s) &= -1.69765\, \alpha_s - 7.45178\, \alpha_s^2
\notag\\
&
-(19.1773+0.021501\, s_2^{\widetilde{C}})\, \alpha _s^3 \,,\nn \\
S_2^{\widetilde C}(\alpha_s) &= 1.03573\, \alpha_s^2 + 2.3245\, \alpha_s^3 \,,\nn\\
S_3^{\widetilde C}(\alpha_s) &= 1.44101\, \alpha_s^2 + 10.299\, \alpha_s^3 \,,\nn\\
S_4^{\widetilde C}(\alpha_s) &= -\,1.46525\, \alpha_s^3 \,,\nn\\
S_5^{\widetilde C}(\alpha_s) &= -\,0.611585\, \alpha_s^3 \,.
\end{align}
Note that $s_2^{\widetilde{C}}$ and $s_3^{\widetilde C}$ are the ${\cal O}(\alpha_s^{2,3})$ coefficients of the
non-logarithmic terms in the series expansion of the logarithm of the position
space C-parameter soft function. The coefficients for the jet function are
\begin{align}
J_{-1}[\as,x]=&
J_{-1}(\as)+\sum_{n=0}^{\infty}\,J_{n}(\as)\,\frac{\ln^{n+1}x}{n+1} 
\,, \nn\\
J_{n}[\as,x]=& \sum_{k=0}^{\infty}\,\frac{(n+k)!}{n!\, k!}
\,J_{n+k}(\as)\,\ln^{k}x\,,
\end{align}
and are known up to ${\cal O}(\alpha_s^3)$ except for the constant $j_3$
term~\cite{Lunghi:2002ju,Bauer:2003pi,Bosch:2004th,Becher:2006qw,Moch:2004pa,Becher:2008cf}. With $n_f=5$ we have
\begin{align} \label{eq:Jncoeff}
J_{-1}(\as) &= 1 - 0.608949\, \as - 2.26795\, \as^2
\notag\\&\qquad + (2.21087 + 0.00100786\, j_3)\,\as^3  \,,\nn \\
J_{0}(\as) &= -\,0.63662\, \as + 3.00401\, \as^2 + 4.45566\, \as^3 \,,\nn\\
J_{1}(\as) &= 0.848826\, \as - 0.441765\, \as^2 - 11.905\, \as^3\,,\nn \\
J_{2}(\as) &= -\,1.0695\, \as^2 + 5.36297\, \as^3\,,\nn\\
J_{3}(\as) &= 0.360253\, \as^2 + 0.169497\, \as^3\,,\nn\\
J_{4}(\as) &= -\,0.469837\, \as^3\,,\nn \\
J_{5}(\as) &= 0.0764481\,\as^3.
\end{align}
The plus-distributions, denoted by $\cL(x)$, are given by
\begin{equation} \label{eq:cLna_def}
\cL_n^a(x) = \biggl[\frac{\theta(x)\ln^n x}{x^{1-a}}\biggr]_+
= \frac{\df^n}{\df a^n}\, \cL^{a}(x) \,\,\, [n \ge 0]\,,
\end{equation}
${\cal L}_{-1}^a(x) = {\cal L}_{-1}(x) = \delta(x)$, and for $a> -1$
\begin{equation}
\label{eq:cLa_def}
\cL^a(x) = \biggl[\frac{\theta(x)}{x^{1-a}} \biggr]_+
= \lim_{\e\to 0}\, \frac{\df}{\df x}
\biggl[ \theta(x - \e)\, \frac{x^a - 1}{a} \biggr] \,.
\end{equation}
\begin{widetext}
In Eq.~(\ref{appendix:P}) we also take advantage of the shorthand for the $V$ coefficients presented in Ref.~\cite{Ligeti:2008ac},
\begin{align} \label{eq:Vkna_def}
\V_k^n(a) &= \begin{cases}
\displaystyle a\, \frac{\df^n}{\df b^n}\,\frac{\V(a,b)}{a+b}\bigg\vert_{b = 0}\,,
&   k=-1\,, \\[10pt]
\displaystyle  a\, \binom{n}{k}   \frac{\df^{n-k}}{\df b^{n-k}}\, \V(a,b)
\bigg\vert_{b = 0} + \delta_{kn} \,, \quad
& 0\le k\le n \,,  \\[10pt]
\displaystyle  \frac{a}{n+1} \,,
& k=n+1  \,,
\end{cases}
\end{align}
and the coefficients
\begin{align} \label{eq:Vkmn_def}
\V_k^{mn} &= \begin{cases}
\displaystyle \frac{\df^m}{\df a^m}\, \frac{\df^n}{\df b^n}\,\frac{\V(a,b)}{a+b}\bigg\vert_{a = b = 0} \,,
& k=-1\,, \\[10pt]
\displaystyle  \sum_{p=0}^m\sum_{q=0}^n\delta_{p+q,k}\,\binom{m}{p} \binom{n}{q}
\frac{\df^{m-p}}{\df a^{m-p}}\, \frac{\df^{n-q}}{\df b^{n-q}} \ \V(a,b)
\bigg\vert_{a = b = 0}\,, \quad
& 0\le k \le m+n \,,\\[15pt]
\displaystyle  \frac{1}{m+1} + \frac{1}{n+1}\,,
& k=m+n+1 \,,
\end{cases}
\end{align}
for
\begin{equation}
\V(a,b) = \frac{\Gamma(a)\,\Gamma(b)}{\Gamma(a+b)} - \frac{1}{a} - \frac{1}{b}
\,.
\end{equation}
We also need the special cases
\begin{align}
&\V_{-1}^{-1}(a) = 1
\,,
&\V_0^{-1}(a) &= a
\,,
&\V_{k \geq 1}^{-1}(a) &= 0
\,,\quad
&\V^{-1,n}_k &= \V^{n,-1}_k = \delta_{nk}
\,.\end{align}

\head{Evolution factors and Anomalous Dimensions}\vspace*{1pt}

The running between scales is encoded in just a few functions. In Eqs.~(\ref{appendix:U_H}),
(\ref{appendix:P}), and (\ref{appendix:E}), we use
\begin{align} \label{eq:w}
\omega(\Gamma,\mi,\mi_0) &=2\!\int_{\alpha_s(\mu_0)}^{\alpha_s(\mu)}\frac{{\rm
d}\,\alpha}{\beta(\alpha)}\,\Gamma(\alpha) \notag\\
&=  -\,\frac{\Gamma_0}{\beta_0}\bigg\{\!\ln \kappa
+\frac{\alpha_s(\mu_0)}{4\pi}\Big(\frac{\Gamma_1}{\Gamma_0}
-\frac{\beta_1}{\beta_0}\Big)(\kappa-1) 
+\frac{1}{2}
\frac{\alpha_s^2(\mu_0)}{(4\pi)^2}\Big(\frac{\beta_1^2}{\beta_0^2}
-\frac{\beta_2}{\beta_0}+\frac{\Gamma_2}{\Gamma_0}
-\frac{\Gamma_1\beta_1}{\Gamma_0\beta_0}\Big)(\kappa^2-1) \notag\\& +\frac{1}{3}
\frac{\alpha_s^3(\mu_0)}{(4\pi)^3}
\bigg[\frac{\Gamma_3}{\Gamma_0}-\frac{\beta_3}{\beta_0}
+\frac{\Gamma_1}{\Gamma_0}\Big(\frac{\beta_1^2}{\beta_0^2}-\frac{\beta_2}{\beta_0}\Big)
-\frac{\beta_1}{\beta_0}\Big(\frac{\beta_1^2}{\beta_0^2}-
2\,\frac{\beta_2}{\beta_0}+\frac{\Gamma_2}{\Gamma_0}\Big) \bigg](\kappa^3-1)\!\bigg\},
\end{align}
and
\begin{align} \label{eq:K}
&K(\Gamma,\gamma,\mi,\mi_0)-\omega\Big(\frac{\gamma}{2},\mi,\mi_0\Big)
= 2\!\int_{\alpha_s(\mu_0)}^{\alpha_s(\mu)}
\frac{{\rm d}\,\alpha}{\beta(\alpha)}\,\Gamma(\alpha)\!
\int_{\alpha_s(\mu_0)}^{\alpha}\frac{{\rm d}\alpha'}{\beta(\alpha')}
\notag\\[3pt]
&\ =\frac{\Gamma_0}{2\beta_0^2}\Bigg\{\frac{4\pi}{\alpha_s(\mu_0)}
\Big(\ln \kappa+\frac{1}{\kappa}-1\Big) +
\Big(\frac{\Gamma_1}{\Gamma_0}-\frac{\beta_1}{\beta_0}\Big)(\kappa-1-\ln \kappa)
-\frac{\beta_1}{2\beta_0}\ln^2 \kappa
+\frac{\alpha_s(\mu_0)}{4\pi}\bigg[
\Big(\frac{\Gamma_1\beta_1}{\Gamma_0\beta_0}-\frac{\beta_1^2}{\beta_0^2}\Big)
(\kappa-1-\kappa \ln \kappa)
\nn\\[3pt]
&\ \ -B_2 \ln \kappa
+\Big( \frac{\Gamma_2}{\Gamma_0}-\frac{\Gamma_1\beta_1}{\Gamma_0\beta_0} +
B_2 \Big)\frac{(\kappa^2\!-\!1)}{2}
+\Big(\frac{\Gamma_2}{\Gamma_0}-\frac{\Gamma_1\beta_1}{\Gamma_0\beta_0}\Big) 
(1\!-\!\kappa)\bigg]
+\frac{\alpha_s^2(\mu_0)}{(4\pi)^2}
\bigg[ \Big[\Big(\frac{\Gamma_1}{\Gamma_0}-\frac{\beta_1}{\beta_0} \Big)B_2
+\frac{B_3}{2}\Big]\frac{(\kappa^2\!-\!1)}{2}
\nn\\
&  \ \ 
+ \Big(\frac{\Gamma_3}{\Gamma_0} -
\frac{\Gamma_2\beta_1}{\Gamma_0\beta_0}
+\frac{B_2\Gamma_1}{\Gamma_0}+B_3\Big) \Big(\frac{\kappa^3-1}{3}-\frac{\kappa^2-1}{2}\Big)
-\frac{\beta_1}{2\beta_0}
\Big(\frac{\Gamma_2}{\Gamma_0}-\frac{\Gamma_1\beta_1}{\Gamma_0\beta_0}+B_2\Big)
\Big(\kappa^2\ln \kappa-\frac{\kappa^2-1}{2}\Big) -\frac{B_3}{2}\ln \kappa
\notag\\
&\ \ 
-B_2\Big(\frac{\Gamma_1}{\Gamma_0}-\frac{\beta_1}{\beta_0} \Big)(\kappa-1)
\bigg]
\Bigg\},
\end{align}
where here $\kappa=\alpha_s(\mu)/\alpha_s(\mu_0)$ requires the known 4-loop running couplings, and
we have defined $B_2 \equiv \beta_1^2/\beta_0^2-\beta_2/\beta_0$ and $B_3 \equiv -\beta_1^3/\beta_0^3+2\beta_1\beta_2/\beta_0^2-\beta_3/\beta_0$.
\end{widetext}
\begin{figure*}[t!]
\subfigure[]
{
\includegraphics[width=0.3\textwidth]{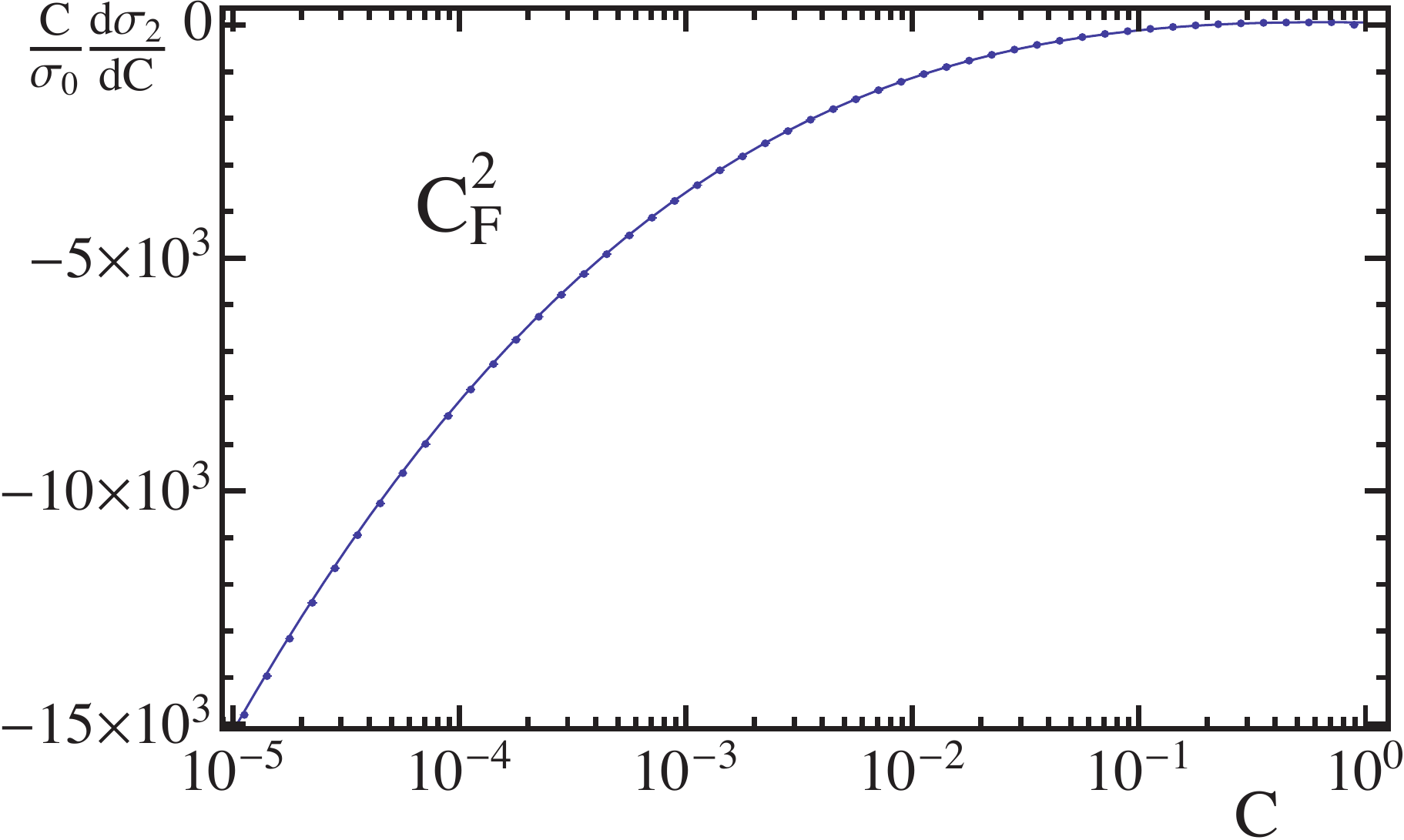}
\label{fig:EVENT2-CF}
}
\subfigure[]{
\includegraphics[width=0.3\textwidth]{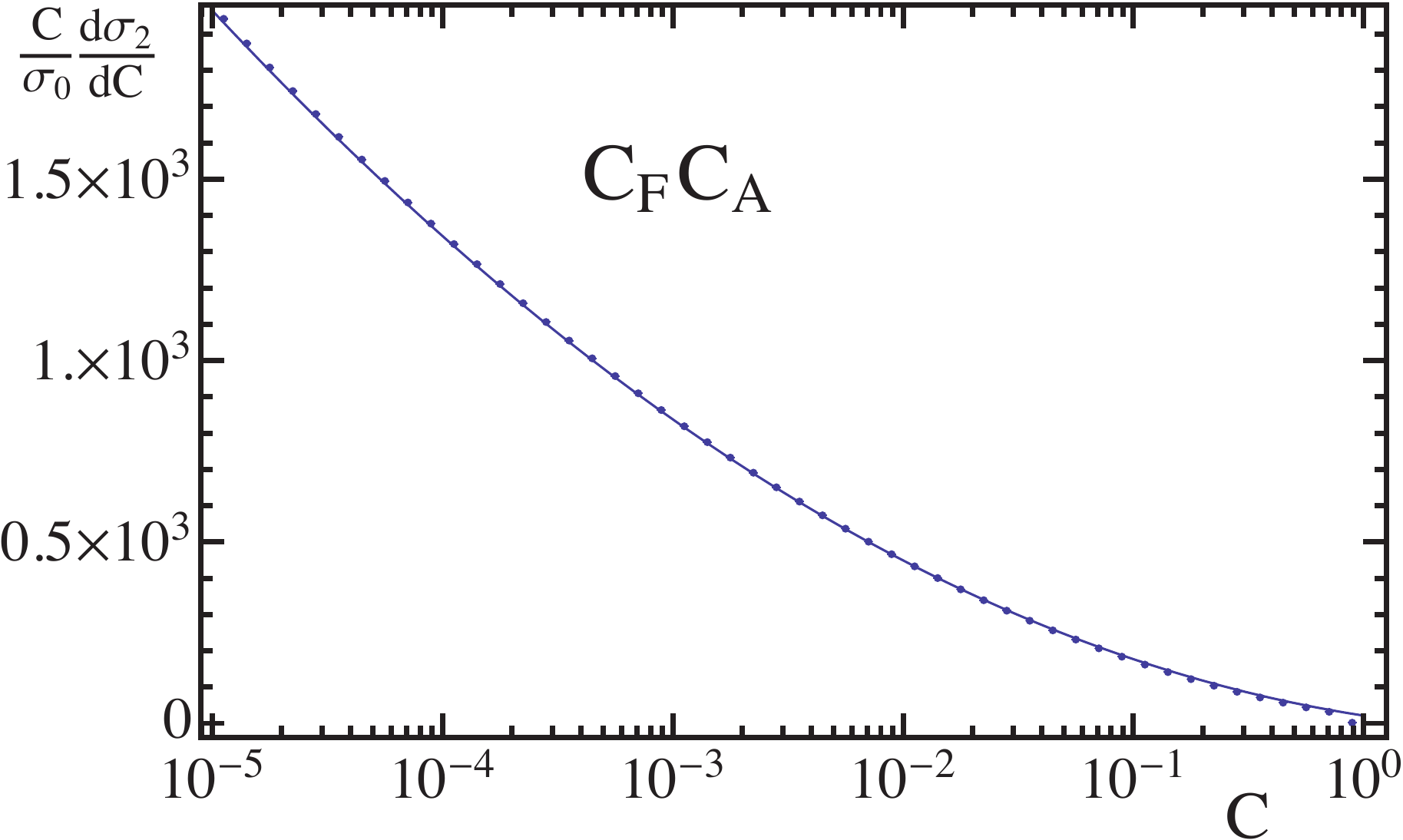}
\label{fig:EVENT2-CA}
}
\subfigure[]{
\includegraphics[width=0.31\textwidth]{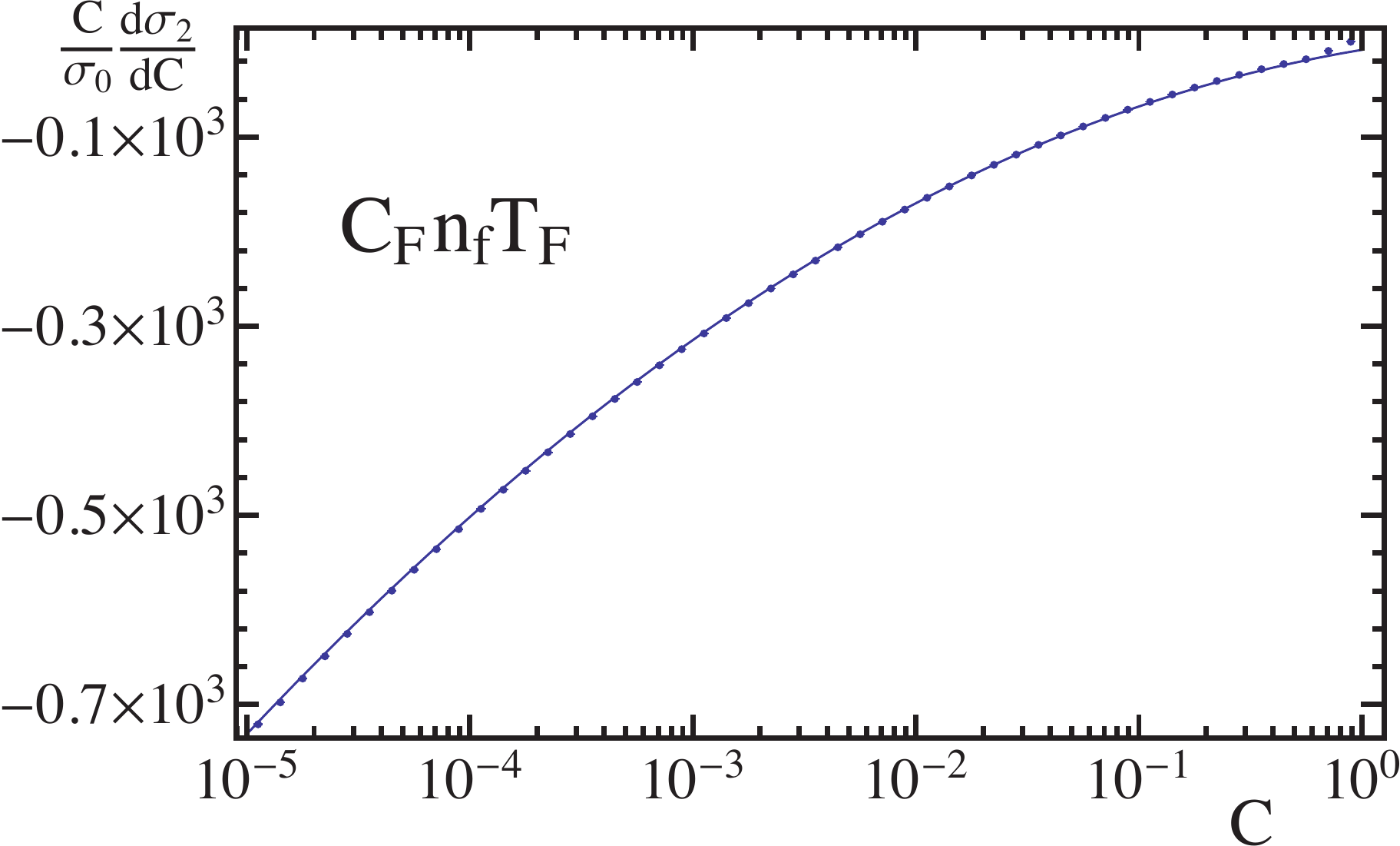}
\label{fig:EVENT2-NF}
}
\caption{ \label{fig:1loop-comparison} Comparison of the fixed-order analytic SCET prediction for the $\mathcal{O}(\alpha_s^2)$ piece with the parton level Monte Carlo EVENT2. The decomposition in the three color structures $C_F^2$, $C_F C_A$, and $C_F n_f T_f$ is shown in panels (a), (b), and (c), respectively. The factor $\alpha_s^2/(2\pi)^2$ has been divided out. We use a log binning in the horizontal axis to emphasize the dijet region.}
\label{fig:EVENT2}%
\end{figure*}
\noindent
These results are expressed in terms of the coefficients of
series expansions of the QCD $\beta$ function $\beta[\alpha_s]$,  $\Gamma[\alpha_s]$ (which is given
by a constant of proportionality times the QCD cusp anomalous dimension) and of
a non-cusp anomalous dimension $\gamma[\alpha_s]$. These expansion coefficients are defined by the equations
\begin{align}
&\beta(\as)=-\,2\,\as\,\sum_{n=0}^{\infty}\beta_n\Big(\frac{\as}{4\pi}\Big)^{\!\!n+1}\!\!,\;\\
&\Gamma(\as)=\sum_{n=0}^{\infty}\Gamma_n\Big(\frac{\as}{4\pi}\Big)^{\!\!n+1}\!\!,\;
\gamma(\as)=\sum_{n=0}^{\infty}\gamma_n\Big(\frac{\as}{4\pi}\Big)^{\!\!n+1}\!\!.\notag
\end{align}
For $n_f=5$, the relevant coefficients are~\cite{Tarasov:1980au, Larin:1993tp,
vanRitbergen:1997va, Korchemsky:1987wg, Moch:2004pa,Czakon:2004bu}
\begin{align}\label{eq:betaCusp}
\beta_0 &= 23/3\,,\quad
\beta_1 = 116/3\,,\quad
\beta_2 = 180.907,
\\
\beta_3 &= 4826.16,
\nn \\
\Gamma^{\rm{cusp}}_0 &= 16/3,\quad
\Gamma^{\rm{cusp}}_1 = 36.8436,\quad
\Gamma^{\rm{cusp}}_2 = 239.208
\,.
\nn
\end{align}
As was mentioned in the text, we use a Pad\'e
approximation for the unknown four-loop cusp anomalous dimension, assigning a large uncertainty to this estimate:
\begin{align}
\Gamma_{3}^{\rm cusp} = (1 \pm 2) \frac{(\Gamma _2^{\rm
cusp}){}^2}{\Gamma_{1}^{\rm cusp}}.
\end{align}
For the hard, jet, and soft functions, the anomalous dimensions are the same as in the thrust case and are given by~\cite{vanNeerven:1985xr,Matsuura:1988sm,Catani:1992ua,Vogt:2000ci,Moch:2004pa,Neubert:2004dd,Moch:2005id,Idilbi:2006dg,Becher:2006mr}
\begin{align}
\Gamma^H_n &= -\,\Gamma^{\rm{cusp}}_n,\quad
& \Gamma^J_n &= 2\,\Gamma^{\rm{cusp}}_n,\quad
& \Gamma^S_n &= -\,\Gamma^{\rm{cusp}}_n,
\nn \\[4pt]
\gamma^H_0 &= -\,8,\quad
&\gamma^H_1 &= 1.14194,\quad
&\gamma^H_2 &= -\,249.388,
\nn \\[4pt]
\gamma^J_0 &= 8,\quad
&\gamma^J_1 &= -\,77.3527,\quad
&\gamma^J_2 &= -\,409.631,
\nn \\
\gamma^S_n &= -\,\gamma^H_n-\gamma^J_n.
\end{align}

For the 4-loop running of the strong coupling constant, we use a form that agrees very well numerically with the solution to the beta function. For $n_f=5$, the value of the coupling is given by
\begin{align} \label{alphas}
\frac{1}{\alpha_s(\mu)} &= \frac{X}{\alpha_s(m_Z)} + 0.401347248\, \ln X 
\\
&+ \frac{\alpha_s(m_Z)}{X} \big[ 0.01165228\, (1-X) + 0.16107961\,
{\ln X}\big]
\nn\\
&+ \frac{\alpha_s^2(m_Z)}{X^2}
\big[ 0.1586117\, (X^2-1) +0.0599722\,(X
\nn\\
&+\ln X-X^2)
+0.0323244\, \{(1-X)^2-\ln^2 X\} \big] ,
\nn
\end{align}
where we have used the values from \eq{betaCusp} for the $\beta_i$ and $X=1+\alpha_s(m_Z) \ln(\mu/m_Z) \beta_0/(2\pi)$.

For the singular cross section, we have implemented the formulas described in this appendix into a Mathematica~\cite{mathematica} code. Additionally, we have created an independent Fortran~\cite{gfortran} code based on a Fourier space implementation (the nonsingular distributions have also been implemented into the two codes independently). These two codes agree with each other at $10^{-6}$ or better.

\section{Comparison to Parton Level Monte Carlos}
\label{ap:MC-comparison}
A useful way to validate our results is to compare the SCET prediction with fixed scales $\mu_H=\mu_J=\mu_S=Q$ to
the fixed-order prediction, for small values of $C$. This also constitutes an important test on the
accuracy of parton level Monte Carlos such as EVENT2 and EERAD3.

At $\mathcal{O}(\alpha_s^2)$ we compare our runs to the EVENT2 parton level Monte Carlo, splitting the output
in the various color structures. We used logarithmically-binned EVENT2 distributions across the entire
spectrum for this comparison. Details on the run parameters (number of events and cutoff parameter) have
been given in Sec.~\ref{sec:nonsingular}. Additionally, Fig.~\ref{fig:EVENT2} clearly shows that for all color structures the
agreement is excellent all the way to $C\sim 10^{-5}$. The very large number of events used in our runs ($3\times 10^{11}$) makes the error bars here essentially invisible. Also note that the cross section shoulder at $C=0.75$ is all in the largest $C$ bin and hence not visible in these plots.

\begin{figure*}[t!]
\subfigure[]
{
\includegraphics[width=0.3\textwidth]{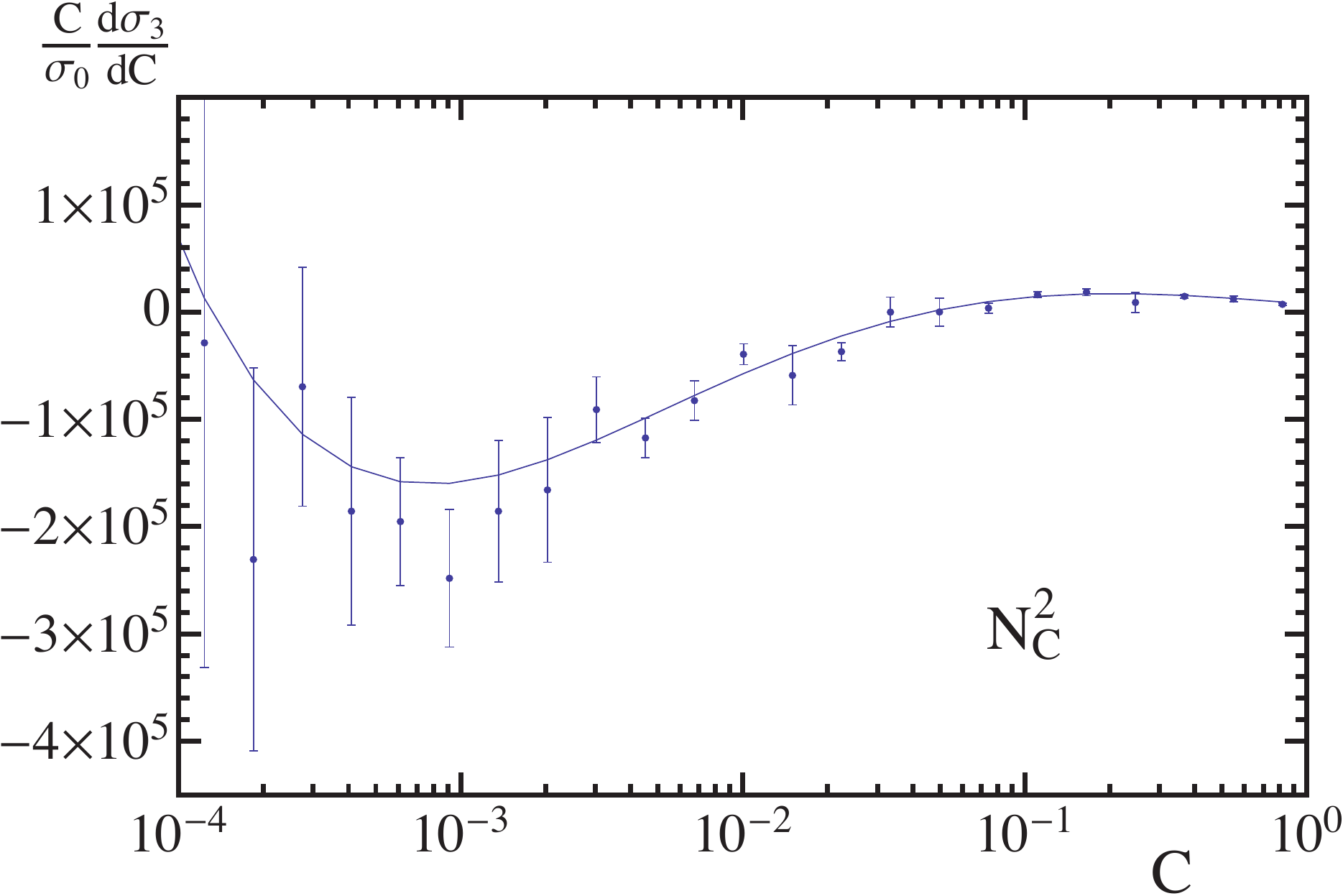}
\label{fig:EERAD-NC2}
}
\subfigure[]{
\includegraphics[width=0.3\textwidth]{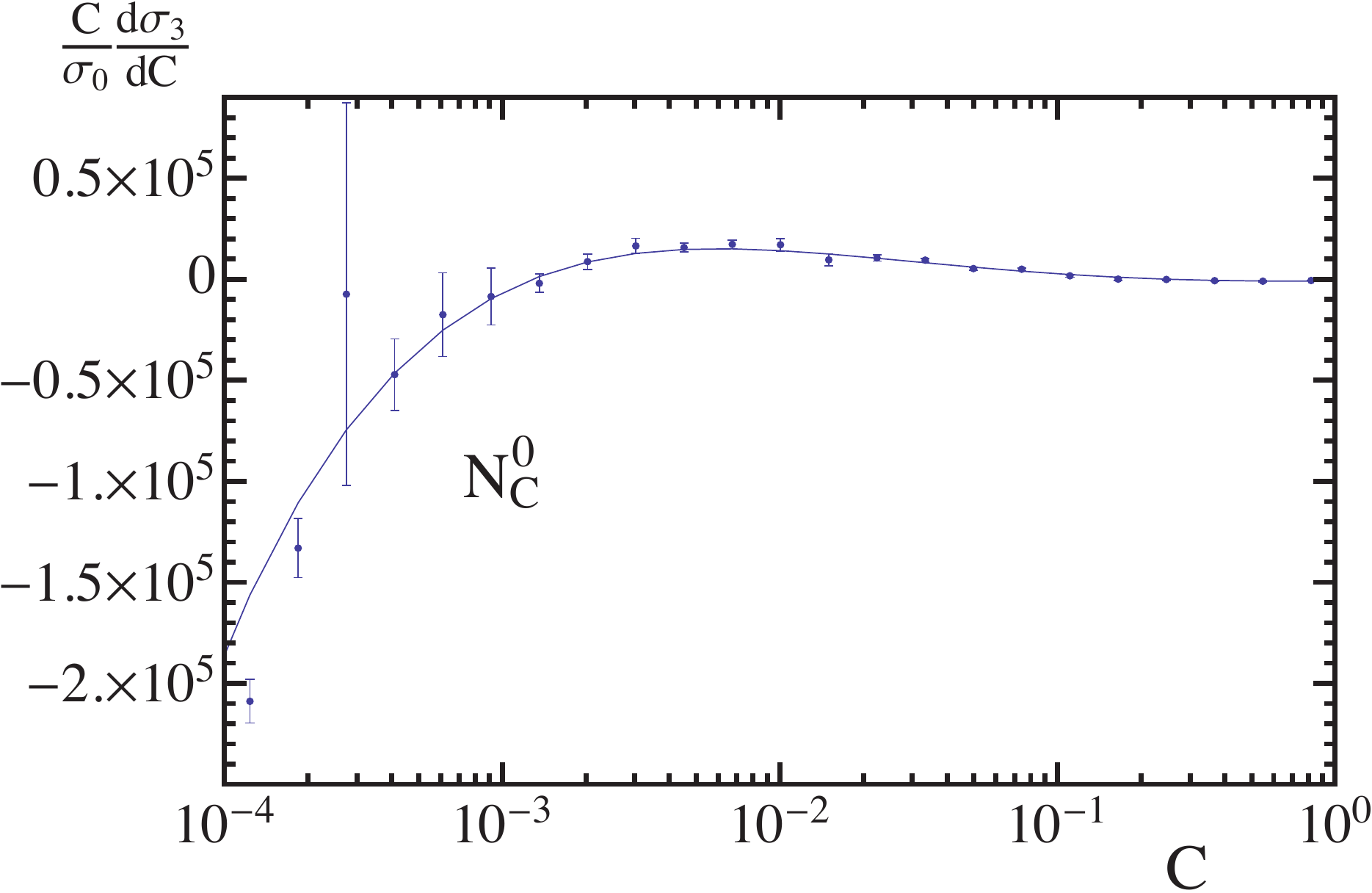}
\label{fig:EERAD-NC0}
}
\subfigure[]{
\includegraphics[width=0.3\textwidth]{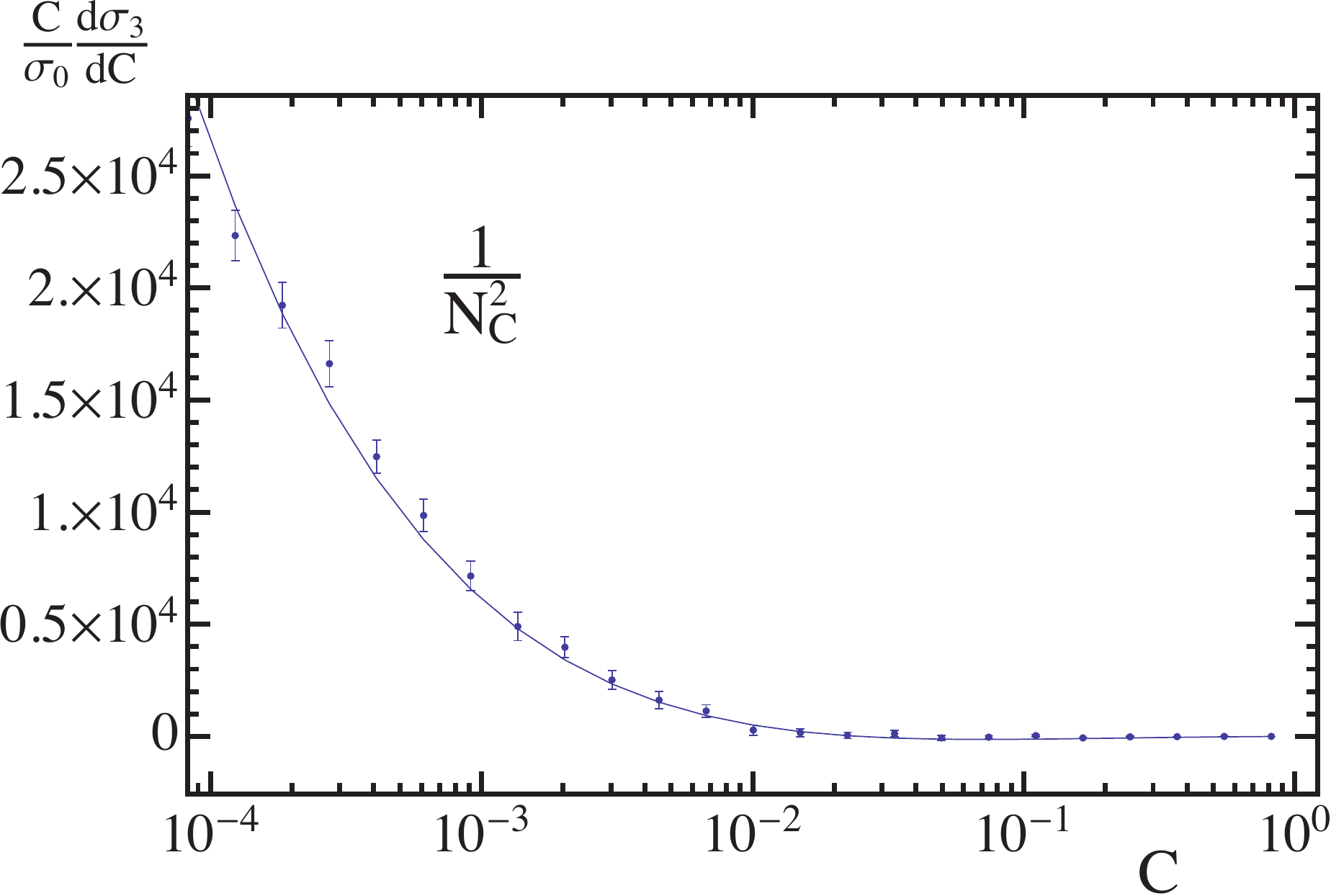}
\label{fig:EERAD-NC-2}
}
\subfigure[]
{
\includegraphics[width=0.3\textwidth]{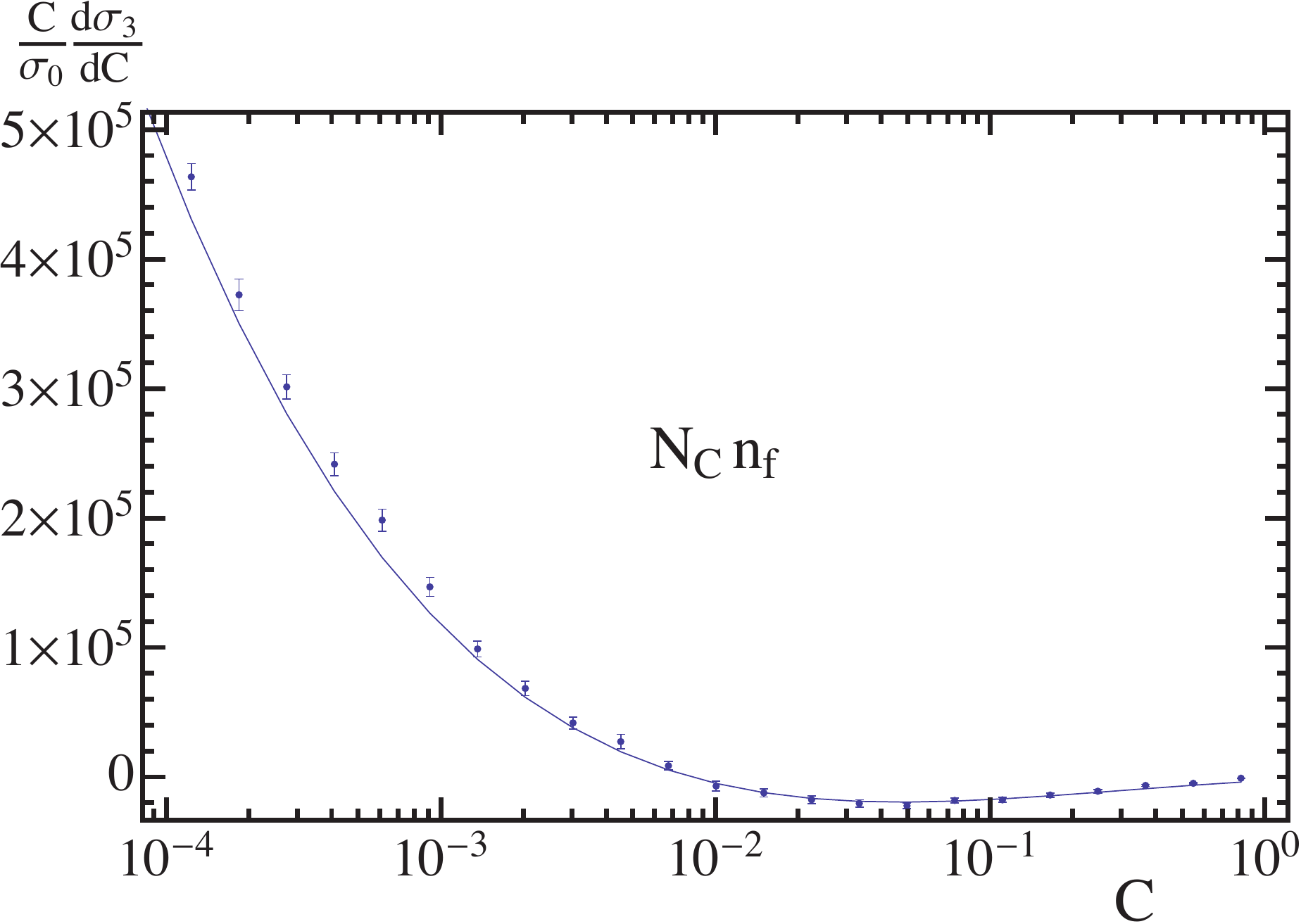}
\label{fig:EERAD-NC-nf}
}
\subfigure[]{
\includegraphics[width=0.3\textwidth]{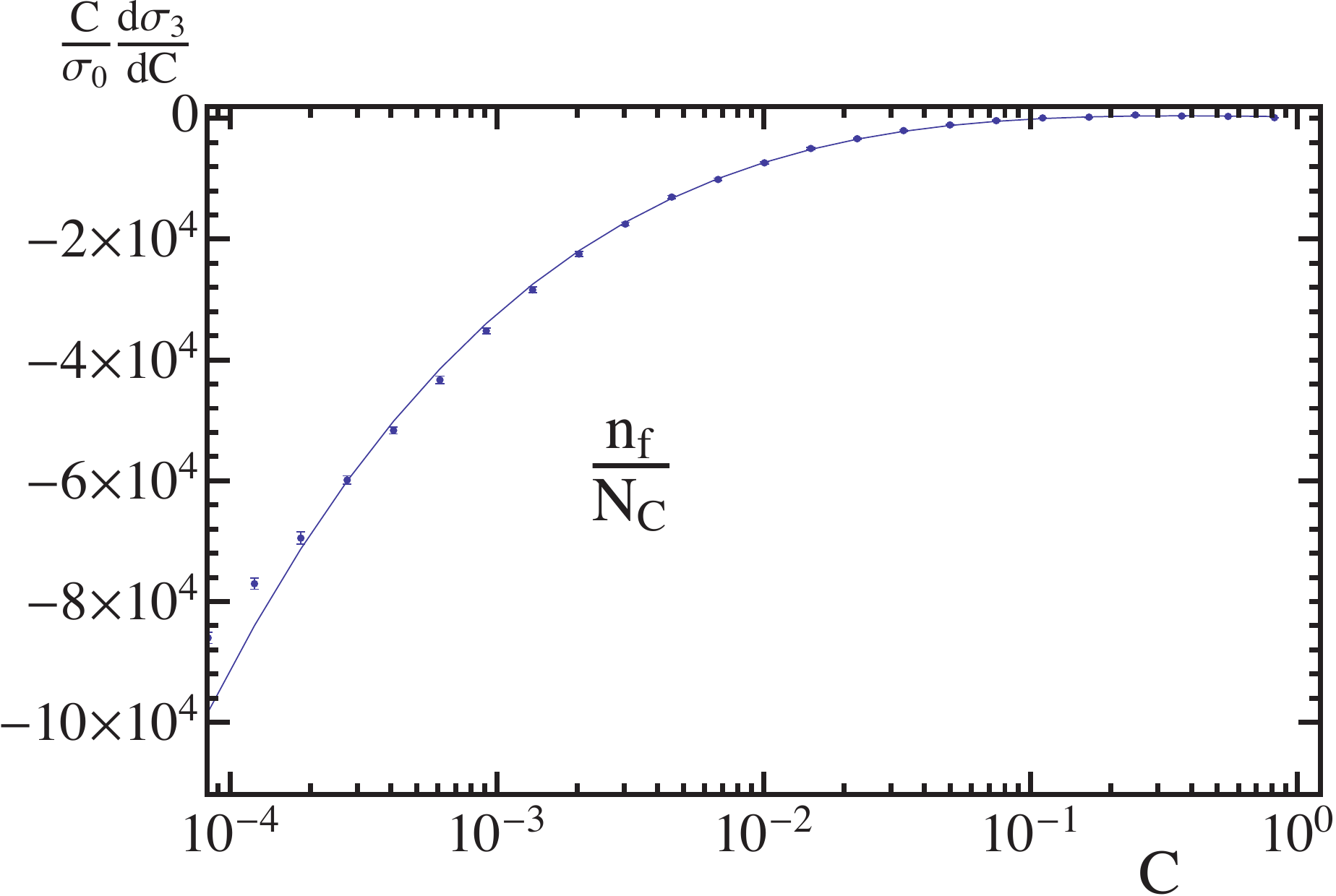}
\label{fig:EERAD-NC-1_nf}
}
\subfigure[]{
\includegraphics[width=0.3\textwidth]{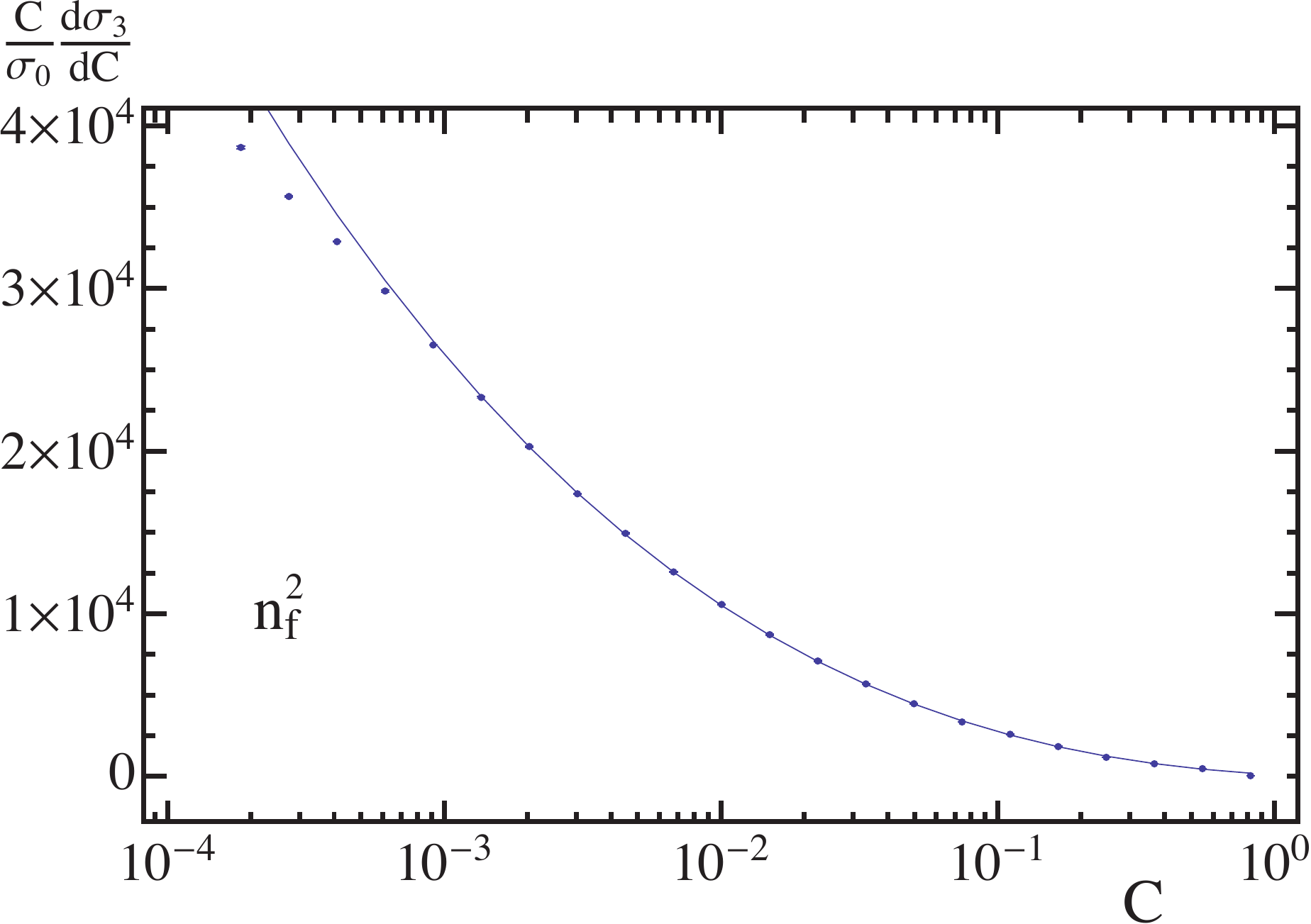}
\label{fig:EERAD-nf-2}
}
\caption{Comparison of the fixed-order analytic SCET prediction for the $\mathcal{O}(\alpha_s^3)$ piece with the parton
level Monte Carlo EERAD3. The decomposition in the three color structures $N_C^2$, $1$, and $N_C^{-2}$ are shown in panels (a), (b), and (c), of the first row, respectively, and $N_C\,n_f$, $n_f/N_C$ and $n_f^2$ are shown in panels (d), (e) and (f),
respectively, on the second row. The factor $\alpha_s^3/(2\pi)^3$ has been divided out. We use a log binning in the 
horizontal axis to emphasize the dijet region.}
\label{fig:EERAD3}%
\end{figure*}
\begin{figure*}[t!]
\includegraphics[width=0.3\textwidth]{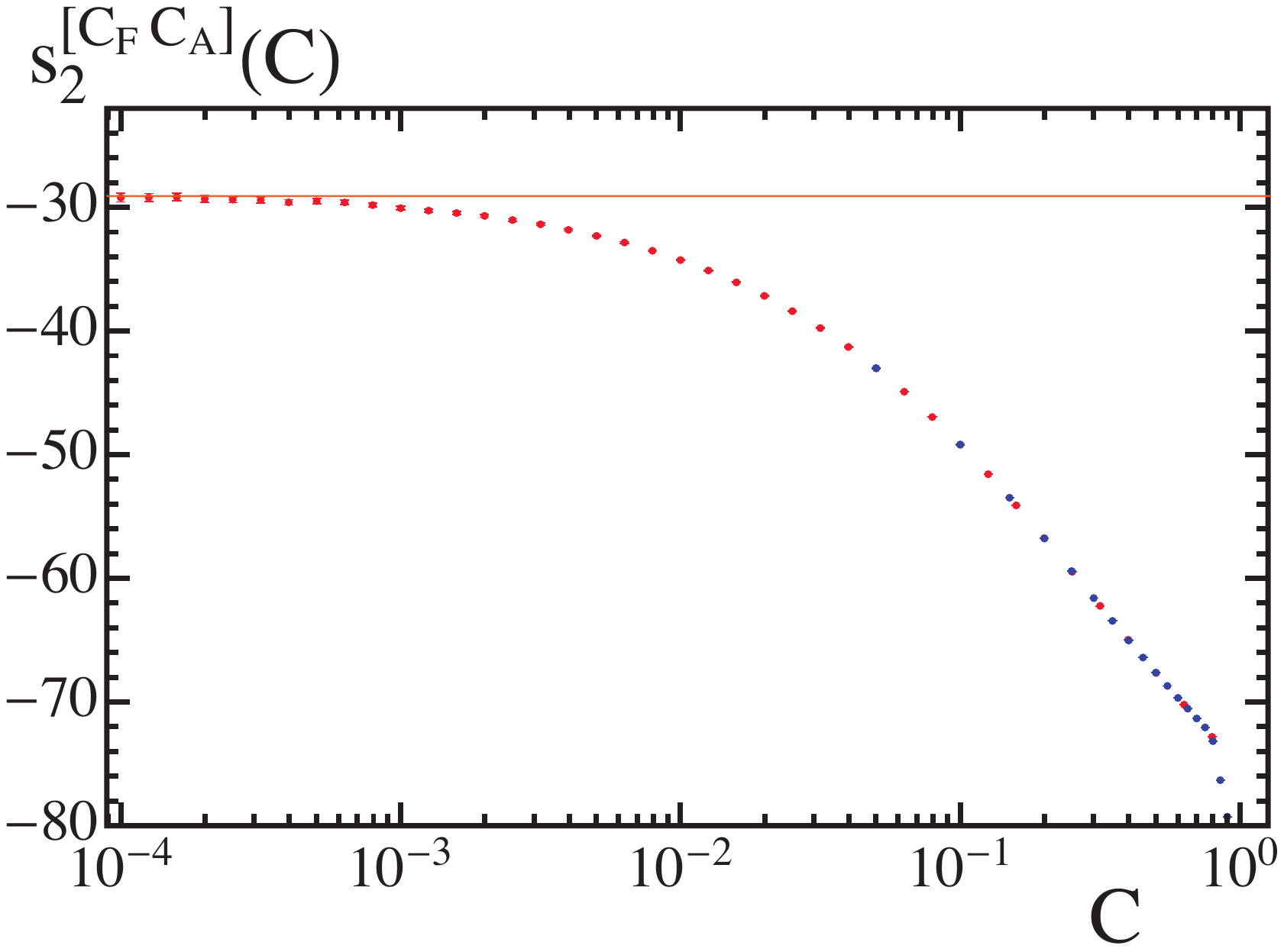}~~~~~~~~
\includegraphics[width=0.3\textwidth]{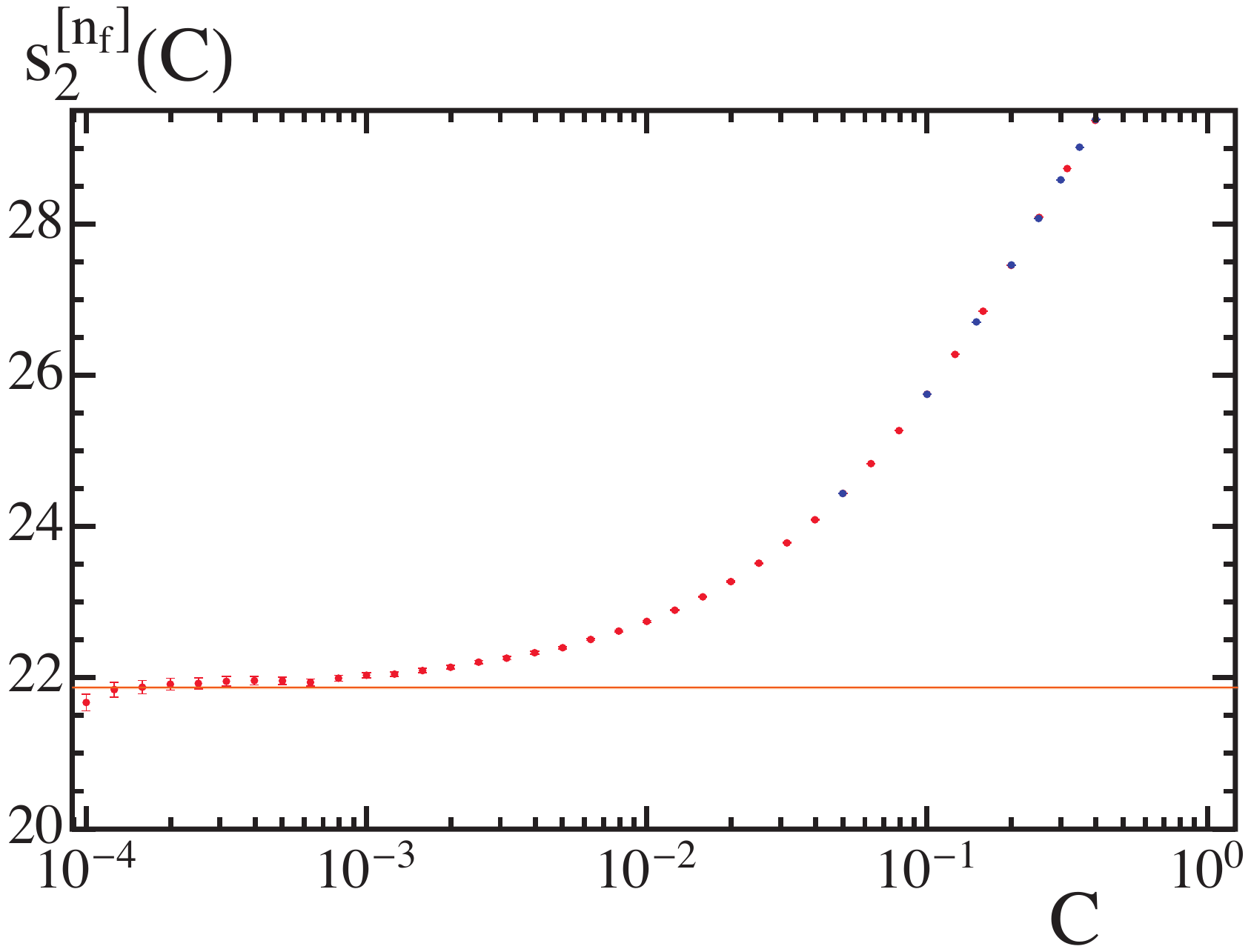}
\caption{Comparison of the determination of the soft function non-logarithmic constants at ${\cal O}(\alpha_s^2)$ as explained in Sec.~\ref{sec:softtwoloop} (shown as horizontal lines), with a determination employing
the method used in Ref.~\cite{Hoang:2008fs} (plateau at $10^{-4}\lesssim C \lesssim 10^{-3}$). The function $s_2^{\widetilde C}(C)$ is defined in Eq.~(\ref{eq:s2-lim}).
The $C_F C_A$ and $C_F n_f T_f$ color structures are shown in left and right panels, respectively.
The logarithmic horizontal axis emphasizes the $C\to 0$ extrapolation. Error bars are included and the blue and red points correspond to data from a different binning.}
\label{fig:s2}%
\end{figure*}

At $\mathcal{O}(\alpha_s^3)$ we compare to the EERAD3 parton level Monte Carlo output, again splitting the results
in the various color structures. Once again we use logarithmically binned distributions for this exercise. The results are shown in \Fig{fig:EERAD3}. The comparison looks very good for the numerically most relevant color structures (which also have the biggest uncertainties) and quite good for other structures. Slight deviations are observed in some cases for $C < 10^{-3}$, presumably indicating systematic uncertainties due the numerical infrared cutoff. The dominant color structures do not have this problem and have larger uncertainties, so one can still use the distribution even for $C\sim 10^{-4}$ as long as all color structures are summed up.

In Fig.~\ref{fig:s2} we compare our numerical determination of $s_2^{\widetilde C}$, that was described above in Sec.~\ref{sec:softtwoloop}, to the alternate method used in Ref.~\cite{Hoang:2008fs} and show that both procedures yield very similar results. The result from Sec.~\ref{sec:softtwoloop} is shown as an orange line (whose width is its uncertainty). The method of Ref.~\cite{Hoang:2008fs} gives the points. In Ref.~\cite{Hoang:2008fs} $s_2^\tau$ is computed from a relation very similar to Eq.~(\ref{eq:s2extractionfinalintegral}), written in the following form for \mbox{C-parameter}:
\begin{align}\label{eq:s2-lim}
s_2^{\widetilde C} &= \dfrac{1}{\sigma_0}\bigg\{\sigma_{\rm had}^{(2)} -
\Sigma_{\rm s}^{(2)}(1)\Big|_{s_2^{\widetilde C} = 0}\\
&-\lim_{C\to 0}\Big[\Sigma_{\rm ns}^{(2)}(1)-\Sigma_{\rm ns}^{(2)}(C)\Big]\bigg\}\nn
\equiv \lim_{C\to 0} s_2^{\widetilde C}(C)\,,
\end{align}
where we have implicitly defined the function $s_2^{\widetilde C}(C)$, and used
\begin{align}
\Sigma_{\rm s,ns}^{(2)}(C)  &= \int_0^C \df C\,\dfrac{\df \sigma_{\rm s,ns}^{(2)}}{\df C}\,.
\end{align}
Eq.~(\ref{eq:s2-lim}) can be broken down into various color factors. The limit in Eq.~(\ref{eq:s2-lim}) has to be taken numerically from the output of EVENT2. This is best achieved if events are distributed in logarithmic bins, such that the $C\to 0$ region is enhanced, as can be seen in Fig.~\ref{fig:s2}. The limit can be identified as the value at which the log-binned distribution reaches a plateau, which in the case of \mbox{C-parameter} happens for $10^{-4}\lesssim C \lesssim 10^{-3}$. Figure~\ref{fig:s2} shows that our determination of $s_2^{\widetilde C}$ as described in Sec.~\ref{sec:softtwoloop}, represented by an orange line, agrees with the plateau for the two nontrivial color structures.

\section{Computation of 1-loop Soft Function}
\label{ap:softoneloop}
In this section we present a general computation of the 1-loop soft function for any
event shape $e$ which can be expressed in the dijet limit as
\begin{align}
e = \frac{1}{Q}\sum_i p_i^\perp f_e(y_i)\,,
\end{align}
where the sum is over all particles in the final state, $p_i$ is the magnitude of the
transverse momentum and $y_i$ is the rapidity of the particle, both measured with
respect to the thrust axis.\,\footnote{For perturbative computations partons are taken as massless and
hence rapidity $y$ and pseudorapidity $\eta$ coincide.} For thrust one has $f_\tau(y) = \exp(-|y|)$, for angularities
one has $f_{\tau^a}(y) = \exp[-(1-a)|y|\,]$, and for C-parameter one has $f_{C}(y) = 3/\cosh y$ and
$f_{\widetilde C}(y) = 1/(2\cosh y)$.

\begin{figure}
\begin{center}
\includegraphics[width=0.95\columnwidth]{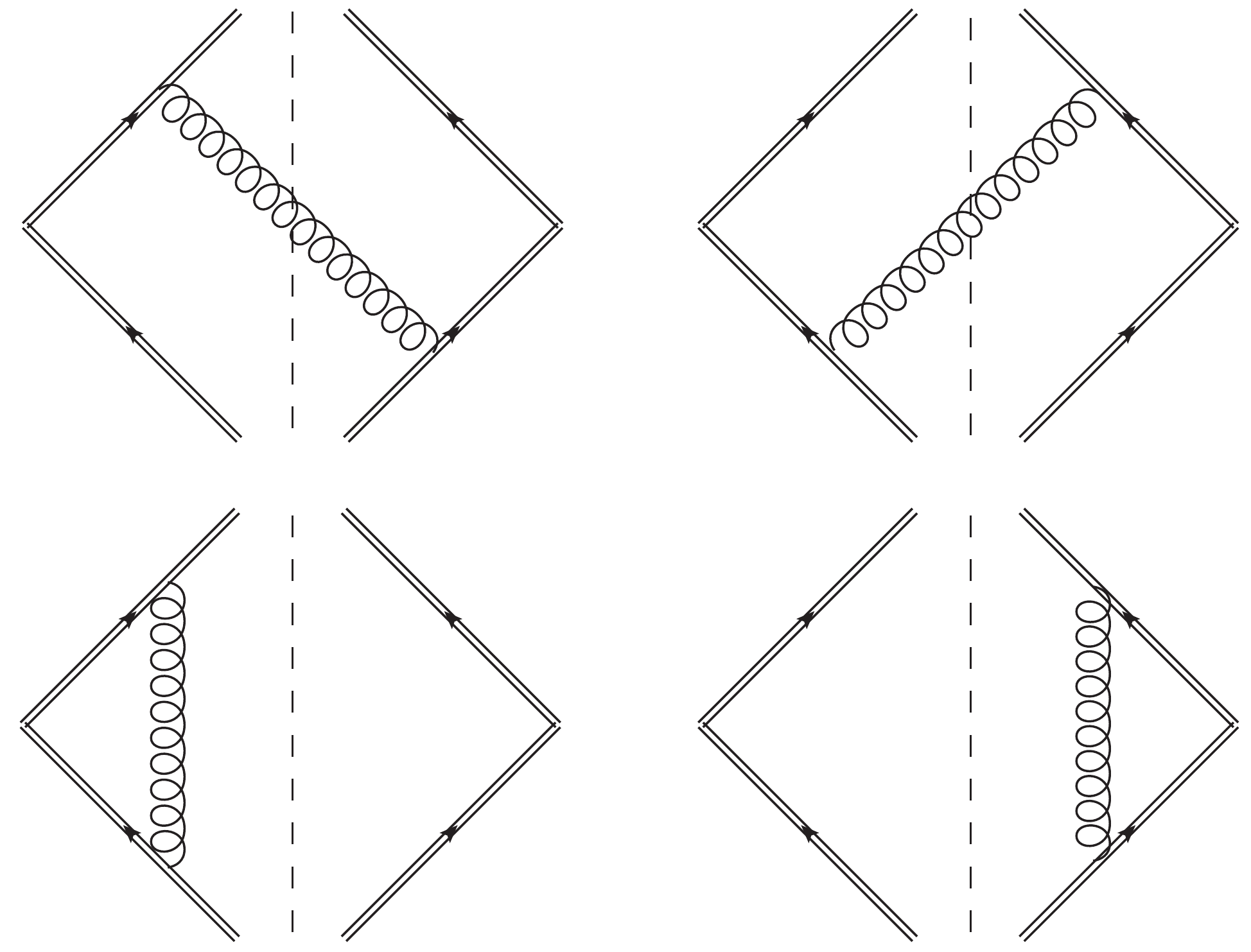}
\caption{Diagrams contributing to the soft function at one loop.}
  \label{fig:soft-one-loop}
\end{center}
\end{figure}
One needs to compute the four diagrams in Fig.~\ref{fig:soft-one-loop} in order to
determine the soft function. The two diagrams on the bottom are scaleless and vanish in
dimensional regularization. They actually convert the IR divergences in the two diagrams
on the top into UV divergences. We take the space-time number of dimensions to be
\mbox{$d = 4 - 2\epsilon$}. A direct computation in momentum space gives
\begin{align}\label{eq:soft-general}
\!\!\!S^{\rm 1-loop}_e(\ell) = 4\, g_s^2 C_F \!\!\int \!
\frac{\df^{3-2\epsilon}\vec{p}}{(2\pi)^{3-2\epsilon}2 |\vec{p}\,|}
\frac{\delta[\ell - p_T f_e(y)]}{p^+ p^-}\,.
\end{align}
After integrating the angular variables, it is convenient to make a change of variables
from $p^\pm$ to ($p_T$, $y$):
\begin{align}\label{eq:chage-variable}
\frac{\df^{3-2\epsilon}{\vec p}}{(2\pi)^{3-2\epsilon}2|\vec{p}\,|} = 
\frac{2}{(4\pi)^{2-\epsilon}}\frac{p_T^{1-2\epsilon}}{\Gamma(1-\epsilon)}
\,\df p_T \df y\,.
\end{align}
Using Eq.~(\ref{eq:chage-variable}) in Eq.~(\ref{eq:soft-general}) and imposing the
on-shell condition $p^+p^- = p_T^2$, where $\vec{p\,}^2 = p_z^2 + p_T^2$, we obtain
\begin{align}\label{eq:soft-momentum}
& S_e^{\rm 1-loop}(\ell)  \\
&= \frac{2\,\alpha_s(\mu) C_F e^{\epsilon \gamma_E}}{\mu\,\pi\,\Gamma(1-\epsilon)}
\!\int \! \df p_T\, \df y\, \Big(\frac{p_T}{\mu}\Big)^{-1-2\epsilon} 
\delta[\ell - p_T f_e(y)] \nonumber\\
&= \frac{2\,\alpha_s(\mu) C_F
e^{\epsilon \gamma_E}}{\mu\,\pi\,\Gamma(1-\epsilon)}
\Big(\frac{\ell}{\mu}\Big)^{-1-2\epsilon}\,I_e(\epsilon)\,,\nonumber
\end{align}
where we have defined
\begin{align}
I_e(\epsilon) = \int_{-\infty}^\infty\! \df y\, [\,f_e(y)\,]^{2\epsilon}\,.
\end{align}
Similarly in Fourier space, one gets
\begin{align}\label{eq:soft-position}
\!\!\!\tilde{S}_e^{\rm 1-loop}(x)  = \frac{2\,\alpha_s C_F}{\pi}
\frac{\Gamma(-2\epsilon)}{\,\Gamma(1-\epsilon)}
(i\, x\, \mu)^{2\epsilon}e^{\epsilon \gamma_E}\,I_e(\epsilon)\,.
\end{align}
For thrust and angularities, one trivially obtains
\begin{align}
I_{\tau}(\epsilon) = \frac{1}{\epsilon}\,,\qquad
I_{\tau^a}(\epsilon) = \frac{1}{1-a}\frac{1}{\epsilon}\,.
\end{align}
For C-parameter it is convenient to perform a change of variables,
\begin{align}
y = \frac{1}{2}\ln\!\Big(\frac{1+x}{1-x}\Big)\,,\quad
\cosh y = \frac{1}{\sqrt{1-x^2}}\,,
\end{align}
obtaining for the integral:
\begin{align}
I_{\widetilde C}(\epsilon) & = 2^{1-2\epsilon}\!\int_0^1\!\df x\, (1-x^2)^{-1+\epsilon}
  = \frac{1}{2}\frac{\Gamma(\epsilon)^2}{\Gamma(2\epsilon)}\,.
\end{align}
Expanding in $\epsilon\to 0$ and upon renormalization in ${\overline{\rm MS}}$, we find for the position space soft
function
\begin{align}
\tilde{S}^{\rm 1-loop}_{\widetilde C} & = -\frac{\alpha_s(\mu)}{4\pi}\,C_F
\Big[\frac{\pi^2}{3}+8\ln^2(ix\mu e^{\gamma_E})\Big]\,,\\
\tilde{S}^{\rm 1-loop}_{\tau^a} & = -\frac{1}{1-a}\frac{\alpha_s(\mu)}{4\pi}\,C_F
\Big[\pi^2+8\ln^2(ix\mu e^{\gamma_E})\Big]\,.\nonumber
\end{align}
Writing the logarithm of the soft function in Fourier space evaluated at the point
$x = -\,i\exp(-\,\gamma_E)/\mu$ in a generic form as
\begin{align}
\ln {\widetilde S_e} = 2\sum_{n=1}^\infty \Big(\frac{\alpha_s(\mu)}{4\pi}\Big)^n s_n^e \,.
\end{align}
we obtain
\begin{align}
s_1^{\widetilde C} = -\,\frac{\pi^2}{6}\,C_F\,,\qquad
s_1^{\tau_a} = -\, \frac{1}{1-a}\frac{\pi^2}{2}\,C_F\,.
\end{align}
Fourier transforming the result, one obtains the renormalized momentum space soft function:
\begin{align}
S^{\rm 1-loop}_{\widetilde C} & = \frac{\alpha_s(\mu)}{4\pi}\,C_F
\bigg(\!\pi^2\,\delta(\ell) -
\frac{16}{\mu}\bigg[\frac{\ln(\ell/\mu)}{\ell/\mu}\bigg]_+\bigg)
  \,,\\
S^{\rm 1-loop}_{\tau_a}   & = \frac{1}{1-a}\frac{\alpha_s(\mu)}{4\pi}\,C_F
\bigg(\!\frac{\pi^2}{3}\,\delta(\ell)-
\frac{16}{\mu}\bigg[\frac{\ln(\ell/\mu)}{\ell/\mu}\bigg]_+\bigg)\,.\nn
\end{align}

\section{$\mathbf{B_i, G_{ij}}$ coefficients}
\label{ap:Gijcoefficients}
The resummed cross section at N$^3$LL$^\prime$ can be used to compute various fixed-order coefficients, as in \Eqs{eq:BG1}{eq:BG2}.  The results for coefficients up to ${\cal O}(\alpha_s^3)$ in perturbation theory are summarized here.

The $B_i$ coefficients read
\begin{align}
B_{1}^{[0]} &=C_F \!\left(\dfrac{2 \pi^2}{3}-1 \right), \\[1.5mm]
B_{2}^{[0]} &=C_F^2\! \left(-\,6 \,\zeta
   _3+1-\frac{17 \pi ^2}{24}+\frac{11 \pi ^4}{36}\right) \nn \\
   &+C_A C_F \!\left(\frac{s_2^{[C_F C_A]}}{4}+\frac{283 \zeta
   _3}{18}-\frac{73 \pi ^4}{360}+\frac{85 \pi
   ^2}{24}+\frac{493}{324}\right) \nn \\
   &+C_F
   n_f T_F\! \left(\frac{s_2^{[n_f]}}{4}-\frac{22 \zeta
   _3}{9}-\frac{7 \pi ^2}{6}+\frac{7}{81}\right), \nn \\[1.5mm]
B_{3}^{[0]} &=C_A^2 C_F \!\left(\frac{620179 \zeta _3}{1944}-\frac{41 \pi
   ^2 \zeta _3}{2}-\frac{284 \zeta _3^2}{9}-\frac{217 \zeta
   _5}{18} \right. \nn \\
   & \left. -\,\frac{51082685}{209952}+\frac{1294933 \pi
   ^2}{34992}-\frac{3641 \pi ^4}{7776}+\frac{4471 \pi
   ^6}{102060}\right) \nn \\
   &+C_A C_F^2\! \left(\frac{5 \pi ^2
   s_2^{[C_F C_A]}}{24}-\frac{s_2^{[C_F C_A]}}{4}-\frac{2273 \zeta
   _5}{9}+\frac{2 \zeta _3^2}{3} \right. \nn \\
   & \left. +\,\frac{248 \pi ^2 \zeta
   _3}{9}-\frac{89 \zeta _3}{27}-\frac{14887 \pi
   ^6}{68040}+\frac{23093 \pi ^4}{19440}+\frac{172585 \pi
   ^2}{3888} \right. \nn \\
   & \left. -\,\frac{185039}{1296}\right)+C_A C_F n_f
   T_F \!\left(-\frac{352 \zeta _3}{3}+\frac{13 \pi ^2 \zeta
   _3}{3}-\frac{2 \zeta _5}{3} \right. \nn \\
   & \left. +\,\frac{1700171}{13122}-\frac{103903 \pi
   ^2}{4374}+\frac{227 \pi ^4}{4860}\right)+C_F^3 \bigg(-\,167\, \zeta
   _3 \nn \\
   & \left. +\,\frac{38 \pi ^2 \zeta _3}{3}-\frac{4 \zeta _3^2}{3}+22 \zeta
   _5-\frac{4679}{96}+\frac{139 \pi ^2}{18}-\frac{109 \pi
   ^4}{40} \right. \nn \\
   & \left. +\,\frac{42757 \pi ^6}{136080}\right)+C_F^2 n_f
   T_F\! \left(\frac{5 \pi ^2
   s_2^{[n_f]}}{24}-\frac{s_2^{[n_f]}}{4}+\frac{368 \zeta
   _5}{9} \right. \nn \\
   & \left. -\,\frac{94 \pi ^2 \zeta _3}{9}+\frac{4324 \zeta
   _3}{81}-\frac{497 \pi ^4}{2430}-\frac{35503 \pi
   ^2}{1944}+\frac{112073}{972}\right) \nn \\
   &+C_F n_f^2 T_F^2\!
   \left(\frac{808 \zeta _3}{243}-\frac{190931}{13122}+\frac{257 \pi
   ^2}{81}+\frac{52 \pi
   ^4}{1215}\right) \nn \\
   &+\,\frac{j_3}{4}+\frac{s_3^{\widetilde C}}{8}, \nn \\[1.5mm]
B_1 &= C_F \left(\frac{2 \pi ^2}{3}-\frac{5}{2}\right), \\
B_2&=C_F^2\! \left(-\,6\, \zeta
   _3+\frac{41}{8}-\frac{41 \pi ^2}{24}+\frac{11 \pi
   ^4}{36}\right)\nn \\
   &+C_A C_F \!\left(\frac{s_2^{[C_F C_A]}}{4}+\frac{481 \zeta
   _3}{18}-\frac{73 \pi ^4}{360}+\frac{85 \pi
   ^2}{24}-\frac{8977}{648}\right)\nn \\
   &+C_F n_f T_F\!
   \left(\frac{s_2^{[n_f]}}{4}-\frac{58 \zeta _3}{9}-\frac{7 \pi
   ^2}{6}+\frac{905}{162}\right), \nn \\[1.5mm]
B_3&=C_A^2 C_F \!\left(\frac{915775 \zeta _3}{1944}-\frac{41 \pi
   ^2 \zeta _3}{2}-\frac{284 \zeta _3^2}{9}+\frac{113 \zeta
   _5}{18}  \right. \nn \\
   & \left. -\,\frac{95038955}{209952}+\frac{1353739 \pi
   ^2}{34992}-\frac{3641 \pi ^4}{7776}+\frac{4471 \pi
   ^6}{102060}\right) \nn \\
   &+C_A C_F^2\! \left(\frac{5 \pi ^2
   s_2^{[C_F C_A]}}{24}-\frac{5 s_2^{[C_F C_A]}}{8}-\frac{3263 \zeta
   _5}{9}+\frac{2 \zeta _3^2}{3} \right. \nn \\
   & \left. +\,\frac{314 \pi ^2 \zeta
   _3}{9}+\frac{67 \zeta _3}{108}-\frac{14887 \pi
   ^6}{68040}+\frac{14503 \pi ^4}{9720}+\frac{56039 \pi
   ^2}{1944}  \right. \nn \\
   & \left. -\,\frac{87719}{1296}\right)+C_A C_F n_f
   T_F \!\left(-\,\frac{1952 \zeta _3}{9}+\frac{13 \pi ^2 \zeta
   _3}{3}-\frac{22 \zeta _5}{3} \right. \nn \\
   & \left. +\,\frac{3585851}{13122}-\frac{109249 \pi
   ^2}{4374}+\frac{227 \pi ^4}{4860}\right)+C_F^3 \bigg(-\,158\, \zeta
   _3  \nn \\
   & \left. +\,\frac{38 \pi ^2 \zeta _3}{3}-\frac{4 \zeta _3^2}{3}+22\, \zeta
   _5-\frac{5093}{96}+\frac{1517 \pi ^2}{144}-\frac{191 \pi
   ^4}{60} \right. \nn \\
   & \left. +\,\frac{42757 \pi ^6}{136080}\right)+C_F^2 n_f
   T_F \!\left(\frac{5 \pi ^2 s_2^{[n_f]}}{24}-\frac{5
   s_2^{[n_f]}}{8}+\frac{728 \zeta _5}{9} \right. \nn \\
   & \left. -\,\frac{118 \pi ^2 \zeta
   _3}{9}+\frac{2839 \zeta _3}{81}-\frac{497 \pi ^4}{2430}-\frac{24973
   \pi ^2}{1944}+\frac{188173}{1944}\right) \nn \\
   &+C_F n_f^2
   T_F^2 \!\left(\frac{4912 \zeta
   _3}{243}-\frac{484475}{13122}+\frac{275 \pi ^2}{81}+\frac{52 \pi
   ^4}{1215}\right) \nn \\
   &+\,\frac{j_3}{4}+\frac{s_3^{\widetilde C}}{8}. \nn
\end{align}
The results for the first few $G_{ij}$ coefficients read
\begin{align}
 G_{12}=& -2\, C_F, \:\:\:\: G_{11}= 3\, C_F,\\[1.5mm]
 G_{23}=&\,C_F \!\left(\frac{4 n_f T_F}{3}-\frac{11 C_A
   }{3}\right), \nn \\[1.5mm]
G_{22}=&C_F \!\left[ C_A\!
   \left(\frac{\pi ^2}{3}- \frac{169}{36}\right)-\frac{4}{3} \pi ^2 C_F+\frac{11 n_f
   T_F}{9}\right], \nn \\[1.5mm]
G_{21}=&\, C_F\!\left[ C_A\! \left(-\,6\, \zeta _3+\frac{57}{4}+\frac{11 \pi
   ^2}{9}\right)\right. \nn \\
   &\left. +\,C_F\! \left(-\,4\, \zeta _3+\frac{3}{4}+\pi
   ^2\right) +n_f
   T_F \!\left(-\,5-\frac{4 \pi ^2}{9}\right) \right],  \nn \\[1.5mm]
 G_{34}=&\,C_F \!\left(-\,\frac{847 C_A^2}{108}+\frac{154}{27} C_A
   n_f T_F-\frac{28}{27} n_f^2
   T_F^2 \right), \nn \\[1.5mm]
   G_{33}=&\,C_F\! \left[ C_A^2\!
   \left(\frac{11 \pi ^2}{9}-\frac{3197}{108}\right) -\frac{22\pi^2}{3} C_A
   C_F \right. \nn \\
   &+C_A
   n_f T_F\! \left(\frac{512}{27}-\frac{4 \pi ^2}{9}\right) +\frac{64 C_F^2 \zeta
   _3}{3}\nn \\
   &\left.+\,C_F n_f T_F\! \left(2+\frac{8 \pi ^2}{3}\right) -\frac{68}{27} n_f^2 T_F^2 \right], \nn \\
     &\nn \\
   G_{32}=&\,C_F\! \left[ C_A^2\! \left(11\, \zeta _3-\frac{11323}{648}+\frac{497 \pi
   ^2}{54}-\frac{11 \pi ^4}{90}\right) \right. \nn \\
   &+C_A C_F \!\left(-\,110\,
   \zeta _3+\frac{11}{8}-\frac{70 \pi ^2}{27}+\frac{4 \pi^4}{9}\right) \nn \\
   &+C_A n_f T_F \!\left(4\, \zeta
   _3+\frac{673}{162}-\frac{152 \pi ^2}{27}\right) \nn \\
   &+C_F^2\!
   \left(\frac{8 \pi ^4}{45}-48 \zeta _3\right)+C_F n_f
   T_F \!\left(32\, \zeta _3+\frac{43}{6}+\frac{8 \pi
   ^2}{27}\right)\nn \\
   &\left.+\,n_f^2 T_F^2\! \left(\frac{70}{81}+\frac{8 \pi ^2}{9}\right) \right]
   , \nn \\[1.5mm]
G_{31}=&\,C_A^2 C_F\! \left(\frac{11 s_2^{[C_F C_A]}}{6}+10\, \zeta
   _5-\frac{361 \zeta _3}{27}-\frac{541 \pi ^4}{540} +\frac{892 \pi
   ^2}{81} \right. \nn \\
   & \left. +\,\frac{77099}{486}\right)+C_A C_F^2\!
   \left(\frac{452 \zeta _3}{9}+2 \pi^2 \zeta _3+30\, \zeta
   _5  +\frac{23}{2} \right. \nn \\
   & \left. +\,\frac{161 \pi^2}{72}-\frac{49 \pi
   ^4}{135}\right)+C_A C_F n_f T_F\!
   \left(-\,\frac{2 s_2^{[C_F C_A]}}{3} \right. \nn \\
   & \left. +\,\frac{11 s_2^{[n_f]}}{6} -\frac{608
   \zeta _3}{27}+\frac{10 \pi ^4}{27}-\frac{520 \pi
   ^2}{81}-\frac{24844}{243}\right) \nn \\
   &+C_F^3 \!\left(53\, \zeta
   _3-\frac{44 \pi^2 \zeta _3}{3}+132\zeta _5+\frac{29}{8}+\frac{5
   \pi ^2}{4}-\frac{8 \pi ^4}{15}\right) \nn \\
   &+C_F^2 n_f T_F\!
   \left(-\,\frac{208 \zeta _3}{9}-\frac{77}{4}-\frac{31 \pi
   ^2}{18}+\frac{8 \pi ^4}{135}\right) \nn \\
   &+C_F n_f^2 T_F^2\!
   \left(\!-\,\frac{2 s_2^{[n_f]}}{3}+\frac{176 \zeta _3}{27}+\frac{64 \pi
   ^2}{81}+\frac{3598}{243}\right). \nn
\end{align}
Note that the entire infinite series of $G_{ij}$ coefficients listed in \Tab{tab:Gijorders} is determined by our resummation results.

\section{R-evolution with and without Hadron Mass Effects}
\label{ap:hadronmassR}
The result for R-evolution in the case of no hadron masses is given in \Eq{eq:DeltaRevolution}.
The resummed $\omega$ appearing in this equation was given in App.~\ref{ap:formulae}. The remaining coefficients and variables that appear in this equation are
\begin{align} \label{eq:Sjnor}
S_0 &= \frac{\gamma_0^R}{2 \beta_0} \,,\,\,
S_1  = \frac{\gamma_1^R}{(2 \beta_0)^2} - (\hat{b}_1 + \hat{b}_2) \frac{\gamma_0^R}{2 \beta_0}\,, \nn \\
S_2 &=  \frac{\gamma_2^R}{(2 \beta_0)^3} - (\hat{b}_1 + \hat{b}_2) \frac{\gamma_1^R}{(2 \beta_0)^2}\,,  \nn \\
\hat{b}_1 &= \frac{\beta_1}{2 \beta_0^2}\,, \;\; \hat{b}_2 = \frac{\beta_1^2 - \beta_0 \beta_2}{4 \beta_0^4}\,,\\
\hat{b}_3 &= \frac{\beta_1^3-2 \beta_0 \beta_1 \beta_2 + \beta_0^2 \beta_3}{8 \beta_0^6}\,,  \nn \\
t_1 &= -\,\frac{2 \pi}{\beta_0 \alpha_s(R)},\,t_0 = -\,\frac{2 \pi}{\beta_0 \alpha_s(R_\Delta)}\,. \nn
\end{align}
\begin{figure}[t!]
\begin{center}
\includegraphics[width=0.95\columnwidth]{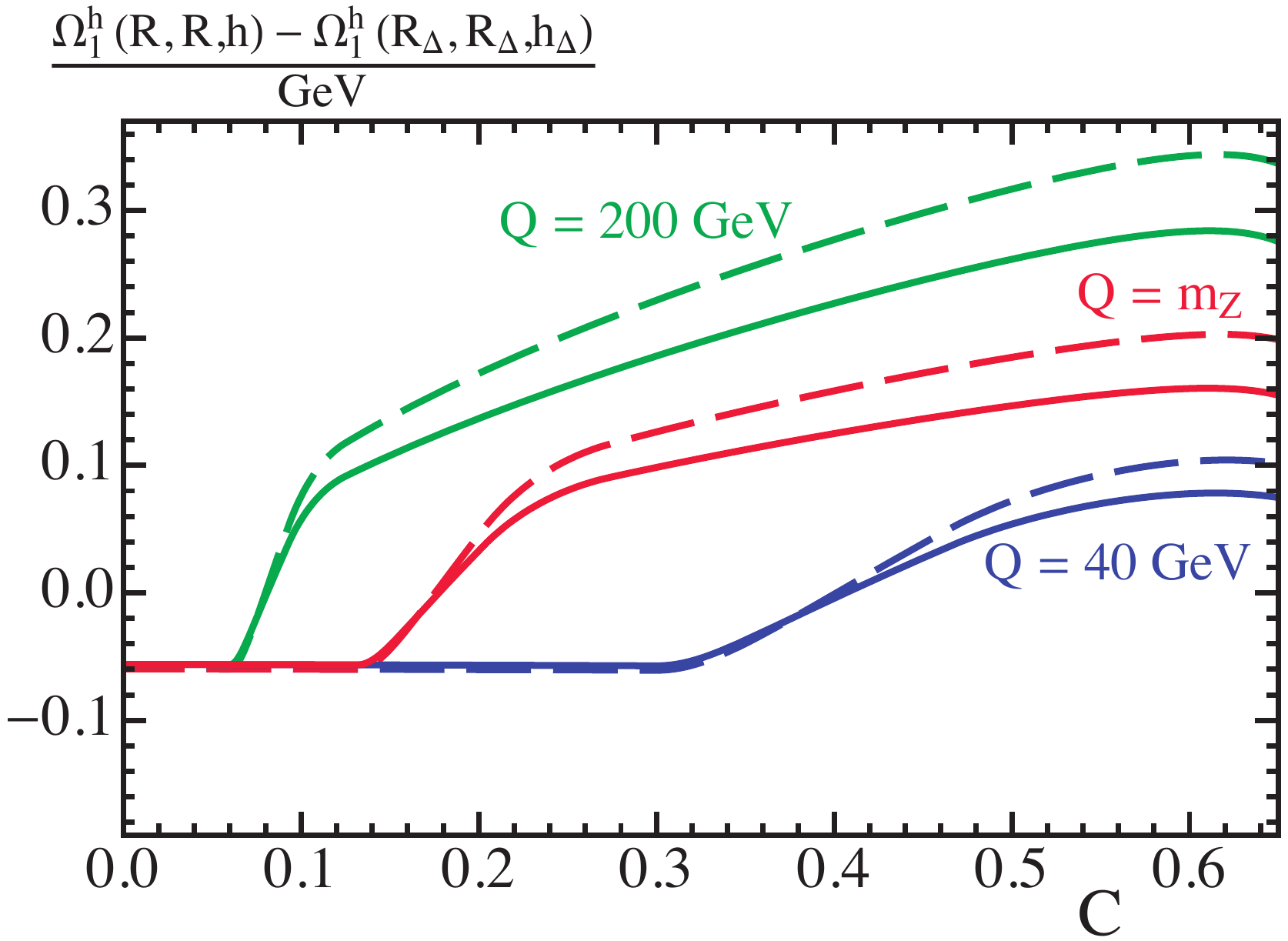}
\caption{Running of the short-distance power correction $\Omega_1^h(R,R,h)$ with respect to the reference value $\Omega_1^h(R_\Delta,R_\Delta,1)$. The scale $R$ is set to the default profile, and the scheme parameter $h$ is set to the function displayed in Eq.~(\ref{eq:arctan}). Red , blue,  and green correspond to center-of-mass energies of $91.2$, $40$, and $200$\,GeV, respectively. The solid lines do not include hadron-mass effects, whereas the dashed ones do.}
  \label{fig:Omega-run}
\end{center}
\end{figure}

The impact of R-evolution on the value of the hadronic parameter $\Omega_1(R,R)$ is shown in \Fig{fig:Omega-run} (with and without hadron-mass effects). Here we have set $\mu = R$ and used the default profile function for $R(C)$. We actually plot the hybrid scheme $\Omega_1^h(R,R,h)$ which accounts for a reasonable treatment of threshold effects at the shoulder $C=0.75$ and which is discussed in detail below in \App{ap:ArcTan}. The  effect of using the hybrid scheme rather than  the Rgap scheme is quite small in the region of this figure but does cause the bending over of the curves that is visible at $C=0.6$. Above the shoulder the hybrid scheme uses the $\msbar$ power correction, which has a flat behavior. 

When we introduce hadron-mass effects, it is necessary to extent these results to account for the value of $r$,
the transverse velocity. Due to the running in \Eq{eq:omega1-r-murunning}, the scheme change result in \Eq{eq:omegaschemechange} now becomes
\begin{align} \label{eq:rrunningschemechange}
g_C(r)\,\Omega_1^C(R,\mu,r) &= g_C(r)\,\overline{\Omega}_1(\mu,r) - \delta (R,\mu,r)\,,
\end{align}
where there is additional $\mu$ dependence from the hadron-mass induced running. Our scheme for $\delta$ now becomes
\begin{align} \label{eq:delta-scheme-littler}
\delta(R,\mu,r) &= \left[ \frac{\alpha_s(\mu)}{\alpha_s(R)}\right]^{\hat{\gamma}_1(r)}\!\delta(R,\mu)\,, 
\end{align}
and the $r$ dependence is encoded in the known one-loop anomalous dimension~\cite{Mateu:2012nk}
\begin{align} \label{eq:gamma1hatdef}
\hat{\gamma}_1(r) &= \frac{2\, C_{\!A}}{\beta_0} \ln (1 - r^2)\,.
\end{align}
With this $r$-dependent scheme change, we can once again derive the equations that govern the $R$ evolution and $\mu$
running for $\Omega_1^C$, which following Ref.~\cite{Hoang:2015zz} become
\begin{align} \label{eq:RandmuRGElittler}
 &  R \frac{\df}{\df R} \Big\{ \Big[ g_C(r)\, \Omega_1^C (R,\mu,r) 
   \big( \alpha_s(\mu) \big)^{-\hat\gamma_1(r)} \Big]_{\mu=R} \Big\}
 \\[4pt]
 &\quad
= -\, R\, \big(\alpha_s(R)\big)^{-\hat{\gamma}_1(r)} \sum_{n=0}^\infty \gamma_{n}^{R\,r}(r) \left( \frac{\alpha_s(R)}{4 \pi} \right)^{\!\!n+1}
\, , \nn \\
 & \mu \frac{\df}{\df\mu} \Big\{  g_C(r)\,\Omega_1^C (R,\mu,r) 
       \big( \alpha_s(\mu) \big)^{-\hat\gamma_1(r)}  \Big\}
\nn\\
 &\quad
 = 2\, R\, e^{\gamma_E} \big(\alpha_s(R)\big)^{-\hat{\gamma}_1(r)}
  \sum_{n=0}^\infty \Gamma_n^\text{cusp}
  \left( \frac{\alpha_s(\mu)}{4 \pi} \right)^{\!\!n+1}
   , \nn
\end{align}
for
\begin{align} \label{eq:gammaRlittler}
\gamma_0^{R\,r}(r) &= \gamma_0^R, \qquad 
\gamma_1^{R\,r}(r) = \gamma_1^R + 2 \beta_0 \hat{\gamma}_1(r)  \gamma_0^R  \,, \nn \\
\gamma_2^{R\,r}(r) &= \gamma_2^R + 2 \beta_0 \hat{\gamma}_1(r) (\gamma_1^R + 2 \beta_0 \gamma_0^R) \,.
\end{align}
Solving this set of equations gives \Eq{eq:omega1R-r-murunning}, with the $S_i^r$ given by
\begin{align} \label{eq:littlerSdef}
S^r_0(r) &= S_0\,,\qquad
S^r_1 (r) = S_1
+ \frac{\hat{\gamma}_1(r) \gamma_0^R}{2 \beta_0}\,, \\
S^r_2(r) & =  S_2 + (\gamma_1^R + 2 \beta_0 \gamma_0^R)\nn \\
 &\ \ \times  \bigg[ (1 + \hat b_1)\, \hat b_2 + \frac{\hat b_2^2 + \hat b_3}{2} \bigg]
\frac{\hat{\gamma}_1(r)}{(2 \beta_0)^2}\,. \nn
\end{align}
Values for the $\gamma_i^R$ anomalous dimensions were given above in Eq.~(\ref{eq:gammaR}).

\section{Gap Scheme in the Fixed-Order Region}
\label{ap:ArcTan}

In the fixed-order (or far-tail) region, there is no longer a hierarchy between the hard, jet, and soft scales, and one sets them equal to reproduce the fixed-order QCD predictions exactly. This is done through our profiles.
In this region the singular and non-singular terms are of similar size, and hence the factorization of the cross section in a hard factor and a convolution of jet and soft functions is no longer relevant.
The issue is further complicated by the analytic structure of the shoulder region, which is located in the fixed-order region and contains the integrable singularity at $C = 0.75$ which has its own logarithmic series (see Fig.~\ref{fig:component-plot} and the accompanying discussion there). Thus, in the far-tail region $C\gtrsim 0.75$, the structure of nonperturbative 
corrections is likely to differ from that of the shape function $F_C$ and thus is unknown at this time. Nevertheless, we do expect a smearing by a function whose width is $\sim\Lambda_{\rm QCD}$, and hence our smearing by $F_C$ is simply a proxy for a more detailed analysis in this region.  Since fits for $\alpha_s$ can be carried out with $C<0.75$, the treatment of this region, and the discussion below, are not relevant for predicting the shape in the fit region.  The region $C\ge 0.75$ does contribute when computing the total cross section from our resummed result, and this motivates us to use a cross section formula that still obtains realistic results in this region.

Due to the shoulder at $C=0.75$, the use of the infrared subtractions $\delta$ in this region can cause an unphysical behavior of the cross section. The reason for this is that these subtractions yield derivatives acting on the partonic cross section. In the shoulder region with the singular discontinuity starting at $\mathcal{O}(\alpha_s^2)$, these derivatives can cause an artificially enhanced singular behavior if the subtraction is not carefully defined in this region. If this is not done, the convolution with the shape function $F_C$ may be insufficient to achieve a smooth cross section near \mbox{$C\simeq 0.75$}. In Ref.~\cite{Catani:1998sf} it was shown that the singularities at $C = 0.75$ can be cured by including soft gluon resummation which makes the cross section smooth. However, this treatment does not resolve the question of the proper field theoretic nonperturbative function for this region, nor any accompanying infrared subtractions due to renormalons. 

To deal with the region $C\gtrsim 0.75$ we take an alternative approach, which is to implement a smooth transition between the Rgap scheme in the dijet region to $\msbar$ in the fixed-order region. This avoids any subtractions at $C=0.75$. To that end we define a new scheme that depends on a continuous parameter $h$ which takes values between $0$ and $1$ and that smoothly switches off the gap subtractions when we get near $C=0.75$. We start by rewriting \Eq{eq:deltasplitting} as
\begin{align} \label{eq:deltasplittingH}
\Delta_C  & =  \frac{3\pi}{2}[\,\bar{\Delta}^h(R,\mu,h) + h\,\delta(R,\mu)\,]\,.
\end{align}
This defines a hybrid short-distance scheme for $\Omega_1$ which we call $\Omega_1^h$,
\begin{align} \label{eq:omegaschemechangeH}
\Omega_1^h(R,\mu,h) &= \overline{\Omega}_1^C - 3\pi\,h\,\delta (R,\mu)\,,
\end{align}
which becomes the $\msbar$ scheme for $h = 0$ and the Rgap scheme for $h = 1$. One can easily derive RGE equations in $\mu$ and $R$, for $\bar{\Delta}^h$, which we write in the convenient form\,\footnote{The right-hand sides of the running
equations for the power correction $\Omega_1^h(R,\mu_S,h)$ are simply $3\pi$ times the right-hand sides of
Eqs.~(\ref{eq:running-h}) and (\ref{eq:change-h}).}
\begin{align}\label{eq:running-h}
R \frac{\df}{\df R}  \bar{\Delta}^h (R,R,h) & = -\,h\,R \, \gamma^R [\alpha_s(R)] \, , \\
\mu \frac{\df}{\df\mu}  \bar{\Delta}^h (R,\mu,h) &= 2\, R\,h\, e^{\gamma_E} \Gamma_{\rm cusp}[\alpha_s(\mu)] \, ,\nn
\end{align}
and a relation to switch from different $h$ schemes:
\begin{align}\label{eq:change-h}
\bar{\Delta}^h (R,\mu,h_1) - \bar{\Delta}^h (R,\mu,h_2) = (h_2 - h_1)\,\delta(R,\mu)\,.
\end{align}
The solution to these three equations is rather simple,
\begin{align}\label{eq:sol-run-h}
\bar{\Delta}^h (R,\mu,h) & = \bar{\Delta}^h (R_\Delta,\mu_\Delta,h_\Delta) - (h - h_\Delta)
\delta(R,\mu) \nn\\
&+ h_\Delta\,\Delta^{\rm diff}(R_\Delta,R,\mu_\Delta,\mu)\,,
\end{align}
where $\Delta^{\rm diff}$ has been defined in \Eq{eq:DeltaRevolution}. We choose to evolve first in $R$
and $\mu$ in the $h_\Delta$ scheme, where above the peak region [see Eq.~(\ref{eq:muRprofile})] there is only a single evolution since $\mu_S(C)=R(C)$. Close to the shoulder region \mbox{$C\sim 0.75$}, we then smoothly transform from the $h_\Delta$ scheme to the $h$ scheme. This implements the transition from the Rgap scheme with ${\cal O}(\Lambda_{\rm QCD})$ renormalon subtraction to the  $\msbar$ scheme where this renormalon is not subtracted. The procedure entails a 
residual dependence on $R$ in the region $C\gtrsim 0.75$ even once $h=0$, which comes from the fact that we are transforming from Rgap to $\msbar$ at the scale $R>R_\Delta$. This residual dependence leads to a somewhat smaller effect of the ${\cal O}(\Lambda_{\rm QCD})$ renormalon even though one employs $\overline{\Omega}_1$ in this region.

In the hybrid scheme described above, the first moment of the shape function reads
\begin{equation}
\!\!\int\! \!\df k \,k\,F_C(k) = \Omega_1^h(R_\Delta,\mu_\Delta,h_\Delta) \,-\,
3\pi\, \bar\Delta^h(R_\Delta,\mu_\Delta,h_\Delta)\,.
\end{equation}
For the practical implementation, we choose $h_\Delta = 1$ and thus identify $\Delta^h(R_\Delta,\mu_\Delta,1) = \Delta(R_\Delta,\mu_\Delta)$ as well as $\Omega_1^{h}(R_\Delta,\mu_\Delta,1)=\Omega_1(R_\Delta,\mu_\Delta)$. In the numerical codes, this amounts to inserting a factor $h$ in front of each $\delta$ and substituting each $\bar{\Delta}(R,\mu)$ of \Eq{eq:DeltaRevolution} appearing in any of the equations shown in the main text by $\bar{\Delta}^h(R,\mu,h)$ of \Eq{eq:sol-run-h}. 
Note again that all of the changes induced by the use of the hybrid scheme rather than the Rgap scheme only influence the shape of the C-parameter cross section for $C\gtrsim 0.75$.

\begin{figure*}[t!]
\begin{center}
\includegraphics[width=0.65\columnwidth]{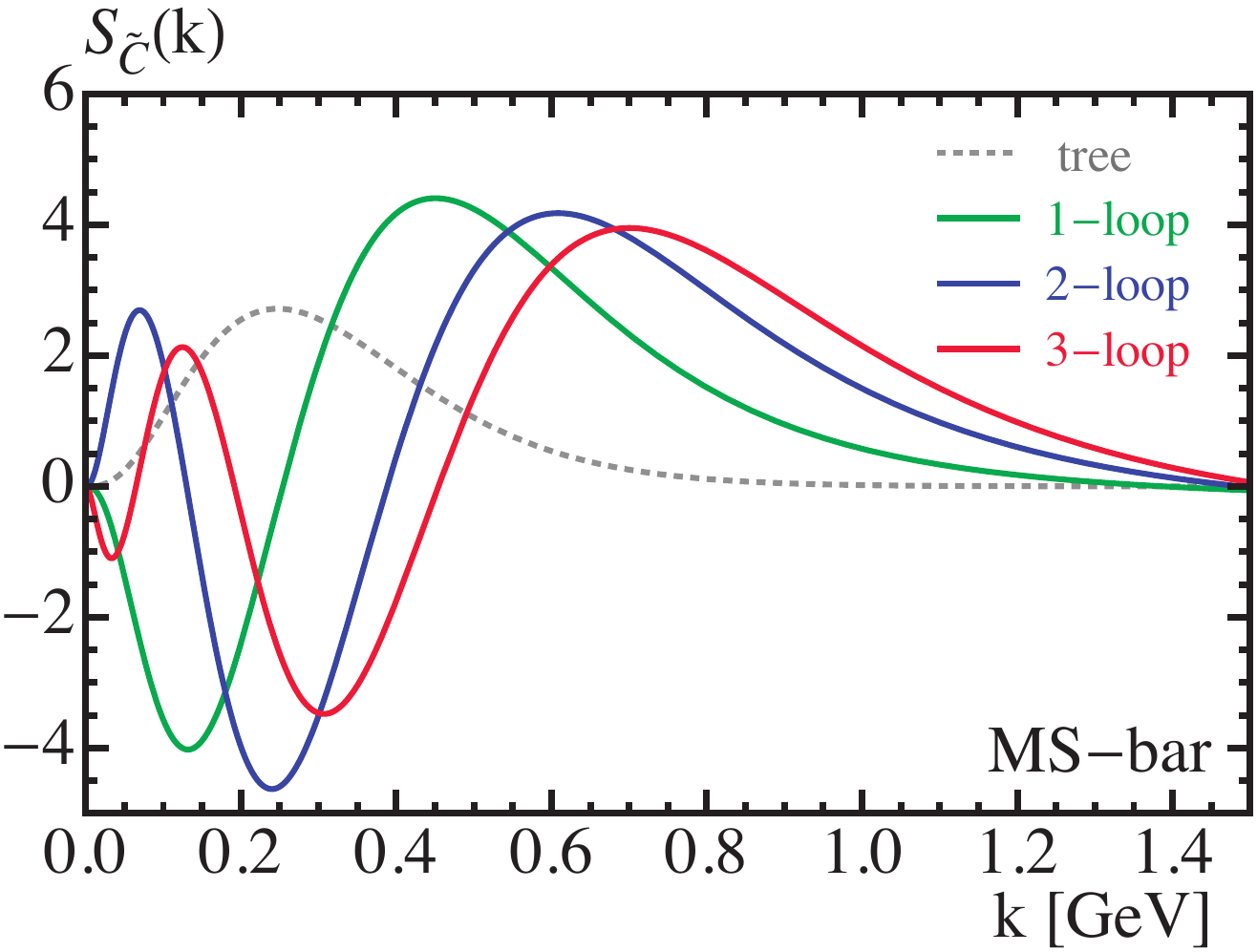}~
\includegraphics[width=0.67\columnwidth]{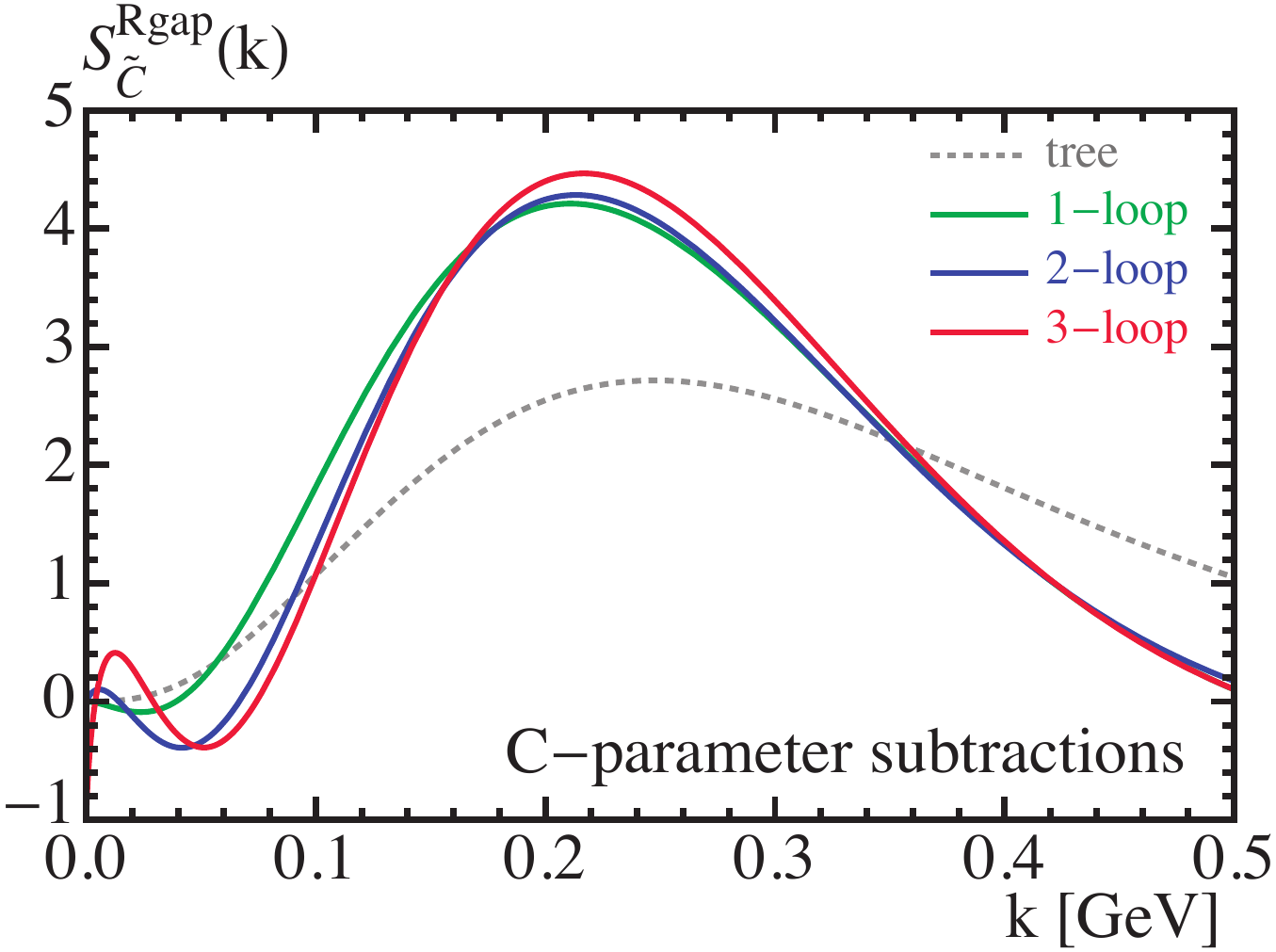}
\includegraphics[width=0.67\columnwidth]{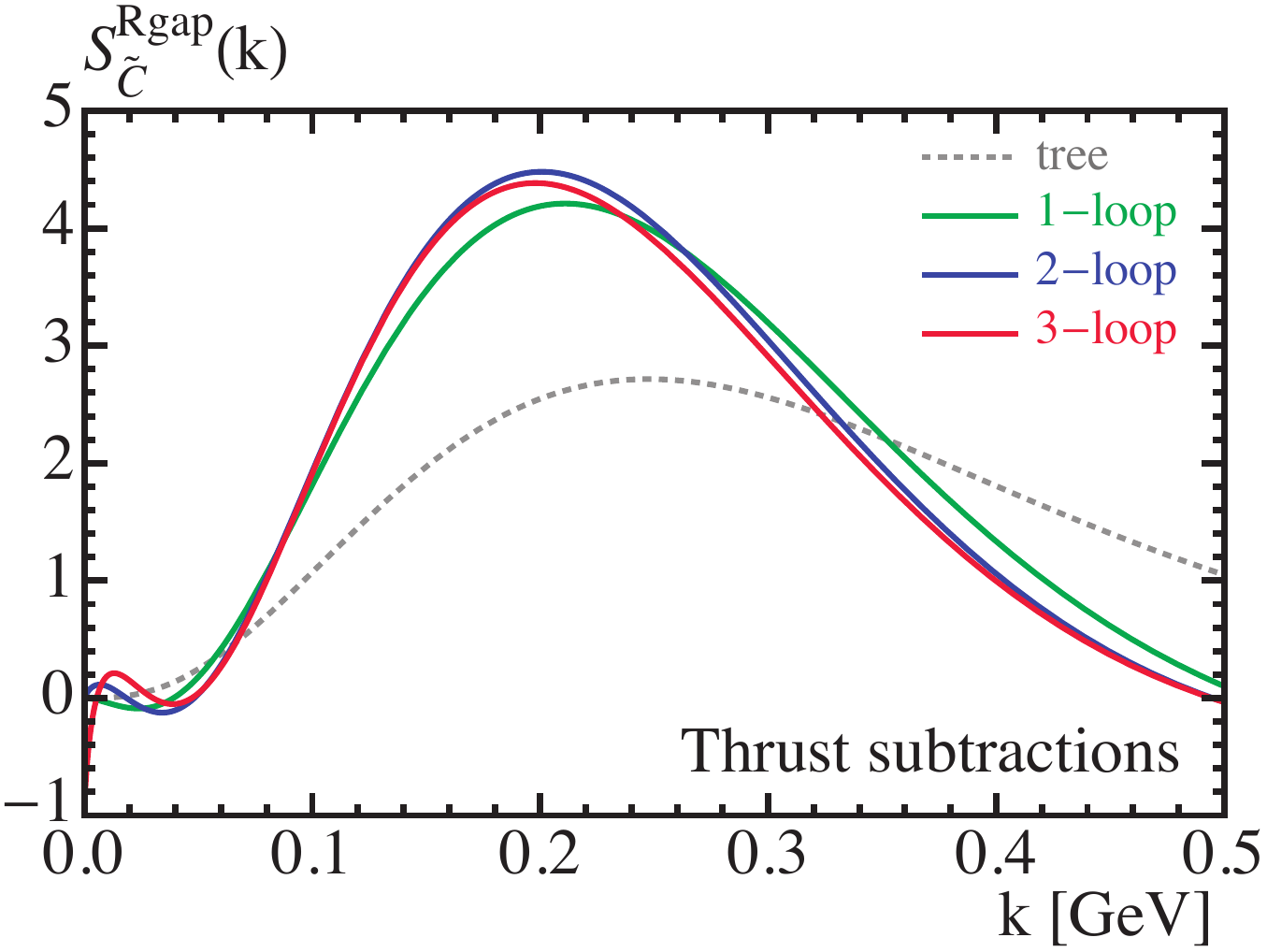}
\caption{Convolution $S = \hat S \otimes F$ combining the perturbative and nonperturbative C-parameter soft functions in the $\overline{\rm MS}$ scheme (left), Rgap with C-parameter subtractions (center), and Rgap with thrust subtractions (right). The results at 1-, 2- and 3-loops are shown in green, blue, and red, respectively, whereas the tree level result is depicted as a gray dotted line. We use $\mu = 1\,$GeV, $R=0.8\,$GeV, $\alpha_s(m_Z) = 0.1141$, and a fixed generic shape function whose first moment is $\Omega_1 = 0.33\,$\,GeV.}
  \label{fig:gap-effects-soft}
\end{center}
\end{figure*}

One can also easily extend the hybrid scheme to account for hadron-mass effects by defining
\begin{align} \label{eq:rrunningschemechangeh}
g_C(r)\,\Omega_1^{C,h}(R,\mu,h,r) &= g_C(r)\,\overline{\Omega}_1^C(\mu,r) - h\,\delta (R,\mu,r)\,,
\end{align}
and the evolution and scheme-transformation equations simply read
\begin{align} \label{eq:RandmuRGElittlerh}
& R \frac{\df}{\df R} \Big\{\Big[g_C(r)\big(\alpha_s(\mu)\big)^{-\hat{\gamma}_1(r)}\Omega_1^{C,h} (R,\mu,h,r)\Big]_{\mu=R}\Big\} \\
&= -\,h \,R \big(\alpha_s(R)\big)^{-\hat{\gamma}_1(r)} \gamma_r^{R}[\alpha_s(R),r] \, , \nn \\
&\mu \frac{\df}{\df\mu} \Big\{g_C(r)\big(\alpha_s(\mu)\big)^{-\hat{\gamma}_1(r)}\Omega_1^{C,h} (R,\mu,h,r)\Big\}\nn\\
&= 2 \, h\, R\, e^{\gamma_E} \big(\alpha_s(R)\big)^{-\hat{\gamma}_1(r)} \Gamma_{\rm cusp}[\alpha_s(\mu)] \, , \nn\\
&g_C(r)\big[\,\Omega_1^{C,h} (R,\mu,h_1,r) - \Omega_1^{C,h} (R,\mu,h_2,r)\big]\nn \\
& = (h_2 - h_1)\,\delta(R,\mu,r)\,.\nn
\end{align}
The solution to these equations is again simple:
\begin{align}\label{eq:sol-run-Omega-h}
&g_C(r)\,\Omega_1^{C,h} (R,\mu,h,r)  = \\
&g_C(r)\!\left[ \frac{\alpha_s(\mu)}{\alpha_s(\mu_\Delta)} \right]^{\hat{\gamma}_1(r)}\!
\Omega_1^{C,h}(R_\Delta, \mu_\Delta,h_\Delta,r)\nn \\
&\!\!\!- (h_\Delta - h)\, \delta(R_\Delta,\mu_\Delta,r)
+ h_\Delta\,\Delta^{\rm diff}(R_\Delta,R,\mu_\Delta,\mu,r)\,.\nn
\end{align}
For phenomenological analyses that consider $C\gtrsim 0.75$ or integrate over $C$ to compute a normalization, we must specify $h=h(C)$. It must be a function of $C$ which smoothly interpolates between the values
$1$ and $0$ as one transitions from the resummation to the fixed-order region near \mbox{$C\sim 0.75$}. To achieve this we use the following simple form:
\begin{align}\label{eq:arctan}
h(C) & = \frac{1}{2} - \frac{1}{\pi} \arctan\,[\,\eta\,(C - C_0)\,]\,,\\
C_0 & = 0.7\,,\; \eta = 30\,.\nn
\end{align}

\section{Rgap Scheme based on the C-Parameter Soft Function}
\label{ap:subtractionchoice}
As an alternative to the Rgap scheme subtraction function used in this work and defined in \Eq{eq:delta-scheme}, one may employ the analog relation based on the C-parameter partonic soft function:
\begin{equation} \label{eq:delta-C-scheme}
\delta_{\widetilde{C}}(R,\mu) = R\, e^{\gamma_E} \frac{\df}{\df \ln(ix)}
\big[ \ln S^{\text{part}}_{\widetilde C}(x,\mu) \big]_{x = (i R e^{\gamma_E})^{-1}}.
\end{equation}
In this scheme the analogue of \Eq{eq:soft-nonperturbative-subtract} reads
\begin{align} \label{eq:soft-nonperturbative-subtractC}
\!\!\!S_C(k,\mu) & = \!\!\int \!\df k' \,e^{-\, 6\,\delta_{\widetilde C} \frac{\partial}{\partial k}}
\hat S_C(k-k',\mu) F_C(k' - 6\bar{\Delta}_{\widetilde{C}})\,,
\end{align}
with 
\begin{align} \label{eq:deltasplittingC}
\Delta_{\widetilde C}  & =  \,\bar{\Delta}_{\widetilde{C}}(R,\mu) + \delta_{\widetilde{C}}(R,\mu)\,\,.
\end{align}
Here the subtraction function $\delta_{\widetilde{C}}$ can be written as
\begin{align} \label{eq:deltaseriesC}
 \delta_{\widetilde{C}}(R,\mu) =
 R \,e^{\gamma_E} \sum_{i=1}^\infty \alpha_s^i(\mu)\, \delta^i_{\widetilde{C}}(R,\mu)\,,
\end{align}
where the coefficients for five light flavors read
\begin{align} \label{eq:d123C}
\delta_{\widetilde{C}}^1(R,\mu) &= -\,1.69765\, L_R \,, \nn \\
\delta_{\widetilde{C}}^2(R,\mu) &= 0.539295 - 0.933259\, L_R - 1.03573\, L_R^2 \,, \nn \\
\delta_{\widetilde{C}}^3(R,\mu) &= 0.493255 + 0.0309077\, s_2^{\widetilde C} + 0.833905\, L_R
  \nn\\&\quad -\,1.55444\, L_R^2 - 0.842522\, L_R^3 \,,
\end{align}
for $L_R = \ln(\mu/R)$.

Figure~\ref{fig:gap-effects-soft} shows the effect of the renormalon subtractions on the soft function $S_C(k,\mu)$ 
from \Eqs{eq:soft-nonperturbative-subtract}{eq:soft-nonperturbative-subtractC} [we will refer to using \Eq{eq:soft-nonperturbative-subtract} as thrust subtractions and to using \Eq{eq:soft-nonperturbative-subtractC} as C-parameter subtractions], which are compared with the result in the $\overline{\rm MS}$ scheme with any subtractions. The key thing to consider is the stability of the soft function when higher orders in perturbation theory are included, illustrated by the green, blue, and red curves at 1-, 2-, and 3-loop orders. In the leftmost panel of Fig.~\ref{fig:gap-effects-soft}, we show the C-parameter soft function in the $\overline{\rm MS}$ scheme. Here the presence of the $\Lambda_{\rm QCD}$ renormalon is apparent from the shifting of the soft function to the right as we increase the perturbative order. The $\overline{\rm MS}$ result also exhibits  a large negative dip at small momentum, which makes predictions for the cross section at small $C$ inaccurate in this scheme. With either the C-parameter subtractions (middle panel) or the thrust subtractions (rightmost panel), one achieves significantly better convergence for the soft function and alleviates most of the negative dip.  Making an even closer comparison of the C-parameter and thrust subtraction results, it becomes evident that the thrust subtractions exhibit better convergence near the peak (comparing the difference between the blue and red lines in the two panels) and also more completely remove the negative dip at small momenta. (Similar conclusions hold for the thrust soft function, where again thrust subtractions are preferred.) This improvement for the thrust subtractions can be traced back to the fact that the sign of the non-logarithmic 2-loop term in the C-parameter subtractions in \Eq{eq:d123C} is positive, which is opposite to the sign of the renormalon. In the resummation region,  $R(\tau)=\mu_S(\tau)$, so $L_R\to 0$, and numerically the subtraction goes in the opposite direction to the renormalon in this scheme at 2-loops. This term has an impact even when the logarithmic terms are active in the small $C$ nonperturbative region, which we can see by taking $R=R_0$ and $\mu_S=\mu_0$.  For the thrust subtractions, this gives $ \pi/2 \{\delta^1, \delta^2, \delta^3\} = \{-\,0.603, -\,0.743, -\,1.621\}$, whereas for the C-parameter subtractions, we have $\{\delta_{\widetilde C}^1, \delta_{\widetilde C}^2, \delta_{\widetilde C}^3\} = \{-\,0.767, -\,0.094, -\,0.861\}$. The small value for this \mbox{2-loop} C-parameter subtraction coefficient is due to the positive constant term.

\bibliography{../thrust3}
\end{document}